\def\figsize{9.5cm}

\def\smtopskip{-1.2cm}
\def\smbotskip{-1.0cm}
\def\rn{}
\def\nn#1 #2{#2. #1}				% Name with 1 initial
\def\nnn#1 #2 #3{#2. #3. #1}			% Name with 2 initials
\def\nnnn#1 #2 #3 #4{#2. #3. #4 #1}		% Name with 3 initials
\def\nnnnn#1 #2 #3 #4 #5{#2. #3. #4 #5. #1}	% Name with 4 initials
\def\dualand{ and\hbox{ }}				
\def\multiand{, and\hbox{ }}				
\def\rf#1;#2;#3;#4;#5 {{\frenchspacing\par\rn#1, #3 {\bf #4}, #5 (#2). \par}}
\def\rg#1;#2;#3;#4;#5;#6 {{\frenchspacing\par\rn#1, #3 {\bf #4}, #5 (#2). \par}}
\def\rfbook#1;#2;#3;#4;#5 {{\frenchspacing\par\rn#1, {\it #3} (#5, #4, #2).\par}}
\def\rfprep#1;#2;#3 {{\par\frenchspacing\rn#1, #3 (#2).\par}}
\def\rfproc#1;#2;#3;#4;#5;#6 {{\frenchspacing\par\rn#1 #2, in {\it #3}, ed. #4 (#5: #6)\par}}
\def\rfprocp#1;#2;#3;#4;#5;#6;#7 {{\frenchspacing\par\rn#1 #2, in {\it #3}, ed. #4 (#5: #6), p#7\par}}
\def\rg#1;#2;#3;#4;#5;#6 {\par\rn#1 #2, {\it #3}, {\bf #4}, #5 (``#6'') \par}
\def\rf#1;#2;#3;#4;#5 {\par\rn#1, {\it #3}, {\bf #4}, #5 (#2)\par}
\def\rfbook#1;#2;#3;#4;#5 {{\frenchspacing\par\rn#1, {\it #3} (#4: #5, #2)\par}}
\def\rfproc#1;#2;#3;#4;#5;#6 {{\frenchspacing\par\rn#1 #2, in {\it #3}, ed. #4 (#5: #6)\par}}
\def\rfprocp#1;#2;#3;#4;#5;#6;#7 {{\frenchspacing\par\rn#1 #2, in {\it #3}, ed. #4 (#5: #6), p#7\par}}
\def\rfprep#1;#2;#3  {{\par\rn#1, #3, #2\par}}
\def\rfprepp#1;#2;#3 {{\par\rn#1 #2, #3\par}}

\def\Mpc{{\rm Mpc}}

\def\expec#1{\langle#1\rangle}

\def\etal{{\frenchspacing\it et al.}}
\def\ie{{\frenchspacing\it i.e.}}
\def\eg{{\frenchspacing\it e.g.}}
\def\etc{{\frenchspacing\it etc.}}
\def\beq#1{\begin{equation}\label{#1}}
\def\eeq{\end{equation}}
\def\beqa#1{\begin{eqnarray}\label{#1}}
\def\eeqa{\end{eqnarray}}
\def\eq#1{equation~(\ref{#1})}

\def\fig#1{Figure~\ref{#1}}
\def\Fig#1{Figure~\ref{#1}}

\def\ParameterTable{1}
\def\WMAPtable{2}
\def\SDSStable{3}
\def\AddinfoTable{4}
\def\ClusterLensTable{5}
\def\InflationTable{6}
\def\SystematicsTable{7}
\def\ChainTable{8}

\def\Sec#1{Section~\ref{#1}}
\def\Sec#1{Section~\ref{#1}}

\def\spose#1{\hbox to 0pt{#1\hss}}
\def\simlt{\mathrel{\spose{\lower 3pt\hbox{$\mathchar"218$}}
     \raise 2.0pt\hbox{$\mathchar"13C$}}}
\def\simgt{\mathrel{\spose{\lower 3pt\hbox{$\mathchar"218$}}
     \raise 2.0pt\hbox{$\mathchar"13E$}}}
\def\simpropto{\mathrel{\spose{\lower 3pt\hbox{$\mathchar"218$}}
     \raise 2.0pt\hbox{$\propto$}}}

\def\ed{\end{document}}

\def\Ob{\Omega_b}

\def\Ok{\Omega_k}
\def\Ol{\Omega_\Lambda}
\def\Om{\Omega_m}
\def\Od{\Omega_d}
\def\On{\Omega_\nu}
\def\ob{\omega_b}

\def\od{\omega_d}

\def\om{\omega_{\rm m}}
\def\on{\omega_\nu}

\def\fn{f_\nu}
\def\ns{{n_s}}
\def\nt{{n_t}}
\def\al{\alpha}
\def\Ot{\Omega_{\rm tot}}
\def\As{A_s}
\def\At{A_t}
\def\Ap{A_p}
\def\Apivot{A_*}
\def\zion{z_{\rm ion}}

\def\age{t_0}
\def\Th{\Theta_s}
\def\Mnu{M_\nu}

\def\beq#1{\begin{equation}\label{#1}}
\def\eeq{\end{equation}}
\def\beqa#1{\begin{eqnarray}\label{#1}}
\def\eeqa{\end{eqnarray}}
\def\eq#1{equation~(\ref{#1})}

\def\w{w}

\def\p{{\bf p}}

\def\d{{\bf d}}

\def\C{{\bf C}}

\def\I{{\bf I}}
\def\R{{\bf R}}
\def\LL{{\bf\Lambda}}

\def\L{{\cal L}}

\def\R{{\bf R}}

\def\zero{{\bf 0}}

\def\kmax{k_{\rm max}}

\def\l{\ell}

\def\ignore#1{}
\def\simless{\mathbin{\lower 3pt\hbox
        {$\,\rlap{\raise 5pt\hbox{$\char'074$}}\mathchar"7218\,$}}} % < or of order
\def\simgreat{\mathbin{\lower 3pt\hbox
        {$\,\rlap{\raise 5pt\hbox{$\char'076$}}\mathchar"7218\,$}}} % > or of order

\documentclass[twocolumn,amsmath,nofootinbib]{revtex4} % For astro-ph
\usepackage{url}
\begin{document}
% Include Rokicki's epsf.sty file for Encapsulated PostScript graphics
\input{epsf.sty}

% FOR APJ:
%\documentstyle[emulateapj,danonecolfloat]{article}
%%\documentstyle[aasms4]{article}
%\def\NoApjSectionMarkInTitle#1{#1.\ }
%%\draft
%\begin{document}
%\twocolumn[%%% Begin front material

%%%%%%%%%%%%%%%%%%%%%%%%%%%%%

%\tighten
%\eqsecnum
%\received{4 August 1988}
%\accepted{23 September 1988}
%\journalid{337}{15 January 1989}
%\articleid{11}{14}

\def\penn{1}
\def\mit{2}
\def\pton{3}
\def\nyu{4}
\def\lanl{5}
\def\fnal{6}
\def\chicago{7}
\def\osu{8}
\def\tucson{9}
\def\drexel{10}
\def\hopkins{11}
\def\pitt{12}
\def\cmu{13}
\def\hawaii{14}
\def\apo{15}
\def\barcelona{15}
\def\sussex{16}
\def\tokyo{17}
\def\flagstaff{18}
\def\michigan{19}
\def\rochester{20}
\def\psu{21}
\def\efi{22}

\def\affilmrk#1{$^{#1}$}
\def\affilmk#1#2{$^{#1}$#2;}

%\submitted{Submitted to ApJ July 2 2001, accepted January 15 2002}
%\submitted{\today. To be submitted to ApJ.}
%\submitted{Submitted to ApJL September 16; accepted February 2}

\title{Cosmological parameters from SDSS and WMAP}

\author{
Max Tegmark\affilmrk{\penn,\mit}, 
Michael A. Strauss\affilmrk{\pton}, 
Michael R. Blanton\affilmrk{\nyu},
Kevork Abazajian\affilmrk{\lanl},
Scott Dodelson\affilmrk{\fnal,\chicago},
Havard Sandvik\affilmrk{\penn},
%H{\aa}vard Sandvik\affilmrk{\penn},
Xiaomin Wang\affilmrk{\penn},
David H. Weinberg\affilmrk{\osu}, 
Idit Zehavi\affilmrk{\tucson},
%%%%%%%%%%%%%%%%%%%%%%%%%%%%%%%%%%%%%%%%%%%%%%%%%%%%%%%%%%%
Neta A. Bahcall\affilmrk{\pton}, 
Fiona Hoyle\affilmrk{\drexel},
David Schlegel\affilmrk{\pton}, 
Roman Scoccimarro\affilmrk{\nyu}, 
Michael S. Vogeley\affilmrk{\drexel},
Andreas Berlind\affilmrk{\chicago},
Tam\'as Budavari\affilmrk{\hopkins}, 
Andrew Connolly\affilmrk{\pitt},
Daniel J. Eisenstein\affilmrk{\tucson},
Douglas Finkbeiner\affilmrk{\pton}, 
Joshua A. Frieman\affilmrk{\chicago,\fnal},
James E. Gunn\affilmrk{\pton}, 
%Andrew J. S. Hamilton\affilmrk{\colorado}, 
Lam Hui\affilmrk{\fnal}, 
Bhuvnesh Jain\affilmrk{\penn},
David Johnston\affilmrk{\chicago,\fnal}, 
Stephen Kent\affilmrk{\fnal},
Huan Lin\affilmrk{\fnal},
Reiko Nakajima\affilmrk{\penn}, 
Robert C. Nichol\affilmrk{\cmu}, 
Jeremiah P. Ostriker\affilmrk{\pton},
Adrian Pope\affilmrk{\hopkins}, 
Ryan Scranton\affilmrk{\pitt},
Uro\v s Seljak\affilmrk{\pton},
Ravi K. Sheth\affilmrk{\pitt}, 
Albert Stebbins\affilmrk{\fnal},
Alexander S. Szalay\affilmrk{\hopkins},
Istv\'an Szapudi\affilmrk{\hawaii}, 
%Licia Verde\affilmrk{\pton},
Yongzhong Xu\affilmrk{\lanl}, 
%Matias Zaldarriaga\affilmrk{\harvard},
%%%%%%%%%%%%%%%%%%%%%%%%%%%%%%%%%%%%%%%%%%%%%%%%%%%%%%%%%%%%
James Annis\affilmrk{\fnal}, 
J. Brinkmann\affilmrk{\apo},
Scott Burles\affilmrk{\mit},
Francisco J. Castander\affilmrk{\barcelona},
Istvan Csabai\affilmrk{\hopkins},
Jon Loveday\affilmrk{\sussex}, 
Mamoru Doi\affilmrk{\tokyo},  
Masataka Fukugita\affilmrk{\tokyo},
Bruce Gillespie\affilmrk{\apo},
Greg Hennessy\affilmrk{\flagstaff},
David W. Hogg\affilmrk{\nyu},
\v Zeljko Ivezi\'c\affilmrk{\pton},
Gillian R. Knapp\affilmrk{\pton},
Don Q. Lamb\affilmrk{\chicago},
Brian C. Lee\affilmrk{\fnal},
Robert H. Lupton\affilmrk{\pton},
Timothy A. McKay\affilmrk{\michigan},
Peter Kunszt\affilmrk{\hopkins},
Jeffrey A. Munn\affilmrk{\flagstaff}, 
Liam O'Connell\affilmrk{\sussex},
John Peoples\affilmrk{\fnal}, 
Jeffrey R. Pier\affilmrk{\flagstaff},
Michael Richmond\affilmrk{\rochester},
Constance Rockosi\affilmrk{\chicago}, 
Donald P. Schneider\affilmrk{\psu}, 
Christopher Stoughton\affilmrk{\fnal}, 
Douglas L. Tucker\affilmrk{\fnal},
Daniel E. Vanden Berk\affilmrk{\pitt},
Brian Yanny\affilmrk{\fnal}, 
Donald G. York\affilmrk{\chicago,\efi}
%for the SDSS Collaboration
}
%%\altaffiltext{1}{Based on observations obtained with the Sloan Digital Sky Survey}
\address{
%{\it 
\parshape 1 -3cm 24cm
%\parbox[l]{18cm}{
\affilmk{\penn}{Department of Physics, University of Pennsylvania,
Philadelphia, PA 19104, USA}
\affilmk{\mit}{Dept. of Physics, Massachusetts Institute of Technology, 
Cambridge, MA 02139}
\affilmk{\nyu}{Center for Cosmology and Particle Physics,
Department of Physics, New York University, 4 Washington
Place, New York, NY 10003}
\affilmk{\pton}{Princeton University Observatory, Princeton, NJ 08544,
USA}
\affilmk{\drexel}{Department of Physics, Drexel University, Philadelphia,
PA
19104, USA}
\affilmk{\osu}{Department of Astronomy, Ohio State University, 
Columbus, OH 43210, USA}
\affilmk{\fnal}{Fermi National Accelerator Laboratory, P.O. Box 500, Batavia,
IL 60510, USA}
\affilmk{\chicago}{Center for Cosmological Physics and Department of Astronomy \& Astrophysics, University of
Chicago, Chicago, IL 60637, USA}
\affilmk{\hopkins}{Department of Physics and Astronomy, The Johns Hopkins
University, 3701 San Martin Drive, Baltimore, MD 21218, USA}
\affilmk{\pitt}{University of Pittsburgh, Department of Physics and
Astronomy, 3941 O'Hara Street, Pittsburgh, PA 15260, USA}
%\affilmk{\columbia}{Department of Physics, Columbia University, New York, NY
%10027, USA}
\affilmk{\tucson}{Department of Astronomy, University of Arizona, 
Tucson, AZ 85721, USA}
%\affilmk{\colorado}{JILA and Dept.~of Astrophysical and Planetary Sciences, 
%U. Colorado, Boulder, CO 80309, USA}
\affilmk{\cmu}{Department of Physics, 5000 Forbes Avenue, Carnegie
Mellon
University, Pittsburgh, PA 15213, USA}
\affilmk{\hawaii}{Institute for Astronomy, University of Hawaii, 2680
Woodlawn Drive, Honolulu, HI 96822, USA}
\affilmk{\apo}{Apache Point Observatory, 2001 Apache Point Rd, 
Sunspot, NM 88349-0059, USA}
\affilmk{\barcelona}{Institut d'Estudis Espacials de Catalunya/CSIC, Gran Capita 2-4, 
08034 Barcelona, Spain}
\affilmk{\sussex}{Sussex Astronomy Centre, University of Sussex, Falmer,
Brighton BN1 9QJ, UK}
\affilmk{\tokyo}{Institute of Astronomy, Univ. of Tokyo, Kashiwa 277-8582, Japan}
%\affilmk{\tokyo}{Inst. for Cosmic Ray Research, Univ. of Tokyo, Kashiwa 277-8582, Japan}
\affilmk{\flagstaff}{U.S. Naval Observatory, 
Flagstaff Station, Flagstaff, AZ 86002-1149, USA}
\affilmk{\michigan}{Dept. of Physics, Univ. of Michigan, 
%\affilmk{\harvard}{Harvard-Smithsonian Center for Astrophysics,
%60 Garden Street, MS46, Cambridge, MA 02138}
Ann Arbor, MI 48109-1120, USA}
\affilmk{\rochester}{Physics Dept., Rochester Inst. of Technology, 
1 Lomb Memorial Dr, Rochester, NY 14623, USA}
\affilmk{\psu}{Dept. of Astronomy and Astrophysics, Pennsylvania State University, 
University Park, PA 16802, USA}
\affilmk{\efi}{Enrico Fermi Institute, University of
Chicago, Chicago, IL 60637, USA}
%\affilmk{\ias}{Institute for Advanced Study, School of Natural
%Sciences, Olden Lane, Princeton, NJ 08540, USA}
\affilmk{\lanl}{Theoretical Division, MS B285, Los Alamos National Laboratory,
Los Alamos, New Mexico 87545, USA}
}

%\date{\today. To be submitted to Phys. Rev. D.}
\date{Submitted to Phys. Rev. D October 27 2003, revised December 22, accepted January 4.}

\begin{abstract}
We measure cosmological parameters using the three-dimensional power spectrum $P(k)$
from over 200,000 galaxies in the Sloan Digital Sky Survey (SDSS) in combination with WMAP and other data.
Our results are consistent with a 
``vanilla'' flat adiabatic $\Lambda$CDM model without 
tilt ($\ns=1$), running tilt, tensor modes or massive neutrinos.
% no tilt (\ie, $\ns=1$) or running tilt. 
Adding SDSS information more than halves the WMAP-only error bars on some parameters,
tightening $1\sigma$ constraints on the Hubble parameter from
$h\approx 0.74^{+0.18}_{-0.07}$
to
$h\approx 0.70^{+0.04}_{-0.03}$,
on the matter density from
$\Om\approx 0.25\pm 0.10$
to
$\Om\approx 0.30\pm 0.04$ $(1\sigma)$
and on neutrino masses from $<11\>$eV to $<0.6\>$eV (95\%).  
SDSS helps even more when dropping prior assumptions about curvature, neutrinos, tensor modes
and the equation of state.  
%Our results are in
%substantial agreement with the joint analysis of WMAP and the 2dF
%Galaxy Redshift Survey, despite using independent redshift survey data
%and analysis techniques.  
Our results are in
substantial agreement with the joint analysis of WMAP and the 2dF
Galaxy Redshift Survey, which is an impressive consistency check with 
independent redshift survey data and analysis techniques. 
In this paper, 
we place particular emphasis on clarifying the physical origin of the constraints,
\ie, what we do and do not know when using different data sets and
prior assumptions.  
For instance, dropping the assumption that space is perfectly flat,
the WMAP-only constraint on 
%the Hubble parameter are tightened 
%from $h\approx 0.50^{+0.16}_{-0.13}$ to $h\approx 0.60^{+0.09}_{-0.06}$ $(1\sigma)$ by adding 
%SDSS and SN Ia data, and 
the measured age of the Universe tightens from 
$t_0\approx 16.3^{+2.3}_{-1.8}$ Gyr
to
$t_0\approx 14.1^{+1.0}_{-0.9}$ Gyr by adding SDSS and SN Ia data. 
% with an additional reionization
%prior.
%MAS: These seem like random and obscure examples to pull out of
%tables 2-4; everyone knows from HST that h = 0.60 is not a very good
%number... 
% MT: OK, LET'S JUST KEEP THE t0 EXAMPLE THEN.
%
%Our cosmological parameters agree with those from the WMAP team, but with 
%a slightly smaller Hubble parameter ($h\approx 0.70^{+0.04}_{-0.03}$) and 
%a slightly larger matter density ($\Om=0.30\pm 0.04$) at $1\sigma$. 
%We place particular emphasis on clarifying the physical origin of the constraints,
%\ie, what we do and do not know when using different data sets and prior assumptions.
%For instance, dropping the assumption that space is perfectly flat,
%the WMAP constraints on the Hubble parameter get tightened 
%from $h\approx 0.50^{+0.16}_{-0.13}$ to $h\approx 0.60^{+0.09}_{-0.06}$ $(1\sigma)$ by adding 
%SDSS and SN Ia data.
%
Including tensors, running tilt, neutrino mass and equation of state in
the list of free parameters, many constraints are still quite weak,
but future cosmological measurements from SDSS and other sources
should allow these to be substantially tightened. 
%so this paper should be viewed not as the final word on SDSS precision
%cosmology,  
%merely as a humble beginning.
\end{abstract}

% ]%%% End front material

% PACS, the Physics and Astronomy Classification Scheme
% http://www.aip.org/pacs/pacs03/pacs0390.html
%\keywords{cosmic microwave background  -- diffuse radiation}
%%-- radiation mechanisms: thermal and non-thermal
%%-- methods: data analysis}
%]%%% End front material
  
\maketitle

%%%%%%%%%%%%%%%%%%%%%%%%%%%%%%%%%%%%%%%%%%%%%%
%%%%%%%%%%%%%%%%%%%%%%%%%%%%%%%%%%%%%%%%%%%%%%

% Can't redefine this until after Zeljko Ivezic's name :)
\def\v{{\bf v}}

\setcounter{footnote}{0}

\section{Introduction}

%Although everybody agrees that 
%the spectacular recent cosmic microwave background (CMB) measurements from the 
%WMAP satellite 
%\cite{Bennett03,Hinshaw03,kogut03,Page3-03,Peiris03,Spergel03,Verde03}
%and other experiments have opened a
%new chapter in cosmology, there is less agreement about how 
%accurate and reliable our constraints on cosmological models really are.
%In their assessments of precision cosmology, 
%recent papers span the spectrum from ``it's all over" to  ``not quite
%yet" \cite{Bridle03}.

The spectacular recent cosmic microwave background (CMB) measurements from the 
Wilkinson Microwave Anisotropy Probe (WMAP)
\cite{Bennett03,Hinshaw03,kogut03,Page3-03,Peiris03,Spergel03,Verde03}
and other experiments have opened a
new chapter in cosmology.  However, as emphasized, e.g., in
\cite{Spergel03} and \cite{Bridle03}, measurements of CMB fluctuations
by themselves do not constrain all cosmological parameters due to a
variety of degeneracies in parameter space.  These degeneracies can be
removed, or at least mitigated, by applying a variety of priors or
constraints on parameters, and combining the CMB data with other
cosmological measures, such as the galaxy power spectrum.  The WMAP
analysis in particular made use of the power spectrum measured from the Two
Degree Field Galaxy Redshift Survey (2dFGRS)
\cite{Colless01,Colless03,Percival01}.  

The approach of the WMAP team \cite{Spergel03,Verde03}, was to apply
Ockham's razor, and ask what minimal model (i.e., with the smallest
number of free parameters) is consistent with the data.  In doing so,
they used reasonable assumptions about theoretical priors and 
external data sets, which allowed them to obtain quite small error
bars on cosmological parameters. 
The opposite approach is to treat all basic cosmological parameters as
free parameters and  constrain them with data using minimal
assumptions. The latter was done both in WMAP accuracy forecasts based on information theory
\cite{parameters,ZSS97,Wang99,parameters2,EfstathiouBond99} and in many pre-WMAP analyses involving 
up to 11 cosmological parameters.
This work showed that because of physically well-understood parameter degeneracies,
accurate constraints on most parameters could only be obtained by 
combining CMB measurements with something else.  
Bridle, Lahav, Ostriker and Steinhardt \cite{Bridle03}
argue that
in some cases (notably involving the matter density $\Om$), you get quite
different answers depending on your choice of ``something else'', implying that the
small formal error bars must be taken with a grain of salt.
For instance, the WMAP team \cite{Spergel03} quote $\Om=0.27\pm 0.04$ from combining 
WMAP with galaxy clustering from the 2dFGRS
and assumptions about spatial
flatness, negligible tensor modes and a reionization prior,
whereas Bridle {\etal} \cite{Bridle03} argue that combining WMAP 
with certain galaxy cluster measurements prefers $\Om\sim 0.17$.
In other words, WMAP has placed the ball in the non-CMB court. Since non-CMB measurements are now less reliable and precise than the CMB, they have emerged 
as the limiting factor and weakest link in the quest for precision cosmology.
Much of the near-term progress in cosmology will therefore be driven by
reductions in statistical and systematic uncertainties of
non-CMB probes.

The Sloan Digital Sky Survey \cite{York00,Stoughton02,Abazajian03} (SDSS) team has recently measured the 
three-dimensional power spectrum $P(k)$
using over 200{,}000 galaxies.
The goal of that measurement \cite{sdsspower}
was to produce the most reliable non-CMB data
%``something else'' 
to date,
in terms of small and well-controlled systematic errors, and the purpose of the present 
paper is to use this measurement to constrain cosmological
parameters.  The SDSS power spectrum analysis is completely
independent of that of the 2dFGRS, and with greater completeness,
more uniform photometric calibration, analytically computed window functions and 
improved treatment of non-linear redshift distortions, it should be less sensitive to
potential systematic errors. 
We emphasize the specific ways in
which large-scale structure data removes degeneracies in the WMAP-only
analysis, and explore in detail the effect of various priors that are
put on the data.  The WMAP analysis using the 2dFGRS data
\cite{Spergel03,Verde03} was carried out with various strong priors:

\bigskip % Just to avoid breaking this list across columns:
\bigskip 
\bigskip 

%Since we have had the luxury of working without the 
%draconian deadlines under which the WMAP team had to complete their analysis, 
%we will indulge in a more thorough and
%detailed exploration of multiparameter constraints and which assumptions 
%affect what. In particular, the WMAP team CMB+2dFGRS analysis \cite{Spergel03,Verde03} assumed 
\begin{enumerate}
\item reionization optical depth $\tau<0.3,$
\item vanishing tensor fluctuations and spatial curvature when constraining
  other parameters, % and fn=0 and w=-1
\item that galaxy bias was known from the 2dFGRS bispectrum \cite{Verde02}, and
\item that galaxy redshift distortions were reliably modeled.
\end{enumerate}
We will explore the effect of dropping these assumptions, and will 
see that the first three make a dramatic difference. 
Note in particular that both the 
spectral index $\ns$ and the tensor amplitude $r$ are motivated as free parameters only by
inflation theory, not by current observational data (which are consistent with $\ns=1$, $r=0$), 
suggesting that one should either include or exclude them both.

\begin{figure*} 
%\vskip\smtopskip
\bigskip
\centerline{\epsfxsize=18cm\epsffile{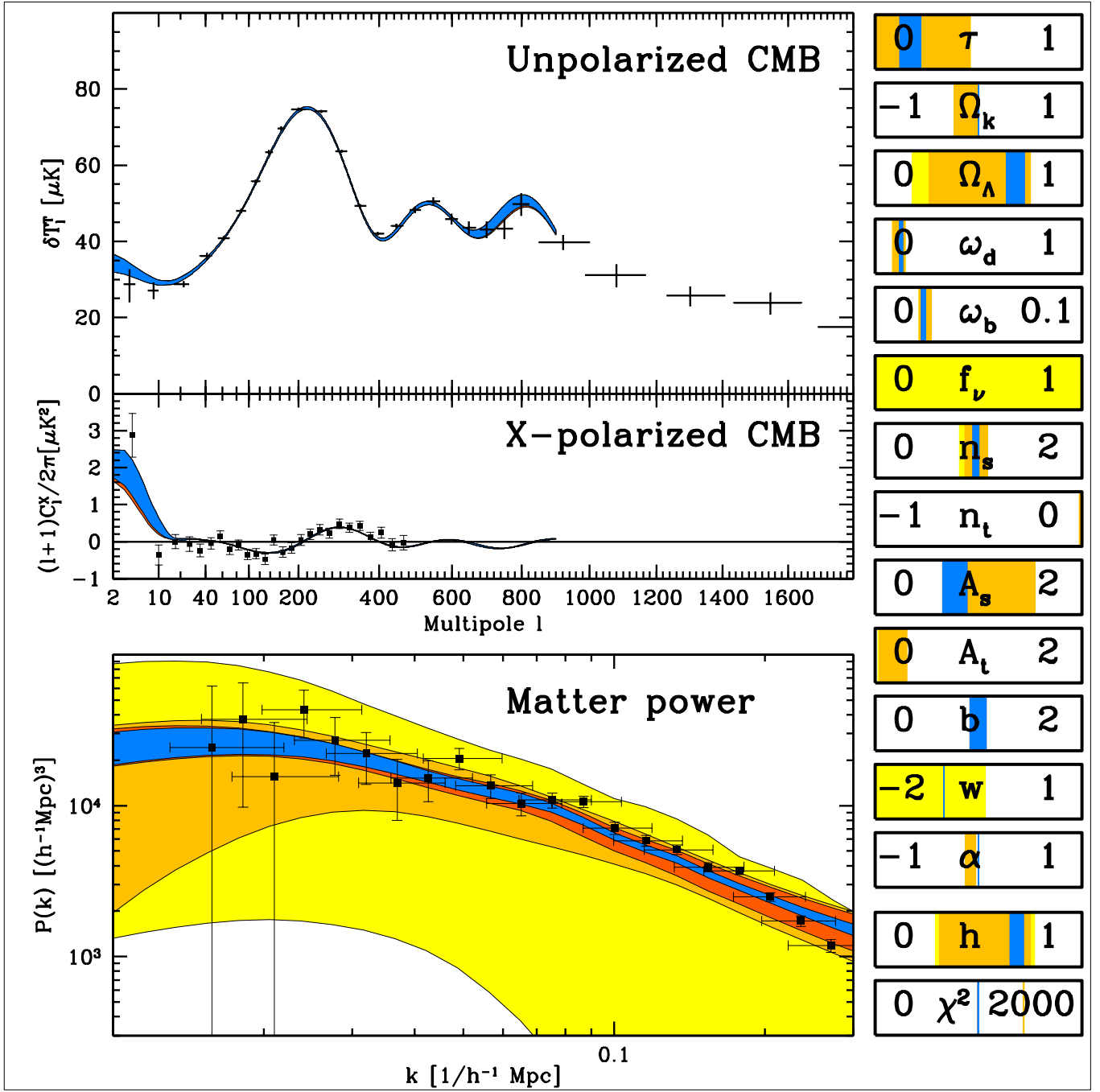}}
%\vskip\smbotskip
\caption[1]{\label{SausageFig}\footnotesize%
Summary of observations and cosmological models.
Data points are for unpolarized CMB experiments combined (top; Appendix A.3 details data used)
cross-polarized CMB from WMAP (middle) and Galaxy power from SDSS (bottom).
Shaded bands show the 1-sigma range of theoretical models from the Monte-Carlo Markov chains,
both for cosmological parameters (right) and for the corresponding power spectra (left).
From outside in, these bands correspond to WMAP with no priors,
adding the prior $\fn=0$, $w=-1$, further adding the priors $\Ok=r=\al=0$, and further
adding the SDSS information, respectively.
These four bands essentially coincide in the top two panels, since 
the CMB constraints were included in the fits.
Note that the $\l$-axis in the upper two panels goes from logarithmic on the
left to linear on the right, to show important features at both ends,
whereas the $k$-axis of the bottom panel is simply logarithmic.
}
\end{figure*}

The basic observational and theoretical situation is summarized
in \fig{SausageFig}. 
Here we have used our Monte Carlo Markov Chains (MCMC, described in detail below)
to show how uncertainty in cosmological parameters (Table~{\ParameterTable}) translates into uncertainty in 
the CMB and matter power spectra. We see that the key reason why SDSS helps so much is that
WMAP alone places only very weak constraints on the matter power spectrum $P(k)$.
As simplifying theoretical assumptions are added, the WMAP $P(k)$ predictions
are seen to tighten into a narrow band whose agreement with the SDSS measurements 
is a striking manifestation of cosmological consistency. Yet even this band is still much wider
than the SDSS error bars, which is why SDSS helps tighten constraints 
(notably on $\Ol$ and $h$) even for this restricted 6-parameter class of models.
 
The rest of this paper is organized as follows.
After presenting our basic results in three tables,
we devote a series of sections to digesting this information one piece at a time,
focusing on what we have and have not learned about the underlying physics, and on
how robust the various conclusions are to the choice of data sets and prior assumptions.
In \Sec{DiscussionSec} we discuss our conclusions and potential systematic uncertainties,
assess the extent to which a robust and consistent cosmological picture emerges,
and comment on upcoming prospects and challenges.
% the road ahead.

%\clearpage

%\clearpage

\clearpage
\section{Basic Results}

\begin{table*}
\noindent 
{\footnotesize
Table~\ParameterTable: Cosmological parameters used. Parameters 14-28 are determined by the first 13.
Our Monte-Carlo Markov Chain assigns a uniform prior to the parameters labeled ``MCMC''.
The last six and those labeled ``Fits'' are closely related to observable power spectrum features 
\protect\cite{Knox03,observables,Kosowsky02} and are helpful 
for understanding the physical origin of the constraints.
\begin{center}
\begin{tabular}{|l|llll|}
\hline
Parameter		&Meaning			&Status			&Use	&Definition\\
\hline
$\tau 	        $	&Reionization optical depth	&Not optional		&	&\\
$\ob   		$       &Baryon density			&Not optional		&MCMC	&$\ob=\Ob h^2 = \rho_b/(1.88\times 10^{-26}$kg$/$m$^3)$\\
$\od   		$       &Dark matter density		&Not optional		&MCMC	&$\od=\Od h^2 = \rho_d/(1.88\times 10^{-26}$kg$/$m$^3)$\\
$\fn	        $	&Dark matter neutrino fraction	&Well motivated		&MCMC	&$\fn=\rho_\nu/\rho_d$\\
$\Ol 		$	&Dark energy density		&Not optional		&MCMC	&\\
$w	        $	&Dark energy equation of state	&Worth testing		&MCMC	&$p_\Lambda/\rho_\Lambda$ (approximated as constant)\\
$\Ok       	$	&Spatial curvature		&Worth testing		&	&\\
$\As            $	&Scalar fluctuation amplitude	&Not optional		&	&Primordial scalar power at $k=0.05$/Mpc\\
$\ns	        $	&Scalar spectral index		&Well motivated		&MCMC	&Primordial spectral index at $k=0.05$/Mpc\\
$\alpha         $	&Running of spectral index	&Worth testing		&MCMC	&$\alpha=d\ln\ns/d\ln k$ (approximated as constant)\\
$r              $	&Tensor-to-scalar ratio		&Well motivated		&MCMC	&Tensor-to-scalar power ratio at $k=0.05$/Mpc\\
$\nt           	$ 	&Tensor spectral index		&Well motivated		&MCMC	&\\
$b	        $	&Galaxy bias factor		&Not optional		&MCMC	&$b=[P_{\rm galaxy}(k)/P(k)]^{1/2}$ (assumed constant for $k<0.2h/$Mpc)\\
\hline
$\zion        	$	&Reionization redshift (abrupt)	&			&	&$\zion\approx 92 (0.03h\tau/\ob)^{2/3}\Om^{1/3}$ (assuming abrupt reionization; \cite{reion})\\
$\om    	$	&Physical matter density	&			&Fits	&$\om=\ob+\od = \Om h^2$\\
$\Om       	$	&Matter density/critical density&			&	&$\Om=1-\Ol-\Ok$\\
$\Ot       	$	&Total density/critical density &			&	&$\Ot=\Om+\Ol=1-\Ok$\\
$\At	        $	&Tensor fluctuation amplitude	&			&	&$\At=r\As$\\
$\Mnu    	$	&Sum of neutrino masses 	&			&	&$\Mnu\approx(94.4\>{\rm eV})\times\od\fn$~~~\cite{KolbTurnerBook}\\
$h	        $	&Hubble parameter		&			&	&$h = \sqrt{(\od+\ob)/(1-\Ok-\Ol)}$\\
$\beta	        $	&Redshift distortion parameter	&			&	&$\beta\approx [\Om^{4/7}+(1+\Om/2)(\Ol/70)]/b$~~~\cite{Carroll92,growl}\\
%$\beta	        $	&Redshift distortion parameter	&			&	&$\beta\approx {1\over b}\left[\Om^{4/7}+\left(1+{\Om\over 2}\right){\Ol\over 70}\right]$\\
% I checked that this is numerically identical to the 2-step expression I have in compute_morepars.x
$t_0    	$	&Age of Universe		&			&	&$t_0\approx$($9.785$ Gyr)$\times h^{-1}\int_0^1 [(\Ol a^{-(1+3w)} + \Ok + \Om/a)]^{-1/2}da$~~~\cite{KolbTurnerBook}\\
%$t_0    	$	&Age of Universe		&			&	&$9.78462$ Gyr$\times h^{-1}\int_0^1 [(\Ol a^{-(1+3w)} + \Ok + \Om/a)]^{-1/2}da$\\
%$\sigma_8       $	&$8 h^{-1}$ Mpc galaxy fluctuation amplitude&			&	&$\sigma_8^2={4\pi}\int_0^\infty \left[{\sin x-x\cos x\over x^3/3}\right]^2 P(k) {k^2 dk\over(2\pi)^3}$, $x\equiv kR$\\
$\sigma_8       $	&Galaxy fluctuation amplitude&			&	&$\sigma_8=\{4\pi\int_0^\infty [{3\over x^3}(\sin x-x\cos x)]^2 P(k) {k^2 dk\over(2\pi)^3}\}^{1/2}$, $x\equiv k\times 8h^{-1}$Mpc\\
\hline
$Z     		$	&CMB peak suppression factor	&			&MCMC	&$Z=e^{-2\tau}$\\
$\Ap	        $	&Amplitude on CMB peak scales	&			&MCMC	&$\Ap=\As e^{-2\tau}$\\
$\Th       	$	&Acoustic peak scale (degrees)	&			&MCMC	&$\Th(\Ok,\Ol,w,\od,\ob)$ given by \protect\cite{observables}\\
%$zlss  = 1008 (1+0.00124\ob^{-0.74})(1 + c_1 * \om^{c_2})$
%$c_1 = 0.0783\ob^{-0.24}/(1+39.5\ob^{0.76})$
%$c_2 = 0.56/(1 + 21.1\ob^{1.8}$
%$\Rlss  = 30\ob/(\zlss/1000)$		% baryon-to-photon density ratio at LSS
%$\rmlss = 0.042(\zlss/1000)/\om$ 	% radiation-to-matter ratio at LSS
% dlss is given by eq (A2), where Omega_tot = Om+Ol = 1-Ok: 	
%$\dlss=((1+\ln[(1-\Ol)^{0.085})^{1+1.14(1+w)})/sqrt((1-\Ok)^{(1-\Ol)^{-0.76}})
%	! lA = pi*D/sstar given by equation (A3):
%	tmp = (sqrt(1.d0+Rlss)+sqrt(Rlss+rmlss*Rlss))/(1.d0+sqrt(rmlss*Rlss))     
%	lA = 172.d0 * d * sqrt(zlss/1.d3)/(log(tmp)/sqrt(Rlss))  
%	Theta = 180.d0/lA      ! Chu et.al. astro-ph/0212466 do it in radians, we do it in degrees
$H_2	        $	&2nd to 1st CMB peak ratio	&			&Fits	&$H_2 = (0.925\om^{0.18} 2.4^{\ns-1})/[1+(\ob/0.0164)^{12\om^{0.52}})]^{0.2}$~~~\cite{observables}\\
$H_3	        $	&3rd to 1st CMB peak ratio	&			&Fits	&$H_3 = 2.17 [1+(\ob/0.044)^2]^{-1} \om^{0.59} 3.6^{\ns-1}/[1 + 1.63(1-\ob/0.071)\om]$\\
$\Apivot      	$	&Amplitude at pivot point	&			&Fits	&$\Apivot=0.82^{\ns-1} \Ap$\\	
\hline
\end{tabular}
\end{center}     
} 
\end{table*}
% H2 replaces ob
% H3 replaces ns
% Apivot replaces Ap

\subsection{Cosmological parameters}

In this paper, we work within the context of a hot Big
Bang cosmology with primordial fluctuations that are adiabatic (\ie,
we do not allow isocurvature modes) and Gaussian, with negligible
generation of fluctuations by cosmic strings, textures, or domain
walls.  Within this framework, we follow \cite{consistent,Spergel03} in parameterizing
our cosmological model in terms of 13 parameters: 
\beq{pEq}
\p\equiv(\tau,\ob,\od,\fn,\Ol,w,\Ok,\As,\ns,\alpha,r,\nt,b).
\eeq
The meaning of these 13 parameters is described in
Table~{\ParameterTable}, together with an additional 16 derived
parameters, and their relationship to the original 13.  
%Table~{\ParameterTable} defines the 28 cosmological parameters discussed in this paper.
%Only 13 of them are independent, and it by now rather
%standard to think of cosmology as parameterized by the set
%%13 parameters
%\beq{pEq}
%\p\equiv(\tau,\ob,\od,\fn,\Ol,w,\Ok,\As,\ns,\alpha,r,\nt,b).
%\eeq
%The table specifies how the remaining parameters are 
%given in terms of these first 13.
%They are the reionization optical depth $\tau$, 
%the primordial amplitudes $\As$, $r\As$ and tilts $\ns$, $\nt$ 
%of scalar and tensor fluctuations, the running of the scalar tilt 
%$\alpha\equiv d\ns/d\ln k$,
%the bias parameter $b$ defined as the ratio between rms 
%galaxy fluctuations and rms matter fluctuations on 
%large scales,
%and parameters giving the contributions to the cosmic matter budget from
%curvature $\Ok$, dark energy $\Ol$, cold dark matter $\Oc$, 
%hot dark matter (neutrinos) $\On$ and baryons $\Ob$.

All parameters are defined just as in version 4.3 of CMBfast \cite{cmbfast}: 
in particular, 
the pivot point unchanged by $\ns$, $\alpha$ and $\nt$ is at $0.05$/Mpc, 
and the tensor normalization convention is such that $r=-8\nt$ for slow-roll models.
$\sigma_8$, the linear
rms mass fluctuation in spheres of radius $8h^{-1}$Mpc, is determined by
the power spectrum, which is in turn determined by $\p$ via CMBfast.
The last six parameters in the table are 
so-called normal parameters \cite{Knox03}, 
which correspond to observable features in the CMB power
spectrum \cite{observables,Kosowsky02}
and are useful for having simpler statistical properties than the underlying cosmological 
parameters as discussed in Appendix A.
Since current $\nt$-constraints are too weak to be interesting, we make the slow-roll assumption $\nt=-r/8$ throughout this
paper rather than treat $\nt$ as a free parameter.
%We also make the usual assumptions that the primordial fluctuations are 
%Gaussian and adiabatic (as opposed to isocurvature), with negligible 
%generation of fluctuations by cosmic strings, textures or domain walls.

\begin{figure*} 
%\vskip\smtopskip
\centerline{\epsfxsize=18cm\epsffile{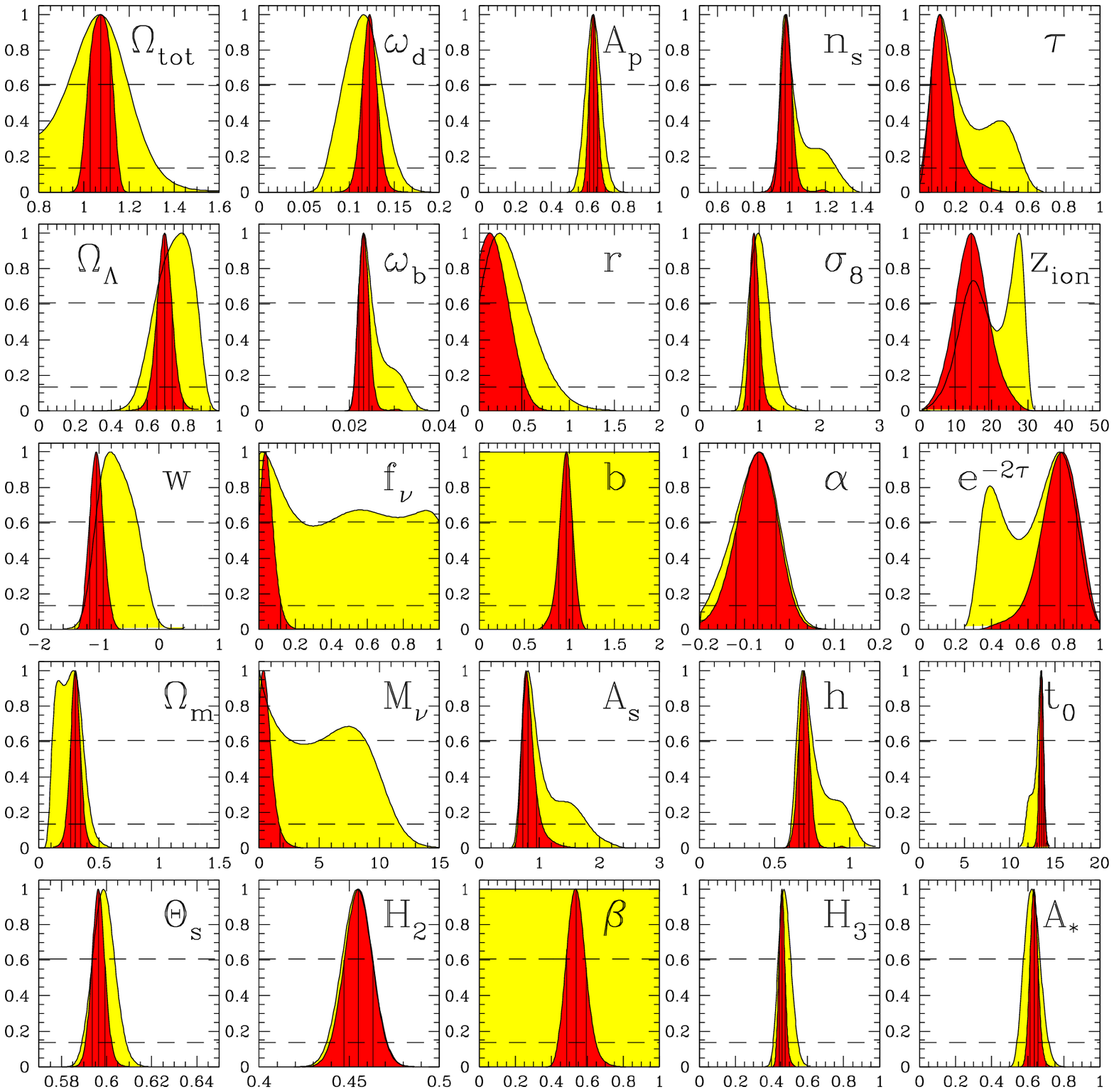}}
%\vskip\smbotskip
\caption[1]{\label{1d_6par_fig}\footnotesize%
Constraints on individual cosmological quantities
using WMAP alone (shaded yellow/light grey distributions)
and including SDSS information 
(narrower red/dark grey distributions).
Each distribution shown has been marginalized over all other quantities
in the class of 6-parameter $(\tau,\Ol,\od,\ob,\As,\ns)$ ``vanilla'' models
as well as over a galaxy bias parameter $b$ for the SDSS case.
%The analyses for the $\Otot$, $r$ and $w$ panels 
%also include curvature, gravity waves and variable equation of state, 
%respectively. The analyses for the 
%$\fn$ and $\Mnu$ panels include massive neutrinos, 
The $\alpha$-distributions are also marginalized over $r$ and $\Ok$.
The parameter measurements and error bars quoted in the tables correspond
to the median and the central 68\% of the distributions, indicated by 
three vertical lines for the WMAP+SDSS case above.
When the distribution peaks near zero (like for $r$), we instead quote an upper limit
at the 95th percentile (single vertical line).
The horizontal dashed lines indicate $e^{-x^2/2}$ for
$x=1$ and $2$, respectively, so if the distribution were Gaussian, 
its intersections with these lines would correspond to 
$1\sigma$ and $2\sigma$ limits, respectively.
}
\end{figure*}

\begin{table*}
\noindent 
{\footnotesize
Table~\WMAPtable:
$1\sigma$ constraints on cosmological parameters using WMAP information
alone.
The columns compare different theoretical priors indicated by numbers in {\it italics}.
The penultimate column has only the six ``vanilla'' parameters 
$(\tau,\Ol,\od,\ob,\As,\ns)$ free and therefore gives the smallest error bars. 
The last column uses WMAP temperature data alone, all others also include WMAP polarization information.
\begin{center}
\begin{tabular}{|l|cccccc|c|}
\hline
&\multicolumn{6}{c|}{Using WMAP temperature and polarization information}																				&No pol.\\ 
					&6par$+\Ok+r+\alpha$ 		&6par+$\Ok$ 			&6par+$r$ 			&6par+$\fn$ 			&6par+$\w$ 			&6par				&6par\\
\hline
$e^{-2\tau}		      $    	&$ 0.52^{+ 0.21}_{- 0.15}$	&$ 0.65^{+ 0.19}_{- 0.32}$	&$ 0.68^{+ 0.13}_{- 0.16}$	&$ 0.75^{+ 0.12}_{- 0.23}$	&$ 0.68^{+ 0.15}_{- 0.21}$	&$ 0.66^{+ 0.17}_{- 0.25}$	&$>0.50\>\> (95\%)$		\\
$\Theta_s		      $    	&$ 0.602^{+ 0.010}_{- 0.006}$	&$ 0.603^{+ 0.015}_{- 0.005}$	&$ 0.5968^{+ 0.0048}_{- 0.0056}$&$ 0.5893^{+ 0.0062}_{- 0.0056}$&$ 0.5966^{+ 0.0066}_{- 0.0105}$&$ 0.5987^{+ 0.0052}_{- 0.0048}$&$ 0.5984^{+ 0.0041}_{- 0.0042}$\\
$\Omega_\Lambda 	      $    	&$ 0.54^{+ 0.24}_{- 0.33}$	&$ 0.53^{+ 0.24}_{- 0.32}$	&$ 0.823^{+ 0.058}_{- 0.082}$	&$ 0.687^{+ 0.087}_{- 0.097}$	&$ 0.64^{+ 0.14}_{- 0.17}$	&$ 0.75^{+ 0.10}_{- 0.10}$	&$ 0.674^{+ 0.086}_{- 0.093}$	\\
$h^2\Omega_d		      $    	&$ 0.105^{+ 0.023}_{- 0.023}$	&$ 0.108^{+ 0.022}_{- 0.034}$	&$ 0.097^{+ 0.021}_{- 0.018}$	&$ 0.119^{+ 0.018}_{- 0.016}$	&$ 0.118^{+ 0.020}_{- 0.020}$	&$ 0.115^{+ 0.020}_{- 0.021}$	&$ 0.129^{+ 0.019}_{- 0.018}$	\\
$h^2\Omega_b		      $    	&$ 0.0238^{+ 0.0035}_{- 0.0027}$&$ 0.0241^{+ 0.0055}_{- 0.0020}$&$ 0.0256^{+ 0.0025}_{- 0.0019}$&$ 0.0247^{+ 0.0029}_{- 0.0016}$&$ 0.0246^{+ 0.0038}_{- 0.0017}$&$ 0.0245^{+ 0.0050}_{- 0.0019}$&$ 0.0237^{+ 0.0018}_{- 0.0013}$\\
$f_\nu			      $    	&${\it   0}$			&${\it   0}$			&${\it   0}$			&No constraint			&${\it   0}$			&${\it   0}$			&${\it   0}$			\\
$n_s			      $    	&$ 0.97^{+ 0.13}_{- 0.10}$	&$ 1.01^{+ 0.18}_{- 0.06}$	&$ 1.064^{+ 0.066}_{- 0.059}$	&$ 0.962^{+ 0.098}_{- 0.041}$	&$ 1.03^{+ 0.12}_{- 0.05}$	&$ 1.02^{+ 0.16}_{- 0.06}$	&$ 0.989^{+ 0.061}_{- 0.031}$	\\
$n_t + 1		      $    	&$ 0.9847^{+ 0.0097}_{- 0.0141}$&${\it   1}$			&$ 0.959^{+ 0.026}_{- 0.037}$	&${\it   1}$			&${\it   1}$			&${\it   1}$			&${\it   1}$			\\
$A_p			      $    	&$ 0.593^{+ 0.053}_{- 0.044}$	&$ 0.602^{+ 0.053}_{- 0.051}$	&$ 0.592^{+ 0.049}_{- 0.046}$	&$ 0.602^{+ 0.045}_{- 0.050}$	&$ 0.637^{+ 0.045}_{- 0.046}$	&$ 0.633^{+ 0.044}_{- 0.041}$	&$ 0.652^{+ 0.049}_{- 0.046}$	\\
$r           	      	      $    	&$<0.90\>\>(95\%)$		&${\it   0}$			&$<0.84\>\>(95\%)$		&${\it   0}$			&${\it   0}$			&${\it   0}$			&${\it   0}$			\\
$b			      $    	&No constraint			&No constraint			&No constraint			&No constraint			&No constraint			&No constraint			&No constraint			\\
$w			      $    	&${\it  -1}$			&${\it  -1}$			&${\it  -1}$			&${\it  -1}$			&$-0.72^{+ 0.34}_{- 0.27}$	&${\it  -1}$			&${\it  -1}$			\\
$\alpha       	              $    	&$-0.075^{+ 0.047}_{- 0.055}$	&${\it   0}$			&${\it   0}$			&${\it   0}$			&${\it   0}$			&${\it   0}$			&${\it   0}$			\\
\hline
%$\Omega_k		      $    	&$-0.095^{+ 0.094}_{- 0.144}$	&$-0.086^{+ 0.057}_{- 0.128}$	&${\it   0}$			&${\it   0}$			&${\it   0}$			&${\it   0}$			&${\it   0}$			\\
$\Ot		      	      $    	&$ 1.095^{+ 0.094}_{- 0.144}$	&$ 1.086^{+ 0.057}_{- 0.128}$	&${\it   0}$			&${\it   0}$			&${\it   0}$			&${\it   0}$			&${\it   0}$			\\
$\Omega_m		      $    	&$ 0.57^{+ 0.45}_{- 0.33}$	&$ 0.55^{+ 0.47}_{- 0.29}$	&$ 0.177^{+ 0.082}_{- 0.058}$	&$ 0.313^{+ 0.097}_{- 0.087}$	&$ 0.36^{+ 0.17}_{- 0.14}$	&$ 0.25^{+ 0.10}_{- 0.10}$	&$ 0.326^{+ 0.093}_{- 0.086}$	\\
$h^2\Omega_m	  	      $    	&$ 0.128^{+ 0.022}_{- 0.021}$	&$ 0.132^{+ 0.021}_{- 0.028}$	&$ 0.123^{+ 0.020}_{- 0.018}$	&$ 0.144^{+ 0.018}_{- 0.016}$	&$ 0.143^{+ 0.020}_{- 0.019}$	&$ 0.140^{+ 0.020}_{- 0.018}$	&$ 0.153^{+ 0.020}_{- 0.018}$	\\
$h			      $    	&$ 0.48^{+ 0.27}_{- 0.12}$	&$ 0.50^{+ 0.16}_{- 0.13}$	&$ 0.84^{+ 0.12}_{- 0.10}$	&$ 0.674^{+ 0.087}_{- 0.049}$	&$ 0.63^{+ 0.14}_{- 0.10}$	&$ 0.74^{+ 0.18}_{- 0.07}$	&$ 0.684^{+ 0.070}_{- 0.045}$	\\
$\tau 			      $    	&$ 0.33^{+ 0.17}_{- 0.17}$	&$ 0.22^{+ 0.34}_{- 0.13}$	&$ 0.19^{+ 0.13}_{- 0.09}$	&$ 0.15^{+ 0.18}_{- 0.07}$	&$ 0.19^{+ 0.18}_{- 0.10}$	&$ 0.21^{+ 0.24}_{- 0.11}$	&$<0.35\>\> (95\%)$		\\
$z_{ion}		      $    	&$25.9^{+ 4.4}_{- 8.8}$		&$20.1^{+ 9.2}_{- 8.3}$		&$17.1^{+ 5.8}_{- 5.8}$	&$15.5^{+ 8.6}_{- 5.6}$			&$18.5^{+ 7.1}_{- 6.6}$ 	&$19.6^{+ 7.8}_{- 7.4}$		&$<25\>\> (95\%)$ 		\\
$A_s    		      $    	&$ 1.14^{+ 0.42}_{- 0.31}$	&$ 0.97^{+ 0.73}_{- 0.23}$	&$ 0.87^{+ 0.28}_{- 0.16}$	&$ 0.81^{+ 0.35}_{- 0.13}$	&$ 0.94^{+ 0.40}_{- 0.18}$	&$ 0.98^{+ 0.56}_{- 0.21}$	&$ 0.80^{+ 0.26}_{- 0.12}$	\\
$A_t			      $    	&$ 0.14^{+ 0.13}_{- 0.10}$	&${\it   0}$			&$ 0.30^{+ 0.22}_{- 0.17}$	&${\it   0}$			&${\it   0}$			&${\it   0}$			&${\it   0}$			\\
$\beta			      $    	&No constraint			&No constraint			&No constraint			&No constraint			&No constraint			&No constraint			&No constraint			\\
$t_0  [$Gyr$]		      $    	&$16.5^{+ 2.6}_{- 3.1}$		&$16.3^{+ 2.3}_{- 1.8}$		&$13.00^{+ 0.41}_{- 0.47}$	&$13.75^{+ 0.36}_{- 0.59}$	&$13.53^{+ 0.52}_{- 0.65}$	&$13.24^{+ 0.41}_{- 0.89}$	&$13.41^{+ 0.29}_{- 0.37}$	\\
$\sigma_8		      $    	&$ 0.90^{+ 0.13}_{- 0.13}$	&$ 0.87^{+ 0.15}_{- 0.13}$	&$ 0.84^{+ 0.17}_{- 0.17}$	&$ 0.32^{+ 0.36}_{- 0.32}$	&$ 0.95^{+ 0.16}_{- 0.14}$	&$ 0.99^{+ 0.19}_{- 0.14}$	&$ 0.94^{+ 0.15}_{- 0.12}$	\\
%$H_1			      $    	&$ 4.8^{+ 3.8}_{- 1.9}$		&$ 7.0^{+ 4.7}_{- 1.6}$		&$ 6.5^{+ 1.5}_{- 1.0}$		&$ 4.77^{+ 0.87}_{- 0.59}$	&$ 7.0^{+ 3.4}_{- 1.7}$ 	&$ 5.5^{+ 1.7}_{- 0.7}$		&$ 5.64^{+ 0.75}_{- 0.60}$	\\
$H_2			      $    	&$ 0.441^{+ 0.013}_{- 0.014}$	&$ 0.4581^{+ 0.0090}_{- 0.0083}$&$ 0.4541^{+ 0.0067}_{- 0.0081}$&$ 0.426^{+ 0.018}_{- 0.010}$	&$ 0.4541^{+ 0.0084}_{- 0.0085}$&$ 0.4543^{+ 0.0083}_{- 0.0085}$&$ 0.4541^{+ 0.0085}_{- 0.0086}$\\
$H_3			      $    	&$ 0.424^{+ 0.043}_{- 0.040}$	&$ 0.455^{+ 0.033}_{- 0.029}$	&$ 0.452^{+ 0.034}_{- 0.033}$	&$ 0.441^{+ 0.039}_{- 0.033}$	&$ 0.477^{+ 0.036}_{- 0.034}$	&$ 0.474^{+ 0.037}_{- 0.033}$	&$ 0.475^{+ 0.032}_{- 0.030}$	\\
$\Apivot		      $    	&$ 0.595^{+ 0.056}_{- 0.048}$	&$ 0.599^{+ 0.055}_{- 0.064}$	&$ 0.584^{+ 0.050}_{- 0.046}$	&$ 0.602^{+ 0.045}_{- 0.046}$	&$ 0.631^{+ 0.047}_{- 0.045}$	&$ 0.624^{+ 0.048}_{- 0.042}$	&$ 0.652^{+ 0.048}_{- 0.046}$	\\
$M_\nu  [$eV$]	      	      $    	&${\it   0}$			&${\it   0}$			&${\it   0}$			&$<10.6\>\>(95\%)$		&${\it   0}$			&${\it   0}$			&${\it   0}$			\\
\hline																			
$\chi^2$/dof			    	&$1426.1/1339$			&$1428.4/1341$			&$1430.9/1341$			&$1431.8/1341$			&$1431.8/1341$			&$1431.5/1342$			&$972.4/893$\\
\hline  
\end{tabular}
\end{center}     
} 
\end{table*}
%$\beta			      $    	&$ 0.73^{+ 0.28}_{- 0.29}$	&$ 0.72^{+ 0.29}_{- 0.24}$	&$ 0.385^{+ 0.089}_{- 0.075}$	&$ 0.526^{+ 0.084}_{- 0.087}$	&$ 0.60^{+ 0.16}_{- 0.13}$	&$ 0.45^{+ 0.11}_{- 0.14}$	\\

\begin{table*}
\noindent 
{\footnotesize
Table~{\SDSStable}:
$1\sigma$ constraints on cosmological parameters combining CMB and SDSS information.
The columns compare different theoretical priors indicated by {\it italics}.
%The first column treats the 9 parameters 
%$(\tau,\Ok,\Ol,\od,\ob,\As,\ns,\alpha,r)$
%as unknown, so the only assumptions are $\fn=0$, $w=-1$.
The second last column drops the polarized WMAP information and the last column drops all WMAP information, replacing it
by pre-WMAP CMB experiments.
The 6par$+w$ column includes SN Ia information.
\begin{center}
\begin{tabular}{|l|cccccc|c|c|}
\hline
					&\multicolumn{6}{c|}{Using SDSS + WMAP temperature and polarization information}														&No pol.			&No WMAP\\ 
					&6par$+\Ok+r+\alpha$ 		&6par+$\Ok$ 			&6par+$r$ 			&6par+$\fn$ 			&6par+$\w$ 			&6par				&6par				&6par\\
\hline
$e^{-2\tau}		      $    	&$ 0.53^{+ 0.22}_{- 0.17}$	&$ 0.69^{+ 0.15}_{- 0.32}$	&$ 0.776^{+ 0.098}_{- 0.116}$	&$ 0.776^{+ 0.095}_{- 0.121}$	&$ 0.80^{+ 0.10}_{- 0.13}$	&$ 0.780^{+ 0.094}_{- 0.119}$	&$>0.63\>\> (95\%)$		&$>0.71\>\> (95\%)$		\\
$\Theta_s		      $    	&$ 0.601^{+ 0.010}_{- 0.006}$	&$ 0.600^{+ 0.013}_{- 0.004}$	&$ 0.5982^{+ 0.0034}_{- 0.0032}$&$ 0.5948^{+ 0.0033}_{- 0.0030}$&$ 0.5954^{+ 0.0037}_{- 0.0038}$&$ 0.5965^{+ 0.0031}_{- 0.0030}$&$ 0.5968^{+ 0.0030}_{- 0.0030}$&$ 0.5977^{+ 0.0048}_{- 0.0045}$\\
$\Omega_\Lambda 	      $    	&$ 0.660^{+ 0.080}_{- 0.097}$	&$ 0.653^{+ 0.082}_{- 0.084}$	&$ 0.727^{+ 0.041}_{- 0.042}$	&$ 0.620^{+ 0.074}_{- 0.087}$	&$ 0.706^{+ 0.032}_{- 0.033}$	&$ 0.699^{+ 0.042}_{- 0.045}$	&$ 0.684^{+ 0.041}_{- 0.046}$	&$ 0.691^{+ 0.039}_{- 0.053}$	\\
$h^2\Omega_d		      $    	&$ 0.103^{+ 0.020}_{- 0.022}$	&$ 0.103^{+ 0.016}_{- 0.024}$	&$ 0.1195^{+ 0.0084}_{- 0.0082}$&$ 0.135^{+ 0.014}_{- 0.012}$	&$ 0.124^{+ 0.012}_{- 0.011}$	&$ 0.1222^{+ 0.0090}_{- 0.0082}$&$ 0.1254^{+ 0.0093}_{- 0.0083}$&$ 0.1252^{+ 0.0088}_{- 0.0076}$\\
$h^2\Omega_b		      $    	&$ 0.0238^{+ 0.0036}_{- 0.0026}$&$ 0.0232^{+ 0.0051}_{- 0.0017}$&$ 0.0242^{+ 0.0017}_{- 0.0013}$&$ 0.0234^{+ 0.0014}_{- 0.0011}$&$ 0.0232^{+ 0.0013}_{- 0.0010}$&$ 0.0232^{+ 0.0013}_{- 0.0010}$&$ 0.0231^{+ 0.0011}_{- 0.0009}$&$ 0.0229^{+ 0.0016}_{- 0.0015}$\\
$f_\nu			      $    	&${\it   0}$			&${\it   0}$			&${\it   0}$			&$<0.12\>\>(95\%)$		&${\it   0}$			&${\it   0}$			&${\it   0}$			&${\it   0}$			\\
$n_s			      $    	&$ 0.97^{+ 0.12}_{- 0.10}$	&$ 0.98^{+ 0.18}_{- 0.04}$	&$ 1.012^{+ 0.049}_{- 0.036}$	&$ 0.972^{+ 0.041}_{- 0.027}$	&$ 0.976^{+ 0.040}_{- 0.024}$	&$ 0.977^{+ 0.039}_{- 0.025}$	&$ 0.973^{+ 0.030}_{- 0.021}$	&$ 1.015^{+ 0.036}_{- 0.033}$	\\
$n_t + 1		      $    	&$ 0.9852^{+ 0.0093}_{- 0.0154}$&${\it   1}$			&$ 0.976^{+ 0.016}_{- 0.021}$	&${\it   1}$			&${\it   1}$			&${\it   1}$			&${\it   1}$			&${\it   1}$			\\
$A_p			      $    	&$ 0.584^{+ 0.045}_{- 0.033}$	&$ 0.584^{+ 0.038}_{- 0.028}$	&$ 0.635^{+ 0.023}_{- 0.021}$	&$ 0.645^{+ 0.029}_{- 0.026}$	&$ 0.637^{+ 0.027}_{- 0.027}$	&$ 0.633^{+ 0.024}_{- 0.022}$	&$ 0.637^{+ 0.025}_{- 0.023}$	&$ 0.588^{+ 0.025}_{- 0.025}$	\\
$r           	      	      $    	&$<0.50\>\>(95\%)$		&${\it   0}$			&$<0.47\>\>(95\%)$		&${\it   0}$			&${\it   0}$			&${\it   0}$			&${\it   0}$			&${\it   0}$			\\
$b			      $    	&$ 0.94^{+ 0.12}_{- 0.10}$	&$ 1.03^{+ 0.15}_{- 0.13}$	&$ 0.963^{+ 0.075}_{- 0.081}$	&$ 1.061^{+ 0.096}_{- 0.105}$	&$ 0.956^{+ 0.075}_{- 0.076}$	&$ 0.962^{+ 0.073}_{- 0.083}$	&$ 1.009^{+ 0.068}_{- 0.091}$	&$ 1.068^{+ 0.066}_{- 0.079}$	\\
$w			      $    	&${\it  -1}$			&${\it  -1}$			&${\it  -1}$			&${\it  -1}$			&$-1.05^{+ 0.13}_{- 0.14}$	&${\it  -1}$			&${\it  -1}$			&${\it  -1}$			\\
$\alpha       	              $    	&$-0.071^{+ 0.042}_{- 0.047}$	&${\it   0}$			&${\it   0}$			&${\it   0}$			&${\it   0}$			&${\it   0}$			&${\it   0}$			&${\it   0}$			\\
\hline
%$\Omega_k		      $    	&$-0.056^{+ 0.045}_{- 0.045}$	&$-0.058^{+ 0.039}_{- 0.041}$	&${\it   0}$			&${\it   0}$			&${\it   0}$			&${\it   0}$			&${\it   0}$			&${\it   0}$			\\
$\Ot     		      $    	&$ 1.056^{+ 0.045}_{- 0.045}$	&$ 1.058^{+ 0.039}_{- 0.041}$	&${\it   0}$			&${\it   0}$			&${\it   0}$			&${\it   0}$			&${\it   0}$			&${\it   0}$			\\
$\Omega_m		      $    	&$ 0.40^{+ 0.10}_{- 0.09}$	&$ 0.406^{+ 0.093}_{- 0.091}$	&$ 0.273^{+ 0.042}_{- 0.041}$	&$ 0.380^{+ 0.087}_{- 0.074}$	&$ 0.294^{+ 0.033}_{- 0.032}$	&$ 0.301^{+ 0.045}_{- 0.042}$	&$ 0.316^{+ 0.046}_{- 0.041}$	&$ 0.309^{+ 0.053}_{- 0.039}$	\\
$h^2\Omega_m	  	      $    	&$ 0.126^{+ 0.019}_{- 0.019}$	&$ 0.126^{+ 0.016}_{- 0.019}$	&$ 0.1438^{+ 0.0084}_{- 0.0080}$&$ 0.158^{+ 0.015}_{- 0.012}$	&$ 0.147^{+ 0.012}_{- 0.011}$	&$ 0.1454^{+ 0.0091}_{- 0.0082}$&$ 0.1486^{+ 0.0095}_{- 0.0084}$&$ 0.1481^{+ 0.0091}_{- 0.0077}$\\
$h			      $    	&$ 0.55^{+ 0.11}_{- 0.06}$	&$ 0.550^{+ 0.092}_{- 0.055}$	&$ 0.725^{+ 0.049}_{- 0.036}$	&$ 0.645^{+ 0.048}_{- 0.040}$	&$ 0.708^{+ 0.033}_{- 0.030}$	&$ 0.695^{+ 0.039}_{- 0.031}$	&$ 0.685^{+ 0.033}_{- 0.028}$	&$ 0.693^{+ 0.038}_{- 0.040}$	\\
$\tau 			      $    	&$ 0.32^{+ 0.19}_{- 0.17}$	&$ 0.18^{+ 0.31}_{- 0.10}$	&$ 0.127^{+ 0.081}_{- 0.059}$	&$ 0.127^{+ 0.085}_{- 0.058}$	&$ 0.113^{+ 0.090}_{- 0.059}$	&$ 0.124^{+ 0.083}_{- 0.057}$	&$<0.23\>\> (95\%)$		&$<0.17\>\> (95\%)$		\\
$z_{ion}		      $    	&$25.3^{+ 4.8}_{- 8.8}$		&$18^{+10}_{- 7}$		&$14.1^{+ 4.8}_{- 4.7}$		&$14.9^{+ 5.4}_{- 4.8}$		&$13.6^{+ 5.7}_{- 5.2}$ 	&$14.4^{+ 5.2}_{- 4.7}$		&$<20\>\> (95\%)$ 		&$<18\>\> (95\%)$ 		\\
$A_s    		      $    	&$ 1.12^{+ 0.43}_{- 0.31}$	&$ 0.86^{+ 0.68}_{- 0.16}$	&$ 0.82^{+ 0.15}_{- 0.10}$	&$ 0.83^{+ 0.16}_{- 0.09}$	&$ 0.80^{+ 0.15}_{- 0.09}$	&$ 0.81^{+ 0.15}_{- 0.09}$	&$ 0.72^{+ 0.15}_{- 0.07}$	&$ 0.64^{+ 0.10}_{- 0.04}$	\\
$A_t			      $    	&$ 0.14^{+ 0.12}_{- 0.09}$	&${\it   0}$			&$ 0.16^{+ 0.15}_{- 0.11}$	&${\it   0}$			&${\it   0}$			&${\it   0}$			&${\it   0}$			&${\it   0}$			\\
$\beta			      $    	&$ 0.633^{+ 0.081}_{- 0.076}$	&$ 0.587^{+ 0.066}_{- 0.062}$	&$ 0.506^{+ 0.056}_{- 0.053}$	&$ 0.554^{+ 0.059}_{- 0.054}$	&$ 0.533^{+ 0.051}_{- 0.048}$	&$ 0.537^{+ 0.056}_{- 0.052}$	&$ 0.529^{+ 0.059}_{- 0.052}$	&$ 0.493^{+ 0.060}_{- 0.051}$	\\
$t_0  [$Gyr$]		      $    	&$15.8^{+ 1.5}_{- 1.8}$		&$15.9^{+ 1.3}_{- 1.5}$		&$13.32^{+ 0.27}_{- 0.33}$	&$13.65^{+ 0.25}_{- 0.28}$	&$13.47^{+ 0.26}_{- 0.27}$	&$13.54^{+ 0.23}_{- 0.27}$	&$13.55^{+ 0.21}_{- 0.23}$	&$13.51^{+ 0.32}_{- 0.31}$	\\
$\sigma_8		      $    	&$ 0.91^{+ 0.11}_{- 0.10}$	&$ 0.86^{+ 0.13}_{- 0.11}$	&$ 0.919^{+ 0.086}_{- 0.073}$	&$ 0.823^{+ 0.098}_{- 0.077}$	&$ 0.928^{+ 0.084}_{- 0.076}$	&$ 0.917^{+ 0.090}_{- 0.072}$	&$ 0.879^{+ 0.088}_{- 0.062}$	&$ 0.842^{+ 0.069}_{- 0.053}$	\\
%$H_1			      $    	&$ 3.9^{+ 1.6}_{- 1.2}$		&$ 5.5^{+ 1.7}_{- 0.6}$		&$ 5.8^{+ 1.0}_{- 0.7}$	&$ 5.04^{+ 0.51}_{- 0.41}$		&$ 4.99^{+ 0.56}_{- 0.45}$	&$ 5.06^{+ 0.46}_{- 0.40}$	&$ 5.46^{+ 0.54}_{- 0.49}$	&$ 6.8^{+ 1.2}_{- 0.9}$ 	\\
$H_2			      $    	&$ 0.441^{+ 0.013}_{- 0.012}$	&$ 0.4577^{+ 0.0086}_{- 0.0082}$&$ 0.4535^{+ 0.0081}_{- 0.0084}$&$ 0.4521^{+ 0.0091}_{- 0.0100}$&$ 0.4545^{+ 0.0087}_{- 0.0090}$&$ 0.4550^{+ 0.0083}_{- 0.0082}$&$ 0.4549^{+ 0.0082}_{- 0.0083}$&$ 0.475^{+ 0.018}_{- 0.020}$	\\
$H_3			      $    	&$ 0.422^{+ 0.027}_{- 0.031}$	&$ 0.444^{+ 0.026}_{- 0.025}$	&$ 0.468^{+ 0.019}_{- 0.017}$	&$ 0.472^{+ 0.022}_{- 0.019}$	&$ 0.461^{+ 0.018}_{- 0.017}$	&$ 0.459^{+ 0.018}_{- 0.016}$	&$ 0.460^{+ 0.017}_{- 0.015}$	&$ 0.485^{+ 0.020}_{- 0.018}$	\\
$\Apivot		      $    	&$ 0.587^{+ 0.049}_{- 0.041}$	&$ 0.582^{+ 0.041}_{- 0.036}$	&$ 0.632^{+ 0.022}_{- 0.021}$	&$ 0.648^{+ 0.028}_{- 0.025}$	&$ 0.639^{+ 0.027}_{- 0.028}$	&$ 0.635^{+ 0.024}_{- 0.022}$	&$ 0.639^{+ 0.024}_{- 0.022}$	&$ 0.586^{+ 0.024}_{- 0.025}$	\\
$M_\nu  [$eV$]	      	      $    	&${\it   0}$			&${\it   0}$			&${\it   0}$			&$<1.74\>\>(95\%)$		&${\it   0}$			&${\it   0}$			&${\it   0}$			&${\it   0}$			\\
\hline
$\chi^2$/dof		      		&$1444.4/1357$			&$1445.4/1359$			&$1446.9/1359$			&$1447.3/1359$			&$1622.0/1531$			&$1447.2/1360$			&$987.8/911$			&$134.6/163$\\
\hline  
\end{tabular}
\end{center}     
} 
\end{table*}

\begin{table*}
\noindent 
{\footnotesize
Table~{\AddinfoTable}:
$1\sigma$ constraints on cosmological parameters as 
progressively more 
information/assumptions are added.
First column uses WMAP data alone and treats the 9 parameters 
$(\tau,\Ok,\Ol,\od,\ob,\As,\ns,\alpha,r)$
as unknown, so the only assumptions are $\fn=0$, $w=-1$.
Moving to the right in the table, we add the assumptions
$r=\alpha=0$, then add SDSS information, then add SN Ia information, 
then add the assumption that $\tau<0.3$.
The next two columns are for 6-parameter vanilla models $(\Ok=r=\alpha=0)$, 
first using WMAP+SDSS data alone, then adding small-scale non-WMAP CMB data. 
The last two columns use WMAP+SDSS alone for 5-parameter models assuming 
$\ns=1$ (``vanilla lite'') and $\ns=0.96$, $r=0.15$ ($V\propto\phi^2$
stochastic eternal inflation), respectively.
\begin{center}
\begin{tabular}{|l|ccccc|cc|c|c|}
\hline
&\multicolumn{5}{c|}{9 parameters $(\tau,\Ok,\Ol,\od,\ob,\As,\ns,\alpha,r)$ free}															&\multicolumn{4}{c|}{WMAP+SDSS, 6 vanilla parameters free}	\\ 
					&WMAP 				&$+r=\alpha=0$ 			&$+$SDSS 			&$+$SN Ia 			&$+\tau<0.3$ 			&				&$+$other CMB			&$+\ns=1$			&$+V(\phi)\propto \phi^2$\\
\hline
$e^{-2\tau}		      $    	&$ 0.52^{+ 0.21}_{- 0.15}$	&$ 0.65^{+ 0.19}_{- 0.32}$	&$ 0.69^{+ 0.15}_{- 0.32}$	&$ 0.44^{+ 0.34}_{- 0.13}$	&$ 0.75^{+ 0.11}_{- 0.12}$	&$ 0.780^{+ 0.094}_{- 0.119}$	&$ 0.813^{+ 0.081}_{- 0.092}$	&$ 0.720^{+ 0.057}_{- 0.049}$   &$ 0.833^{+ 0.063}_{- 0.059}$\\
$\Theta_s		      $    	&$ 0.602^{+ 0.010}_{- 0.006}$	&$ 0.603^{+ 0.015}_{- 0.005}$	&$ 0.600^{+ 0.013}_{- 0.004}$	&$ 0.606^{+ 0.011}_{- 0.010}$	&$ 0.5971^{+ 0.0034}_{- 0.0034}$&$ 0.5965^{+ 0.0031}_{- 0.0030}$&$ 0.5956^{+ 0.0025}_{- 0.0026}$&$ 0.5979^{+ 0.0024}_{- 0.0024}$&$ 0.5953^{+ 0.0021}_{- 0.0022}$\\
$\Omega_\Lambda 	      $    	&$ 0.54^{+ 0.24}_{- 0.33}$	&$ 0.53^{+ 0.24}_{- 0.32}$	&$ 0.653^{+ 0.082}_{- 0.084}$	&$ 0.725^{+ 0.039}_{- 0.044}$	&$ 0.695^{+ 0.034}_{- 0.037}$	&$ 0.699^{+ 0.042}_{- 0.045}$	&$ 0.691^{+ 0.032}_{- 0.040}$	&$ 0.707^{+ 0.031}_{- 0.039}$   &$ 0.685^{+ 0.032}_{- 0.041}$\\
$h^2\Omega_d		      $    	&$ 0.105^{+ 0.023}_{- 0.023}$	&$ 0.108^{+ 0.022}_{- 0.034}$	&$ 0.103^{+ 0.016}_{- 0.024}$	&$ 0.090^{+ 0.028}_{- 0.016}$	&$ 0.115^{+ 0.012}_{- 0.012}$	&$ 0.1222^{+ 0.0090}_{- 0.0082}$&$ 0.1231^{+ 0.0075}_{- 0.0068}$&$ 0.1233^{+ 0.0089}_{- 0.0079}$&$ 0.1233^{+ 0.0082}_{- 0.0071}$\\
$h^2\Omega_b		      $    	&$ 0.0238^{+ 0.0035}_{- 0.0027}$&$ 0.0241^{+ 0.0055}_{- 0.0020}$&$ 0.0232^{+ 0.0051}_{- 0.0017}$&$ 0.0263^{+ 0.0042}_{- 0.0036}$&$ 0.0230^{+ 0.0013}_{- 0.0011}$&$ 0.0232^{+ 0.0013}_{- 0.0010}$&$ 0.0228^{+ 0.0010}_{- 0.0008}$&$ 0.0238^{+ 0.0006}_{- 0.0006}$&$ 0.0226^{+ 0.0006}_{- 0.0006}$\\
$f_\nu			      $    	&${\it   0}$			&${\it   0}$			&${\it   0}$			&${\it   0}$			&${\it   0}$			&${\it   0}$	  		&${\it   0}$			&${\it   0}$  			&${\it   0}$\\
$n_s			      $    	&$ 0.97^{+ 0.13}_{- 0.10}$	&$ 1.01^{+ 0.18}_{- 0.06}$	&$ 0.98^{+ 0.18}_{- 0.04}$	&$ 1.10^{+ 0.11}_{- 0.13}$	&$ 0.979^{+ 0.036}_{- 0.029}$	&$ 0.977^{+ 0.039}_{- 0.025}$	&$ 0.966^{+ 0.025}_{- 0.020}$	&${\it   1}$  			&${\it 0.96}$\\
$n_t + 1		      $    	&$ 0.9847^{+ 0.0097}_{- 0.0141}$&${\it   1}$			&${\it   1}$			&${\it   1}$			&${\it   1}$			&${\it   1}$	  		&${\it   1}$			&${\it   1}$  			&${\it 0.993}$\\
$A_p			      $    	&$ 0.593^{+ 0.053}_{- 0.044}$	&$ 0.602^{+ 0.053}_{- 0.051}$	&$ 0.584^{+ 0.038}_{- 0.028}$	&$ 0.582^{+ 0.043}_{- 0.025}$	&$ 0.613^{+ 0.034}_{- 0.033}$	&$ 0.633^{+ 0.024}_{- 0.022}$	&$ 0.631^{+ 0.020}_{- 0.019}$	&$ 0.642^{+ 0.023}_{- 0.022}$   &$ 0.629^{+ 0.021}_{- 0.019}$\\
$r           	      	      $    	&$<0.50\>\>(95\%)$		&${\it   0}$			&${\it   0}$			&${\it   0}$			&${\it   0}$			&${\it   0}$	  		&${\it   0}$			&${\it   0}$  			&${\it  0.15}$\\
$b			      $    	&${\it   1}$			&${\it   1}$			&$ 1.03^{+ 0.15}_{- 0.13}$	&$ 0.93^{+ 0.10}_{- 0.08}$	&$ 0.998^{+ 0.098}_{- 0.088}$	&$ 0.962^{+ 0.073}_{- 0.083}$	&$ 0.990^{+ 0.060}_{- 0.062}$	&$ 0.918^{+ 0.036}_{- 0.033}$   &$ 1.006^{+ 0.043}_{- 0.039}$\\
$w			      $    	&${\it  -1}$			&${\it  -1}$			&${\it  -1}$			&${\it  -1}$			&${\it  -1}$			&${\it  -1}$	  		&${\it  -1}$			&${\it  -1}$  			&${\it  -1}$\\
$\alpha       	              $    	&$-0.075^{+ 0.047}_{- 0.055}$	&${\it   0}$			&${\it   0}$			&${\it   0}$			&${\it   0}$			&${\it   0}$	  		&${\it   0}$			&${\it   0}$  			&${\it   0}$\\
\hline
%$\Omega_k		      $    	&$-0.095^{+ 0.094}_{- 0.144}$	&$-0.086^{+ 0.057}_{- 0.128}$	&$-0.058^{+ 0.039}_{- 0.041}$	&$-0.054^{+ 0.048}_{- 0.041}$	&$-0.012^{+ 0.018}_{- 0.022}$	&${\it   0}$	  		&${\it   0}$			&${\it   0}$  			&${\it   0}$\\
$\Ot      		      $    	&$ 1.095^{+ 0.094}_{- 0.144}$	&$ 1.086^{+ 0.057}_{- 0.128}$	&$ 1.058^{+ 0.039}_{- 0.041}$	&$ 1.054^{+ 0.048}_{- 0.041}$	&$ 1.012^{+ 0.018}_{- 0.022}$	&${\it   0}$	  		&${\it   0}$			&${\it   0}$  			&${\it   0}$\\
$\Omega_m		      $    	&$ 0.57^{+ 0.45}_{- 0.33}$	&$ 0.55^{+ 0.47}_{- 0.29}$	&$ 0.406^{+ 0.093}_{- 0.091}$	&$ 0.328^{+ 0.050}_{- 0.049}$	&$ 0.317^{+ 0.053}_{- 0.045}$	&$ 0.301^{+ 0.045}_{- 0.042}$	&$ 0.309^{+ 0.040}_{- 0.032}$	&$ 0.293^{+ 0.039}_{- 0.031}$   &$ 0.315^{+ 0.041}_{- 0.032}$\\
$h^2\Omega_m	  	      $    	&$ 0.128^{+ 0.022}_{- 0.021}$	&$ 0.132^{+ 0.021}_{- 0.028}$	&$ 0.126^{+ 0.016}_{- 0.019}$	&$ 0.117^{+ 0.024}_{- 0.013}$	&$ 0.138^{+ 0.012}_{- 0.012}$	&$ 0.1454^{+ 0.0091}_{- 0.0082}$&$ 0.1459^{+ 0.0077}_{- 0.0071}$&$ 0.1471^{+ 0.0090}_{- 0.0080}$&$ 0.1459^{+ 0.0084}_{- 0.0073}$\\
$h			      $    	&$ 0.48^{+ 0.27}_{- 0.12}$	&$ 0.50^{+ 0.16}_{- 0.13}$	&$ 0.550^{+ 0.092}_{- 0.055}$	&$ 0.599^{+ 0.090}_{- 0.062}$	&$ 0.660^{+ 0.067}_{- 0.064}$	&$ 0.695^{+ 0.039}_{- 0.031}$	&$ 0.685^{+ 0.027}_{- 0.026}$	&$ 0.708^{+ 0.023}_{- 0.024}$   &$ 0.680^{+ 0.022}_{- 0.024}$\\
$\tau 			      $    	&$ 0.33^{+ 0.17}_{- 0.17}$	&$ 0.22^{+ 0.34}_{- 0.13}$	&$ 0.18^{+ 0.31}_{- 0.10}$	&$ 0.41^{+ 0.17}_{- 0.28}$	&$ 0.143^{+ 0.089}_{- 0.066}$	&$ 0.124^{+ 0.083}_{- 0.057}$	&$ 0.103^{+ 0.060}_{- 0.047}$	&$ 0.165^{+ 0.035}_{- 0.038}$   &$ 0.092^{+ 0.036}_{- 0.036}$\\
$z_{ion}		      $    	&$25.9^{+ 4.4}_{- 8.8}$		&$20.1^{+ 9.2}_{- 8.3}$		&$18^{+10}_{- 7}$		&$26.7^{+ 3.2}_{-12.4}$		&$15.6^{+ 5.1}_{- 5.0}$		&$14.4^{+ 5.2}_{- 4.7}$	  	&$12.8^{+ 4.3}_{- 4.2}$ 	&$17.0^{+ 2.2}_{- 2.6}$  	&$11.9^{+ 2.9}_{- 3.4}$\\
$A_s    		      $    	&$ 1.14^{+ 0.42}_{- 0.31}$	&$ 0.97^{+ 0.73}_{- 0.23}$	&$ 0.86^{+ 0.68}_{- 0.16}$	&$ 1.30^{+ 0.50}_{- 0.51}$	&$ 0.82^{+ 0.14}_{- 0.11}$	&$ 0.81^{+ 0.15}_{- 0.09}$	&$ 0.777^{+ 0.100}_{- 0.072}$	&$ 0.893^{+ 0.051}_{- 0.053}$   &$ 0.758^{+ 0.050}_{- 0.050}$\\
$A_t			      $    	&$ 0.14^{+ 0.13}_{- 0.10}$	&${\it   0}$			&${\it   0}$			&${\it   0}$			&${\it   0}$			&${\it   0}$	  		&${\it   0}$			&${\it   0}$  			&$ 0.1137^{+ 0.0075}_{- 0.0074}$\\
$\beta			      $    	&$ 0.73^{+ 0.28}_{- 0.29}$	&$ 0.72^{+ 0.29}_{- 0.24}$	&$ 0.587^{+ 0.066}_{- 0.062}$	&$ 0.577^{+ 0.062}_{- 0.063}$	&$ 0.530^{+ 0.050}_{- 0.045}$	&$ 0.537^{+ 0.056}_{- 0.052}$	&$ 0.534^{+ 0.044}_{- 0.046}$	&$ 0.553^{+ 0.054}_{- 0.047}$   &$ 0.525^{+ 0.052}_{- 0.045}$\\
$t_0  [$Gyr$]		      $    	&$16.5^{+ 2.6}_{- 3.1}$		&$16.3^{+ 2.3}_{- 1.8}$		&$15.9^{+ 1.3}_{- 1.5}$		&$15.6^{+ 1.4}_{- 1.8}$		&$14.1^{+ 1.0}_{- 0.9}$		&$13.54^{+ 0.23}_{- 0.27}$	&$13.62^{+ 0.20}_{- 0.20}$	&$13.40^{+ 0.13}_{- 0.12}$  	&$13.67^{+ 0.12}_{- 0.12}$\\
$\sigma_8		      $    	&$ 0.90^{+ 0.13}_{- 0.13}$	&$ 0.87^{+ 0.15}_{- 0.13}$	&$ 0.86^{+ 0.13}_{- 0.11}$	&$ 0.948^{+ 0.089}_{- 0.101}$	&$ 0.882^{+ 0.094}_{- 0.084}$	&$ 0.917^{+ 0.090}_{- 0.072}$	&$ 0.894^{+ 0.060}_{- 0.055}$	&$ 0.966^{+ 0.046}_{- 0.050}$   &$ 0.879^{+ 0.041}_{- 0.046}$\\
%$H_1			      $    	&$ 4.8^{+ 3.8}_{- 1.9}$		&$ 7.0^{+ 4.7}_{- 1.6}$		&$ 5.5^{+ 1.7}_{- 0.6}$		&$ 6.1^{+ 2.1}_{- 1.2}$		&$ 5.04^{+ 0.42}_{- 0.39}$	&$ 5.06^{+ 0.46}_{- 0.40}$	&$ 4.98^{+ 0.39}_{- 0.39}$	&$ 5.14^{+ 0.40}_{- 0.34}$  	&$ 4.84^{+ 0.37}_{- 0.35}$\\
$H_2			      $    	&$ 0.441^{+ 0.013}_{- 0.014}$	&$ 0.4581^{+ 0.0090}_{- 0.0083}$&$ 0.4577^{+ 0.0086}_{- 0.0082}$&$ 0.4585^{+ 0.0086}_{- 0.0093}$&$ 0.4558^{+ 0.0082}_{- 0.0083}$&$ 0.4550^{+ 0.0083}_{- 0.0082}$&$ 0.4552^{+ 0.0087}_{- 0.0079}$&$ 0.4543^{+ 0.0081}_{- 0.0081}$&$ 0.4556^{+ 0.0081}_{- 0.0081}$\\
$H_3			      $    	&$ 0.424^{+ 0.043}_{- 0.040}$	&$ 0.455^{+ 0.033}_{- 0.029}$	&$ 0.444^{+ 0.026}_{- 0.025}$	&$ 0.457^{+ 0.020}_{- 0.021}$	&$ 0.449^{+ 0.021}_{- 0.021}$	&$ 0.459^{+ 0.018}_{- 0.016}$	&$ 0.454^{+ 0.013}_{- 0.012}$	&$ 0.467^{+ 0.012}_{- 0.011}$   &$ 0.451^{+ 0.011}_{- 0.010}$\\
$\Apivot		      $    	&$ 0.595^{+ 0.056}_{- 0.048}$	&$ 0.599^{+ 0.055}_{- 0.064}$	&$ 0.582^{+ 0.041}_{- 0.036}$	&$ 0.567^{+ 0.058}_{- 0.028}$	&$ 0.616^{+ 0.033}_{- 0.032}$	&$ 0.635^{+ 0.024}_{- 0.022}$	&$ 0.634^{+ 0.020}_{- 0.018}$	&$ 0.642^{+ 0.023}_{- 0.022}$   &$ 0.634^{+ 0.021}_{- 0.019}$\\
$M_\nu  [$eV$]	      	      $    	&${\it   0}$			&${\it   0}$			&${\it   0}$			&${\it   0}$			&${\it   0}$			&${\it   0}$	  		&${\it   0}$			&${\it   0}$  			&${\it   0}$\\
\hline
$\chi^2/$dof			     	&$1426.1/1339$			&$1428.4/1341$			&$1445.4/1359$			&$1619.6/1530$			&$1621.8/1530$			&$1447.2/1360$			&$1475.6/1395$			&$1447.9/1359$   		&$1447.1/1395$\\
\hline  
\end{tabular}
\end{center}     
} 
\end{table*}

\subsection{Constraints}

%We constrain theoretical models using the 
%Monte Carlo Markov Chain based 
% technique \cite{Metropolis,Hastings,Gilks96,GelmanRubin92,Christensen01,Lewis02,Slosar03}
%described in Appendix A.
We constrain theoretical models using the Monte Carlo Markov Chain 
method \cite{Metropolis,Hastings,Gilks96,GelmanRubin92,Christensen01,Lewis02,Slosar03}
implemented as described in Appendix A.
Unless otherwise stated, we use the WMAP temperature and cross-polarization power spectra 
\cite{Bennett03,Hinshaw03,kogut03,Page3-03}, evaluating likelihoods with the software
provided by the WMAP team \cite{Verde03}. % Explain the required hack in appendix
When using SDSS information, we fit the nonlinear theoretical power spectrum $P(k)$ 
approximation of \cite{Smith03} to the observations 
reported by the SDSS team \cite{sdsspower}, assuming an unknown
scale-independent linear bias $b$ % WMAP team assume this too
to be marginalized over. This means that we use only the shape of the measured SDSS power spectrum,
not its amplitude.
We use only the measurements with $k\le 0.2h/\Mpc$ as suggested by \cite{sdsspower}.
The WMAP team used this same $k$-limit when analyzing the 2dFGRS \cite{Verde03};
we show in \Sec{DataRobustnessSec} that cutting back to $k\le 0.15h/\Mpc$ causes a
negligible change in our best-fit model.
To be conservative, we do not use the SDSS measurement of redshift space distortion parameter 
$\beta$ \cite{sdsspower},
nor do we use any other information (``priors'') whatsoever unless explicitly stated.
When using SN Ia information, we 
employ the 172 SN Ia redshifts and corrected magnitudes compiled and uniformly 
analyzed by Tonry {\etal} \cite{Tonry03}, 
evaluating the likelihood with the software provided by their team, which 
marginalizes over the corrected SN Ia ``standard candle'' absolute magnitude.
Note that this is an updated and expanded data set from that available
to the WMAP team when they carried out their analysis\cite{Spergel03}.

Our constraints on individual cosmological parameters are given 
in Tables~{\WMAPtable}-{\AddinfoTable} and illustrated in \fig{1d_6par_fig},
both for WMAP alone and when including additional information such as that from the SDSS.
To avoid losing sight of the forest for all the threes (and other digits), we will 
spend most of the remainder of this paper digesting this voluminous information one 
step at a time, focusing on what WMAP and SDSS do and don't tell us about 
the underlying physics. 
The one-dimensional constraints in the tables and \fig{1d_6par_fig} 
fail to reveal important information hidden in parameter correlations and degeneracies,
so a powerful tool will be studying the joint constraints on key 2-parameter 
pairs. We will begin with a simple 6-parameter space of models, 
then gradually introduce additional parameters to quantify both how accurately we can
measure them and to what extent they 
weaken the constraints on the other parameters.

\section{Vanilla $\Lambda$CDM models}
\label{VanillaSec}

In this section, we explore constraints on six-parameter ``vanilla'' models
that have no spatial curvature ($\Ok=0$), no gravity waves ($r=0$), no running tilt ($\al=0$),
negligible neutrino masses ($\fn=0$) and dark energy corresponding to a pure cosmological
constant ($w=-1$). These vanilla $\Lambda$CDM models are thus determined by merely 
six parameters: the matter budget ($\Ol,\od,\ob$), 
the initial conditions ($\As,\ns$) and the reionization optical depth
$\tau$. (When including SDSS information, we bring in the bias parameter $b$ as well.)

Our constraints on individual cosmological parameters are shown in 
Tables~{\WMAPtable}-{\AddinfoTable}
and \fig{1d_6par_fig}
both for WMAP alone and when including SDSS information. 
Several features are noteworthy.

First of all, as emphasized by the WMAP team \cite{Spergel03}, error
bars have shrunk dramatically compared to the situation before WMAP,
and it is therefore quite impressive that {\it any} vanilla model is
still able to fit both the unpolarized 
and polarized CMB data.
The best fit model (Table~\WMAPtable) has $\chi^2\sim 1431.5$ for $899+449-6=1342$ effective degrees of 
freedom, \ie,
% WMAP T:    899 points, chi2=972.6  (1.88 sigma)
% WMAP X:    449 points
% SDSS P:    19 points
% SN Ia:     172 points
% Other CMB:  35 points
% All CMB:   151 points
% (1431.5-1342)/sqrt(2*1342) ~ 1.728
about $1.7\sigma$ high if taken at face value. The 
WMAP team provide an extensive discussion of possible origins of this slight excess, 
and argue that it comes mainly from three unexplained ``blips''
\cite{Verde03,Lewis03}, 
deviations from the model fit over a narrow range of $\l$, in the
measured temperature power spectrum. 
They argue that these blips have nothing to do with features in 
any standard cosmological models, since adding the above-mentioned non-vanilla parameters does
not reduce $\chi^2$ substantially --- we confirm this below, and will
not dwell further on these sharp features.  
Adding the 19 SDSS data points increases the effective degrees of
freedom by $19-1=18$ (since this requires the addition of the bias
parameter $b$), yet raises the best-fit $\chi^2$ by only $15.7$. Indeed, \fig{SausageFig} shows that even 
the model best fitting WMAP alone does a fine job at fitting the SDSS
data with no further parameter tuning.

\begin{figure} 
%\vskip\smtopskip
\centerline{\epsfxsize=\figsize\epsffile{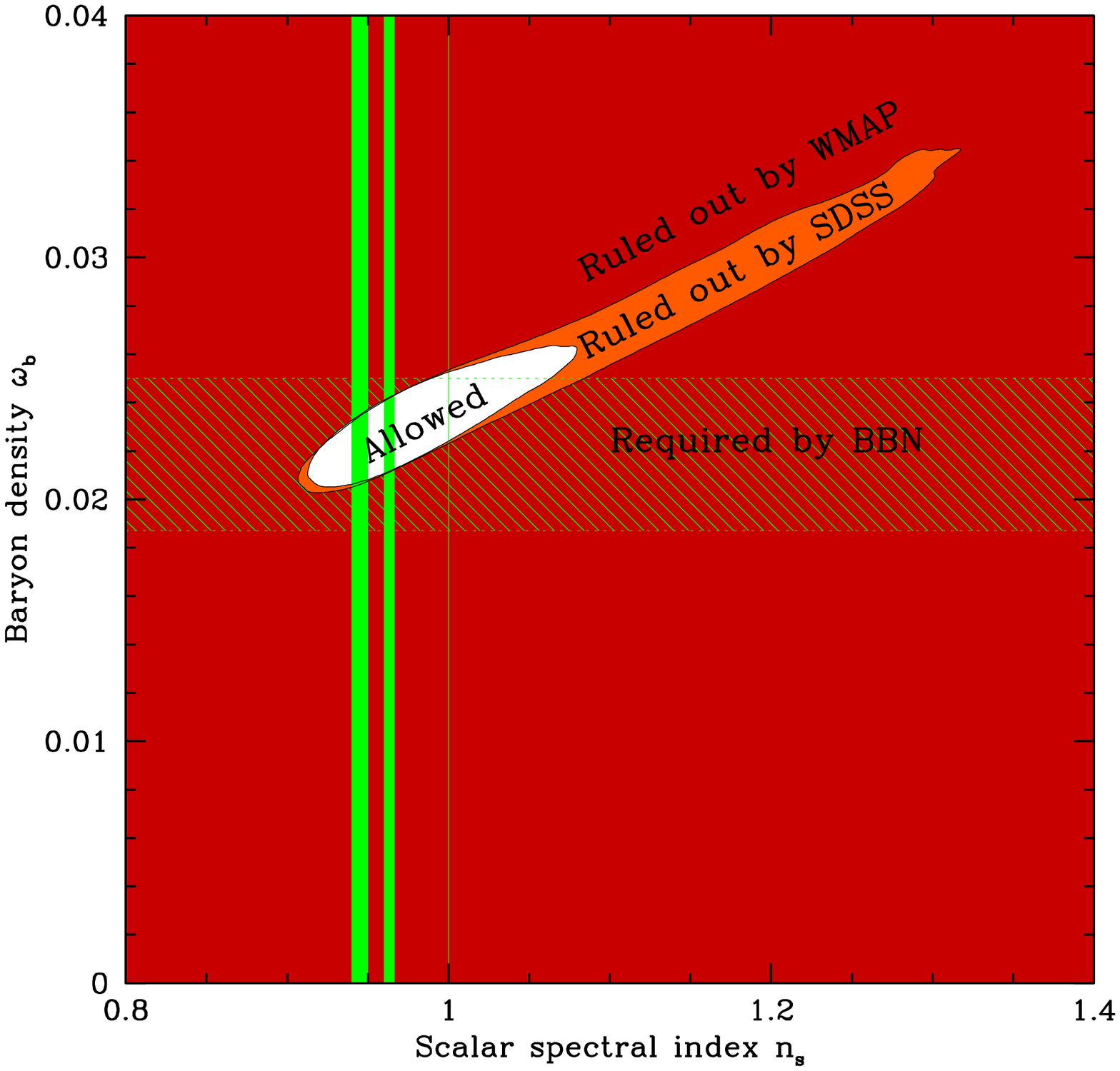}}
%\vskip\smbotskip
\caption[1]{\label{2d_nsob_6par_fig}\footnotesize%
95\% constraints in the $(\ns,\ob)$ plane. 
The shaded dark red/grey region is ruled out by WMAP alone
for 6-parameter ``vanilla'' models, leaving the long degeneracy 
banana discussed in the text. 
The shaded light red/grey region is ruled out when adding SDSS information.
The hatched band is required by Big Bang Nucleosynthesis (BBN).
From right to left, the three vertical bands correspond
to a scale-invariant Harrison-Zel'dovich spectrum and to the 
common inflationary predictions $\ns=1-2/N\sim 0.96$ and $\ns=1-3/N\sim 0.94$
(Table~\protect\InflationTable), 
assuming that the number of e-foldings
between horizon exit of the observed fluctuations and 
the end of inflation is $50<N<60$.
}
\end{figure}

\begin{figure} 
%\vskip\smtopskip
\centerline{\epsfxsize=\figsize\epsffile{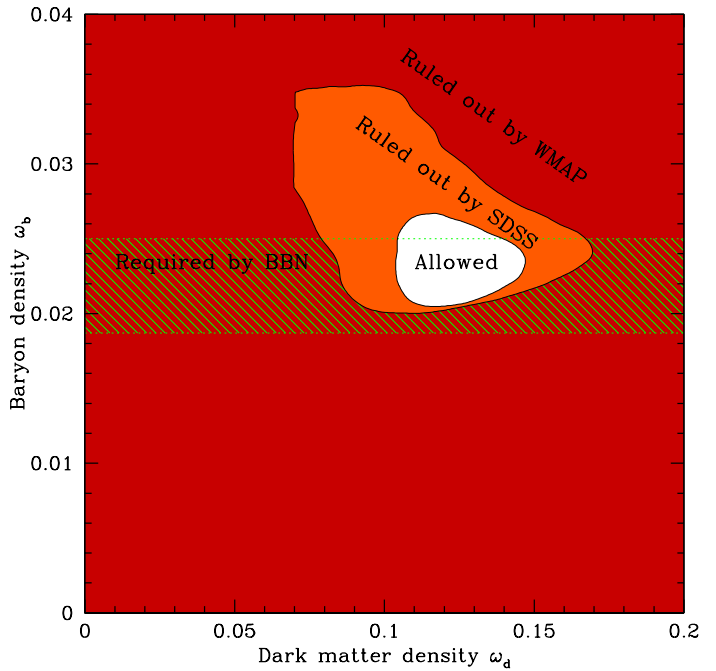}}
%\vskip\smbotskip
\caption[1]{\label{2d_odob_6par_fig}\footnotesize%
95\% constraints in the $(\od,\ob)$ plane. 
Shaded dark red/grey region is ruled out by WMAP alone
for 6-parameter ``vanilla'' models. 
The shaded light red/grey region is ruled out when adding SDSS information.
The hatched band is required by Big Bang Nucleosynthesis (BBN).
}
\end{figure}

\begin{figure} 
%\vskip\smtopskip
\centerline{\epsfxsize=\figsize\epsffile{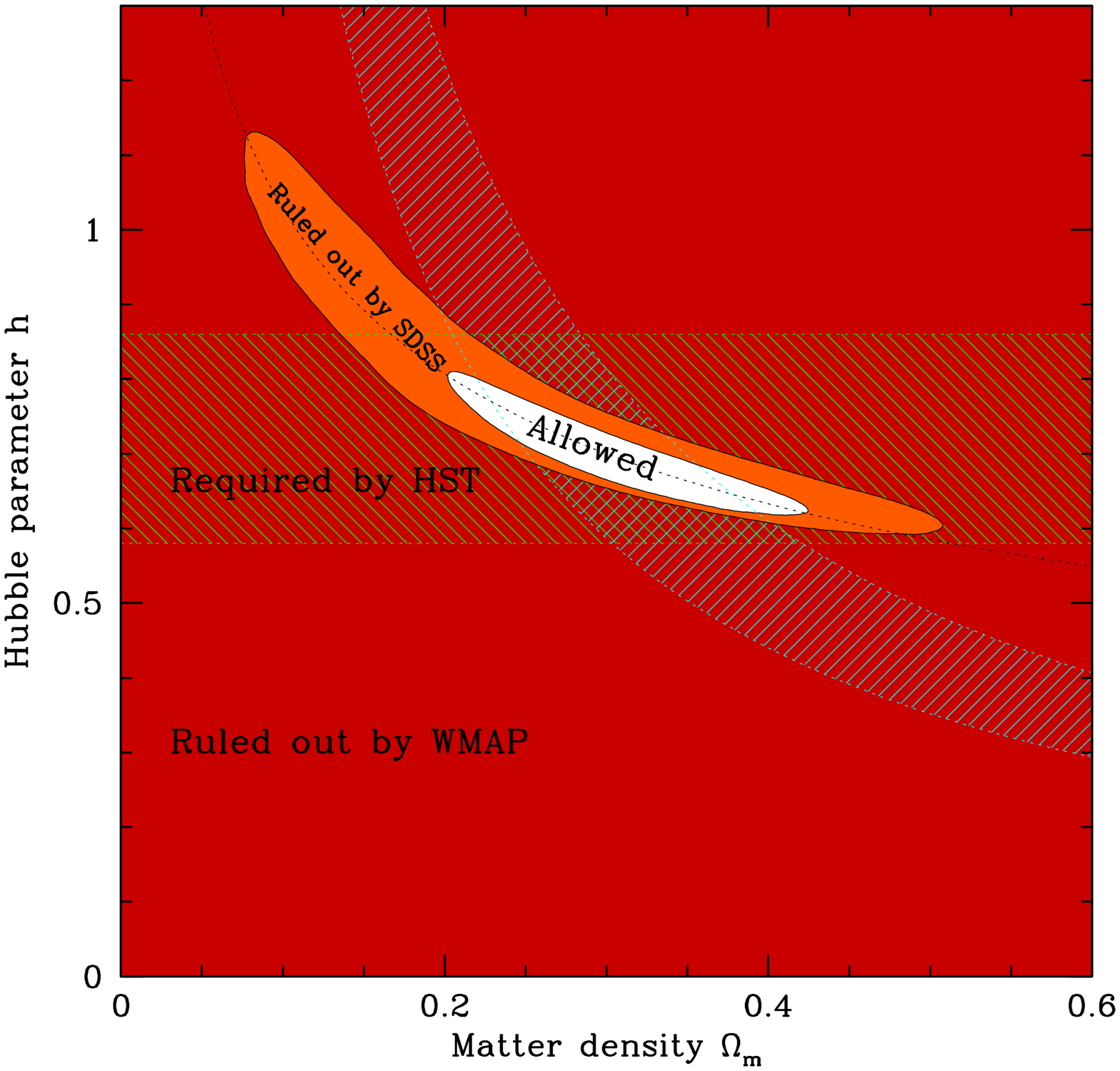}}
%\vskip\smbotskip
\caption[1]{\label{2d_Omh_6par_fig}\footnotesize%
95\% constraints in the $(\Om,h)$ plane. 
Shaded dark red/grey region is ruled out by WMAP alone
for 6-parameter ``vanilla'' models, leaving the long degeneracy 
banana discussed in the text. 
The shaded light red/grey region is ruled out when adding SDSS information,
which can be understood as SDSS accurately measuring the $P(k)$ ``shape 
parameter'' $h\Omega_m=0.21\pm 0.03$ at $2\sigma$ (sloping hatched band).
The horizontal hatched band is required by the HST key project
\protect\cite{Freedman01}. 
The dotted line shows the fit
$h = 0.7(\Om/0.3)^{-0.35}$, explaining the origin of the
accurate constraint $h(\Om/0.3)^{0.35} = 0.70\pm 0.01$ $(1\sigma)$.
%&$ 0.697^{+ 0.012}_{- 0.011}$
}
\end{figure}

\subsection{The vanilla banana}
\label{VanillaBananaSec}

Second, our WMAP-only constraints are noticeably weaker than those
reported by \cite{Spergel03}, mostly because we did not place a prior
on the value of the reionization optical depth $\tau$, and 
adding SDSS information helps rather dramatically with all of our six basic parameters,
roughly halving the $2\sigma$ error bars.
The physical explanation for both of these facts is that the allowed subset of our 6-dimensional parameter
space forms a rather elongated banana-shaped region. In the 2-dimensional projections shown
(Figures \ref{2d_nsob_6par_fig}, \ref{2d_odob_6par_fig}, \ref{2d_Omh_6par_fig} and \ref{2d_Oms8_6par_fig}), 
this is most clearly seen in Figures \ref{2d_nsob_6par_fig} and \ref{2d_Omh_6par_fig}.
Moving along this degeneracy banana, all six parameters
$(\tau,\Ol,\od,\ob,\As,\ns)$ increase together, as does $h$.

There is nothing physically profound about this one-dimensional degeneracy.
Rather, it is present because we are fitting six parameters to only five 
basic observables: the heights of the first three acoustic peaks, 
the large-scale normalization and the angular peak location.
Within the vanilla model space, all models fitting these five 
observables will do a decent job at fitting the power spectra everywhere
that WMAP is sensitive \cite{observables}.
As measurements improve and include additional peaks, this 
approximate degeneracy will go away.

Here is how the banana degeneracy works in practice: increasing $\tau$ and $\As$ in such a way
that $\Ap\equiv\As e^{-2\tau}$ stays constant, the peak heights remain unchanged and the only effect is 
to increase power on the largest scales. The large-scale power relative to the first peak can  
be brought back down to the observed value by increasing $\ns$, after which the second peak
can be brought back down by increasing $\ob$. Adding WMAP polarization information actually lengthens
rather than shortens the degeneracy banana, by stretching out the range of preferred
$\tau$-values --- the largest-scale polarization measurement prefers very 
high $\tau$ (\fig{SausageFig}) while the unpolarized measurements prefer $\tau=0$.
This banana degeneracy was also discussed in numerous accuracy forecasting papers 
% CITE? 9par?
and older parameter constraint papers \cite{parameters,ZSS97,parameters2,EfstathiouBond99}.

Since the degeneracy involves all the parameters, essentially any extra piece of information will break it.
The WMAP team break it by imposing a prior (assuming $\tau<0.3$), which cuts off
much of the banana.  Indeed, \fig{1d_6par_fig} shows that the distribution for several parameters
(notably the reionization redshift $\zion$) are bimodal, so this prior eliminates the 
rightmost of the two bumps.
In the present paper, we wish to keep assumptions to a minimum and 
therefore break the degeneracy using the SDSS measurements instead.
\Fig{2d_Omh_6par_fig} illustrates the physical reason that this works so well:
SDSS accurately measures the $P(k)$ ``shape 
parameter'' $\Gamma\equiv h\Omega_m=0.21\pm 0.03$ at $2\sigma$ \cite{sdsspower}, 
%which crudely speaking shifts $P(k)$ horizontally, 
which crudely speaking determines the horizontal position of $P(k)$
and this allowed region in the $(\Om,h)$-plane 
intersects the CMB banana at an angle.
Once $E$-polarization results from WMAP become available, they should provide another powerful way of 
breaking this degeneracy from WMAP alone, by directly constraining $\tau$ --- from our WMAP+SDSS analysis, we 
make the prediction $\tau<0.29$ at 95\% confidence for what this measurement should find.
% That's tau < 0.28636
(Unless otherwise specified, we quote $1\sigma$ limits in text and tables, 
whereas the 2-dimensional figures show $2\sigma$ limits.)

\Fig{2d_Omh_6par_fig} shows that the banana is well fit by 
$h = 0.7(\Om/0.3)^{-0.35}$, so even from WMAP+SDSS alone, 
we obtain the useful precision constraint
$h(\Om/0.3)^{0.35} = 0.697^{+0.012}_{-0.011}$ (68\%).

\subsection{Consistency with other measurements}

\Fig{2d_nsob_6par_fig} shows that the WMAP+SDSS allowed value of the baryon density 
$\ob=0.023\pm 0.001$ 
agrees well with the latest measurements $\ob=0.022\pm 0.002$ from 
Big Bang Nucleosynthesis \cite{Cyburt03,Cuoco03,Coc03}.
It is noteworthy that the WMAP+SDSS preferred value is higher than the BBN preferred value 
$\ob=0.019\pm 0.001$ of a few years ago \cite{BurlesTytler98}, so the excellent agreement 
hinges on improved reaction rates in the theoretical BBN predictions \cite{Cuoco03}
and a slight decrease in observed deuterium abundance.
This is not to be confused with the more dramatic drop in inferred
deuterium abundance in preceding years as data improved, which
raised the $\ob$ prediction from $\ob=0.0125\pm 0.00125$ \cite{Walker91,Smith93}.

The existence of dark matter could be inferred from CMB alone only as recently as 
2001 \cite{consistent}, cf. \cite{concordance}, yet 
\fig{2d_odob_6par_fig} shows that WMAP alone requires dark matter at very high 
significance, refuting the suggestion of \cite{McGaugh00} that 
an alternative theory of gravity with no dark matter can explain CMB observations.

Table~{\SDSStable} shows that once WMAP and SDSS are combined, 
the constraints on three of the six vanilla parameters 
($\ob$, $\od$ and $\ns$)
are quite robust to the choice of theoretical priors on the other parameters.
This is because the CMB information that constrains them is mostly the relative heights of
the first three acoustic peaks, which are left unaffected by all the other parameters
except $\alpha$.
The four parameters $(\Ok,r,w,\fn)$ that are fixed by priors in many published
analyses cause only a horizontal shift of the peaks ($\Ok$ and $w$)
and modified CMB power on larger angular scales
(late ISW effect from $\Ok$ and $w$, tensor power from $r$).

\Fig{2d_Omh_6par_fig} illustrates that two of the most basic cosmological parameters,
%most frequently discussed in the astronomy literature, 
$\Om$ and $h$, are %unfortunately
not well constrained by WMAP alone even for vanilla models, 
uncertain by factors of about two and five, respectively (at 95\% confidence). 
After including the SDSS information, however, the constraints are seen to
shrink dramatically, giving Hubble parameter constraints
$h\approx 0.70^{+0.04}_{-0.03}$ that are even tighter than (and in good agreement with)
those from the HST project, $h=0.72\pm 0.07$ \cite{Freedman01},
which is of course a completely independent measurement based on entirely different 
physics. (But see the next section for the crucial caveats.) 
Our results also agree well with those from the WMAP team, who obtained 
$h\approx 0.73\pm 0.03$ \cite{Spergel03} by combining WMAP with the
2dFGRS. 
Indeed, our value for $h$ is about 1 $\sigma$ lower.  This is because 
%The reason that we obtain marginally lower $h$-values is that 
the SDSS power spectrum 
has a slightly bluer slope than that of 2dFGRS, favoring slightly higher $\Om$-values
(we obtain $\Om=0.30\pm 0.04$ as compared to the WMAP+2dFGRS value 
$\Om=0.26\pm 0.05$).
% Spergel et al table 7, WMAPext+2dFGRS gives 
% Om_min = (0.134-0.006)/(0.72+0.05)**2 ~ 0.216
% Om_max = (0.134+0.006)/(0.72-0.05)**2 ~ 0.312
% Om ~ 0.264 +/- 0.048
As discussed in more detail in \Sec{DiscussionSec}, this slight difference may be linked to 
differences in modeling of non-linear redshift space distortions and
bias.
For a thorough and up-to-date review of recent $h-$ and $\Om$-determinations, see \cite{Spergel03}.

Whereas the constraints of $\ob$, $\od$ and $\ns$ are rather robust,
we will see in the following section that our constraints on 
$h$ and $\Om$ hinge crucially on the assumption that space is perfectly flat, and become 
substantially weaker when dropping that assumption.

The last columns of Table~{\SDSStable} demonstrate excellent consistency
with pre-WMAP CMB data (Appendix A.3), which involves not only independent experiments but also
partly independent physics, with much of the information coming from 
small angular scales $\l\simgt 600$ where WMAP is insensitive.
In other words, our basic results and error bars
still stand even if we discard either WMAP or pre-WMAP data.
Combining WMAP and smaller-scale CMB data (Table~{\AddinfoTable}, 3rd last column)
again reflects this consistency,
tightening the error bars around essentially the same central values.

\def\sss{\hglue3mm}
\begin{table}
\label{SystematicsTable}
\noindent 
{\footnotesize
Table~{\ClusterLensTable}: Recent constraints in the $(\Om,\sigma_8)$-plane.
%{\bf Table 3} -- Recent measurements of the power spectrum normalization $\sigma_8$.
%For results quoted as fitting formulas involving $\Om$, we have set $\Om=0.3$.
%Error bars are $1\sigma$ --- for \cite{Hoekstra02a} and \cite{Jarvis02}, 
%we have simply divided their quoted 95\% errors by 2.
\begin{center}
{\footnotesize
\begin{tabular}{lll}
\hline
Analysis&&Measurement\\
\hline
{\bf Clusters:}\\
\sss Voevodkin \& Vikhlinin '03         &\cite{Voevodkin03}     &$\sigma_8=0.60+0.28\Om ^{0.5}\pm0.04$\\
\sss Bahcall \& Bode '03, 		$z<0.2$	&\cite{BahcallBode03}	&$\sigma_8(\Om/0.3)^{0.60} = 0.68\pm 0.06$\\  		%  sigma_8*Omega_m^0.6=0.33 +/-0.03 (68%)  
\sss Bahcall \& Bode '03, $z>0.5$	&\cite{BahcallBode03}	&$\sigma_8(\Om/0.3)^{0.14} = 0.92\pm 0.09$\\  		%  sigma_8*Omega_m^0.6=0.33 +/-0.03 (68%)  
\sss Pierpaoli {\etal} '02        	&\cite{Pierpaoli02}     &$\sigma_8=0.77^{+0.05}_{-0.04}$\\
%\sss Allen {\etal} '03			&\cite{Allen0208394}	&$\sigma_8\Om^{0.253} = 0.508\pm 0.026$\\  		% sigma_8 = (0.508\pm0.019) Omega_m^-(0.253\pm0.024)  (68%)
%\sss Allen {\etal} '03			&\cite{Allen0208394}	&$\sigma_8(\Om/0.3)^{0.253} = 0.689\pm 0.019$\\  	% sigma_8 = (0.508\pm0.019) Omega_m^-(0.253\pm0.024)  (68%)
\sss Allen {\etal} '03			&\cite{Allen0208394}	&$\sigma_8(\Om/0.3)^{0.25} = 0.69\pm 0.04$\\  		% sigma_8 = (0.508\pm0.019) Omega_m^-(0.253\pm0.024)  (68%)
\sss Schuecker{\etal} '02         	&\cite{Schuecker02}     &$\sigma_8=0.711^{+0.039}_{-0.031}$\\
%\sss Viana {\etal} '01            	&\cite{Viana02}         &$\sigma_8=0.61\pm 0.05$\\
\sss Viana {\etal} '02            	&\cite{Viana02b}        &$\sigma_8=0.78_{-0.03}^{+0.15}$ (for $\Om=0.35$)\\ %at 95 per cent confidence for $\Omega_0 = 0.35 \sigma_8=0.78_{-0.06}^{+0.30}
\sss Seljak '02           		&\cite{Seljak02}        &$\sigma_8(\Om/0.3)^{0.44} = 0.77\pm 0.07$\\ % *(Gamma/0.2)**0.08, Gamma=0.2
%\sss Reiprich \& B\"ohringer '01  	&\cite{Reiprich01}      &$\sigma_8(\Om/0.3)^{0.38} = 0.68^{+0.08}_{-0.06}$\\ % Error bars from where?
%\sss Reiprich \& B\"ohringer '02  	&\cite{Reiprich01}      &$\sigma_8(\Om/0.3)^{0.38} \approx 0.68$\\ 	% Error bars from where?
\sss Reiprich \& B\"ohringer '02  	&\cite{Reiprich01}      &$\sigma_8=0.96^{+0.15}_{-0.12} $\\%(90% c.l.)
%\sss Borgani {\etal} '01          	&\cite{Borgani01}       &$\sigma_8=0.76^{+0.08}_{-0.05}$\\
\sss Borgani {\etal} '01          	&\cite{Borgani01}       &$\sigma_8=0.66\pm -0.06$\\
%\sss Pierpaoli {\etal} '01        	&\cite{Pierpaoli01}     &$\sigma_8\Om^{0.60}=0.495^{+0.034}_{-0.037}$\\  	% sigma_8 Omega_M^{0.60}= 0.495^{+0.034}_{-0.037} for Gamma=0.23 (68%)
\sss Pierpaoli {\etal} '01        	&\cite{Pierpaoli01}     &$\sigma_8(\Om/0.3)^{0.60}=1.02^{+0.070}_{-0.076}$\\  	% sigma_8 Omega_M^{0.60}= 0.495^{+0.034}_{-0.037} for Gamma=0.23 (68%)
\hline
{\bf Weak lensing:}\\
%\sss Heymans {\etal} '03           	&\cite{COMBO03}         &$\sigma_8(\Om/0.27)^{0.6} =0.71\pm0.11$\\ 
%\sss Heymans {\etal} '03           	&\cite{COMBO03}         &$\sigma_8(\Om/0.27)^{0.6} =0.67\pm0.10$\\ 
\sss Heymans {\etal} '03           	&\cite{COMBO03}         &$\sigma_8(\Om/0.3)^{0.6} =0.67\pm0.10$\\ 
\sss Jarvis {\etal} '02           	&\cite{Jarvis02}        &$\sigma_8(\Om/0.3)^{0.57}=0.71^{+0.06}_{-0.08}$\\		% 95% is 0.71^{+0.12}_{-0.16}
\sss Brown {\etal} '02            	&\cite{Brown02}         &$\sigma_8(\Om/0.3)^{0.50}=0.74\pm 0.09$\\	 		% 68%, PLOTTED
\sss Hoekstra {\etal} '02         	&\cite{Hoekstra02}      &$\sigma_8(\Om/0.3)^{0.52}=0.86^{+0.05}_{-0.07}$\\   	% 68%, PLOTTED.
%\sss Refregier{\etal} '02         	&\cite{Refregier02}     &$\sigma_8=0.94\pm 0.14$\\
\sss Refregier{\etal} '02         	&\cite{Refregier02}     &$\sigma_8(\Om/0.3)^{0.44}=0.94^{+0.24}_{-0.24}$\\
\sss Bacon {\etal} '02            	&\cite{Bacon02}         &$\sigma_8(\Om/0.3)^{0.68} = 0.97\pm0.13$\\			% 68%, PLOTTED
%\sss Van Waerbeke {\etal} (2002)      &\cite{Waerbeke02}      &$\sigma_8=0.97\pm 0.06$\\
\sss Van Waerbeke {\etal} '02      &\cite{Waerbeke02}      &$\sigma_8(\Om/0.3)^{(0.24\mp 0.18)\Om-0.49}$\\
                                       &                      &$=0.94^{+0.14}_{-0.12}$\\
%\sss Hamana {\etal} '02            	&\cite{Hamana02}        &$\sigma_8=(0.50_{-0.08}^{+0.18})\Om^{-0.37}$\\  	%95% confidence	sigma_8=(0.50_{-0.16}^{+0.35})Omega_m^{-0.37}			       
\sss Hamana {\etal} '02            	&\cite{Hamana02}        &$\sigma_8(\Om/0.3)^{-0.37}=(0.78_{-0.12}^{+0.27})$\\  	%95% confidence	sigma_8=(0.50_{-0.16}^{+0.35})Omega_m^{-0.37}			       
\hline
\end{tabular}
}
\end{center}     
} 
\end{table}

\begin{figure} 
%\vskip\smtopskip
\centerline{\epsfxsize=\figsize\epsffile{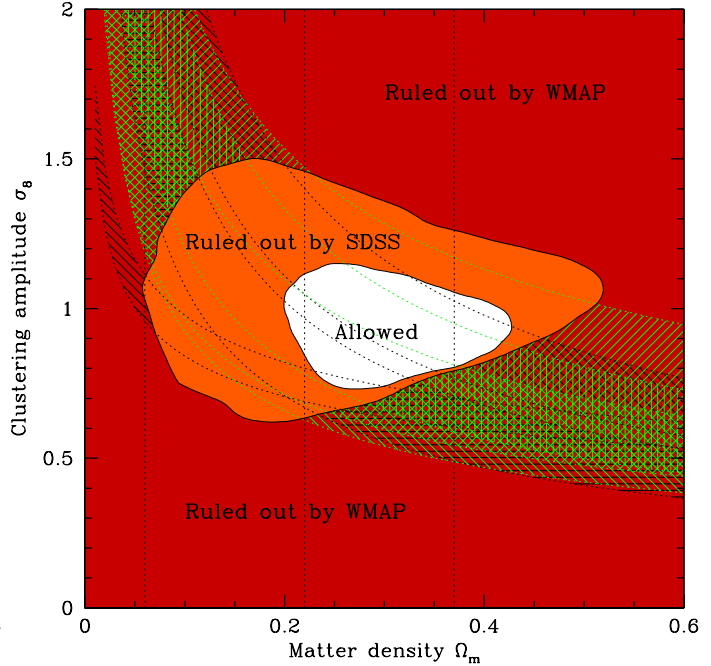}}
%\vskip\smbotskip
\caption[1]{\label{2d_Oms8_6par_fig}\footnotesize%
95\% constraints in the $(\Om,\sigma_8)$ plane. 
Shaded dark red/grey region is ruled out by WMAP alone
for 6-parameter ``vanilla'' models.
The shaded light red/grey region is ruled out when adding SDSS information.
The 95\% confidence regions are hatched for various recent 
cluster (black) and lensing (green/grey) analyses as discussed in the 
text. 
The vertical lines indicate the constraints described in \cite{Bridle03} 
from mass-to-light ratios in galaxies and clusters ($0.06\simlt\Om\simlt 0.22$)
and from cluster baryon fractions ($0.22\simlt\Om\simlt0.37$).
}
\end{figure}

\Fig{2d_Oms8_6par_fig} compares various constraints on the
linear clustering amplitude $\sigma_8$. Constraints from both
galaxy clusters \cite{Pierpaoli01,BahcallBode03,Allen0208394}
(black) 
and 
weak gravitational lensing \cite{Brown02,Hoekstra02,Bacon02} (green/grey) are shown as shaded bands in the
$(\Om,\sigma_8)$-plane for the recent measurements listed in Table~{\ClusterLensTable} and are seen to 
all be consistent 
with the WMAP+SDSS allowed region.
However, we see that there is no part of the allowed region that simultaneously 
matches all the cluster constraints, indicating that cluster-related systematic 
uncertainties such as the mass-temperature relation may still not have been 
fully propagated into the quoted cluster error bars.

Comparing \fig{2d_Oms8_6par_fig} with Figure 2 from \cite{Contaldi03} demonstrates excellent 
consistency with an analysis combining the weak lensing data of \cite{Hoekstra02} 
(Table~{\ClusterLensTable}) with WMAP, small-scale CMB data and an $\ob$-prior from 
Big Bang Nucleosynthesis. \Fig{2d_Oms8_6par_fig} also shows good consistency with
$\Om$-estimates from cluster baryon fractions \cite{Bridle03}, which in turn are
larger than estimates based on mass-to-light ratio techniques reported in \cite{Bridle03}
(see \cite{Ostriker03} for a discussion of this).

The constraints on the bias parameter $b$ in Tables~{\SDSStable}
and~{\AddinfoTable} refer to the clustering amplitude of 
SDSS $L_*$ galaxies at the effective redshift of the survey
relative to the clustering amplitude of dark matter at $z=0$.
If we take $z\sim 0.15$ as the effective redshift based on 
Figure 31 in \cite{sdsspower}, then the ``vanilla lite'' model 
(second last column of Table~{\AddinfoTable})
gives dark matter fluctuations $0.925$ times their present value
and hence a physical bias factor $b_*=b/0.925=0.918/0.925\approx 0.99$,
in good agreement with the completely independent 
measurement $b_*=1.04\pm 0.11$ \cite{Verde02} based on the bispectrum
of $L_*$ 2dFGRS galaxies. A thorough discussion of such bias
cross-checks is given by \cite{Lahav02}.

\begin{figure} 
%\vskip\smtopskip
\centerline{\epsfxsize=\figsize\epsffile{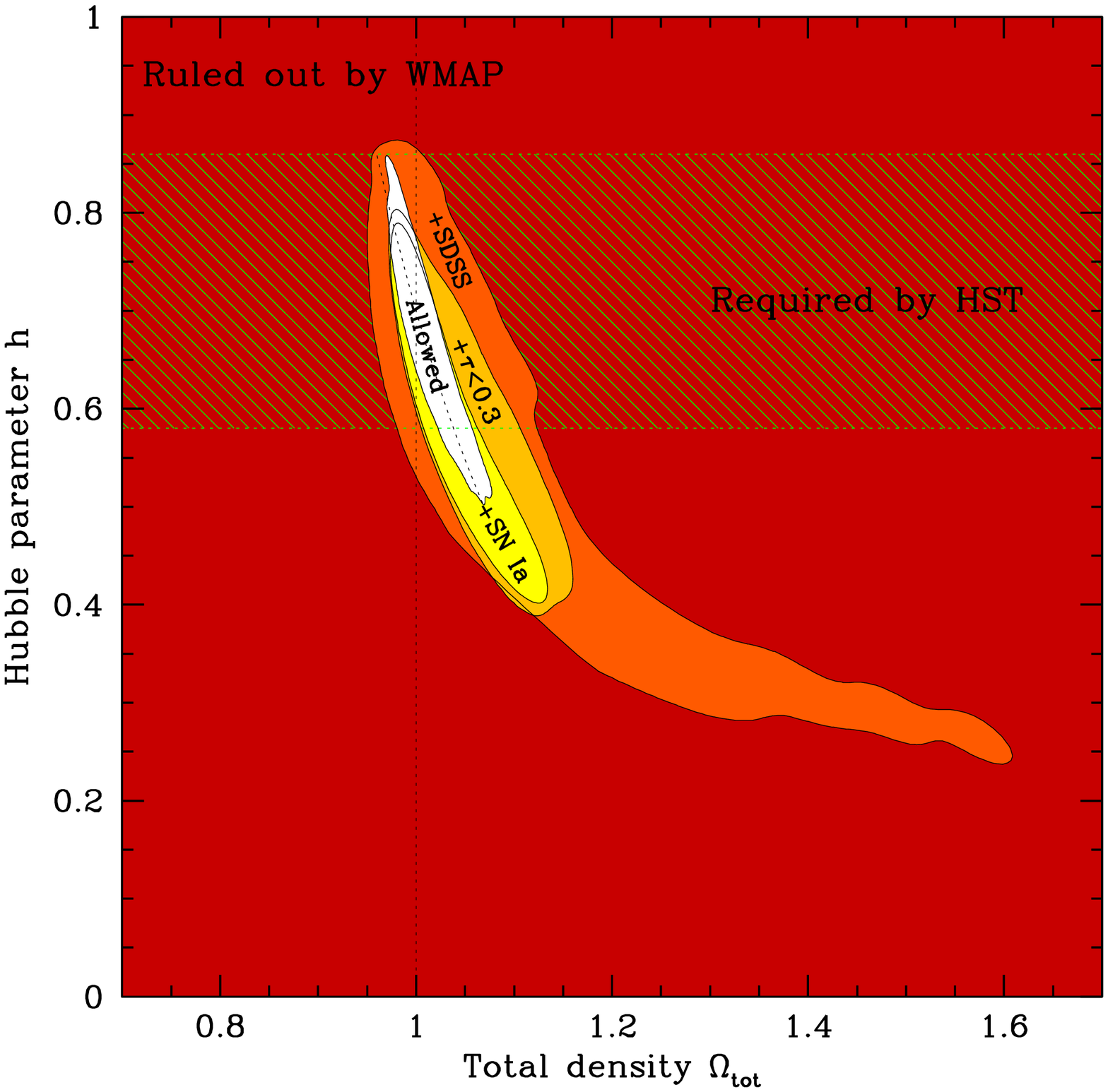}}
%\vskip\smbotskip
\caption[1]{\label{2d_Oth_7par_fig}\footnotesize%
95\% constraints in the $(\Ot,h)$ plane. 
Shaded dark red/grey region is ruled out by WMAP alone
for 7-parameter curved models, showing that 
%why one cannot claim that
CMB fluctuations alone do not simultaneously show space to be flat and measure the Hubble parameter.
The shaded light red/grey region is ruled out when adding SDSS information.
Continuing inwards, the next two regions are ruled out when 
adding the $\tau<0.3$ assumption and when adding SN Ia information as well.
The light hatched band is required by the HST key project
\protect\cite{Freedman01}. 
The dotted line shows the fit
$h = 0.7\Ot^{-5}$, explaining the origin of the
accurate constraints
$h\Ot^5 = 0.703^{+ 0.029}_{- 0.024}$ and
$\Ot(h/0.7)^{0.2}=1.001^{+ 0.008}_{- 0.007}$ $(1\sigma)$.
}
\end{figure}

\begin{figure} 
%\vskip\smtopskip
\centerline{\epsfxsize=\figsize\epsffile{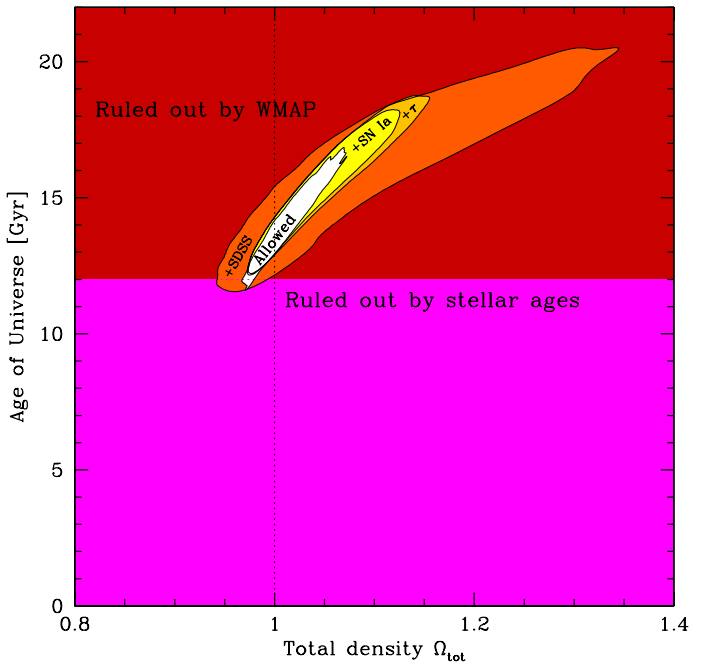}}
%\vskip\smbotskip
\caption[1]{\label{2d_Ott0_7par_fig}\footnotesize%
95\% constraints in the $(\Ot,t_0)$ plane. 
Shaded dark red/grey region is ruled out by WMAP alone
for 7-parameter curved models, showing that 
%why one cannot claim that
CMB fluctuations do not simultaneously show space to be flat and measure the age of 
the Universe.
The shaded light red/grey region is ruled out when adding SDSS information.
Continuing inwards, the next two regions are ruled out when 
adding the $\tau<0.3$ assumption and when adding SN Ia information as well.
Stellar age determinations (see text) rule out $t_0<12$ Gyr.
%The dotted line show the fit
%$h = 0.7\Ot^4$, explaining the origin of the
%accurate constraint $h\Ot^{-4} = 0.70\pm 0.0??$ $(1\sigma)$.
}
\end{figure}

\section{Curved models}
\label{CurvatureSec}

Let us now spice up the vanilla model space by adding spatial curvature $\Ok$
as a free parameter, both to constrain the curvature and to quantify how other
constraints get weakened when dropping the flatness assumption. 

%It is well-known that CMB shows space to be roughly flat 
%($|\Ok|\simlt 30\%$). 
% There is a persistent myth that
Figures~\ref{2d_Oth_7par_fig} and~\ref{2d_Ott0_7par_fig} show that
there is a strong degeneracy between the curvature of the universe
$\Ok \equiv 1 - \Ot$ and both the Hubble parameter $h$ and the age of the universe
$t_0$, when constrained by WMAP alone (even with only the seven parameters
we are now considering allowed to change); without further information or priors, one
cannot simultaneously demonstrate spatial flatness and measure $h$ or
$t_0$.  
%According to some popular accounts,
%CMB alone {\it both} shows spatial flatness and 
%accurately measures quantities such as the Hubble parameter and the age of
%the Universe. Figures~\ref{2d_Oth_7par_fig} and~\ref{2d_Ott0_7par_fig}
%show that this is incorrect.
% indeed a mere myth.
We see that although WMAP alone abhors open models, requiring 
$\Ot\equiv\Om+\Ol = 1-\Ok\simgt 0.9$ (95\%), 
closed models with $\Ot$ as large as 1.4 are still marginally allowed provided
that the Hubble parameter $h\sim 0.3$ and the age of the Universe $\age\sim 20$ Gyr.
%To claim that CMB observations alone have nailed down $\Ok$, $h$ and $\age$ jointly, one would thus need 
%to fold in theoretical prejudice and make the following rather dubious argument:
%``CMB shows that space is roughly flat (to within 40\% or so), so space
%must be exactly flat, so $h$ and $\age$ are tightly constrained.''
Although most inflation models do predict space to be flat and 
closed inflation models require particularly ugly fine-tuning \cite{Linde0303245}, 
a number of recent papers on other subjects have considered nearly-flat models 
either to explain the low CMB quadrupole \cite{Efstathiou03}
or for anthropic reasons \cite{Linde95,Vilenkin97,Q}, 
so it is clearly interesting and worthwhile to test the flatness assumption observationally.
In the same spirit, measuring the Hubble parameter $h$
independently of theoretical assumptions about curvature and
measurements of galaxy distances at low redshift provides a powerful consistency check 
on our whole framework.  

\begin{figure} 
%\vskip\smtopskip
\centerline{\epsfxsize=\figsize\epsffile{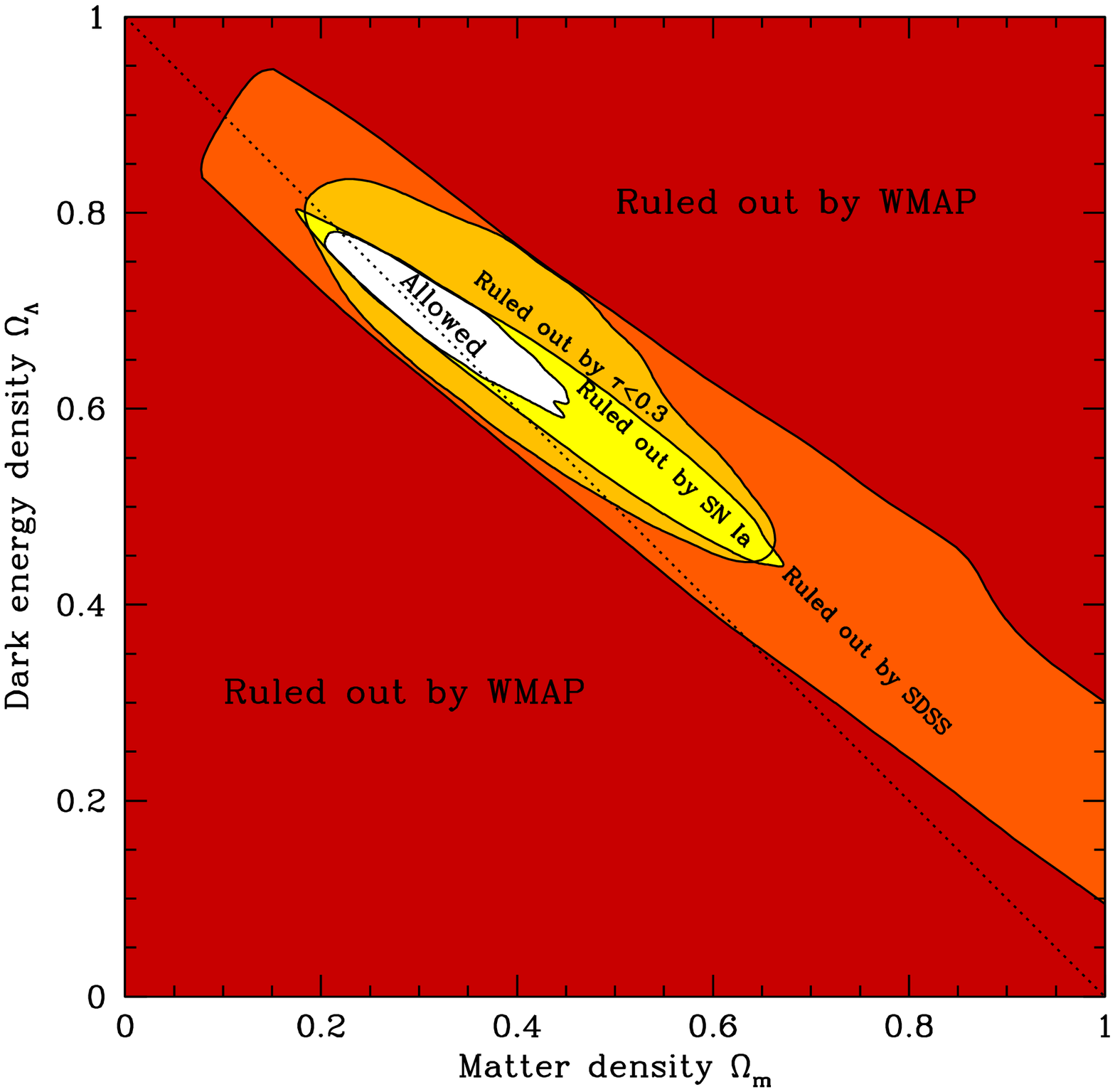}}
%\vskip\smbotskip
\caption[1]{\label{2d_OmOl_7par_fig}\footnotesize%
95\% constraints in the $(\Om,\Ol)$ plane. 
Shaded dark red/grey region is ruled out by WMAP alone
for 7-parameter curved models, illustrating the well-known 
geometric degeneracy between models that all have the same 
acoustic peak locations.
The shaded light red/grey region is ruled out when adding SDSS information.
Continuing inwards, the next two regions are ruled out when 
adding the $\tau<0.3$ assumption and when including SN Ia information as well.
Models on the diagonal dotted line are flat, those below are
open and those above are closed.
The constraints in this plot agrees well with those in Figure 13 from \cite{Spergel03}
when taking the $\tau$ prior into account.
}
\end{figure}

Including SDSS information is seen to reduce the curvature uncertainty by
about a factor of three. 
We also show the effect of adding the above-mentioned prior $\tau<0.3$ and SN Ia information
from the 172 SN Ia compiled by \cite{Tonry03}, which is seen to further
tighten the curvature constraints to $\Ot=1.01\pm 0.02$ ($1\sigma$), providing a
striking vindication of the standard inflationary prediction $\Ot=1$.
Yet even with all these constraints, a strong degeneracy is seen 
to persist between curvature and $h$, and curvature and $\age$, so that the HST key project \cite{Freedman01}
remains the most accurate 
measurement of $h$.  If we add the additional assumption that
space is {\it exactly} flat, then uncertainties shrink by factors around 
3 and 4 for $h$ and $\age$, respectively, still in beautiful agreement with other measurements.
The age limit $\age>12$ Gyr shown in \fig{2d_Ott0_7par_fig} is the 95\% 
lower limit from
white dwarf ages by \cite{Hansen02}; for a thorough reviews of 
recent age determinations, see \cite{Spergel03,Lineweaver99}.

This curvature degeneracy is also seen in 
\fig{2d_OmOl_7par_fig}, which illustrates that the existence of dark energy
$\Ol>0$ is only required at high significance when augmenting WMAP 
with either galaxy clustering information or SN Ia information (as
also pointed out by \cite{Spergel03}).
This stems from the well-known geometric degeneracy where 
$\Ok$ and $\Ol$ can be altered so as to leave the acoustic peak locations unchanged, which has
been exhaustively discussed in the pre-WMAP literature 
--- see, \eg, \cite{WhiteComplementarity,parameters,ZSS97,parameters2,EfstathiouBond99}.

In conclusion, we obtain sharp constraints on spatial curvature and
interesting constraints on $h$, $\age$ and $\Ol$, but only when combining WMAP with
SDSS and/or other data.
In other words, within the class of almost flat models, 
the WMAP-only constraints on $h$, $\age$ and $\Ol$ are weak, and 
including SDSS gives a huge improvement in precision.

Since the constraints on $h$ and $t_0$ are further tightened by a large factor
if space is exactly flat, can one justify the convenient assumption $\Ot=1$?
% However, WMAP alone does show that $\Ot$ is close to unity
% MT: I think it's a bit of a stretch to call 1.5 "close to unity".
Although WMAP alone marginally allows $\Ot=1.5$ (\fig{2d_Oth_7par_fig}),
WMAP+SDSS shows that $\Ot$ is within 15\% of unity.
It may therefore
be possible to bolster the case for perfect spatial flatness by
demolishing competing theoretical explanations of the observed approximate flatness --- for instance, 
it has been argued that if the near-flatness is due to an anthropic
selection effect, then one expects departures from $\Ot\sim 1$ of order unity 
\cite{Linde95,Vilenkin97}, perhaps larger than we now observe.
This approach is particularly promising if one uses a prior on $h$.
Imposing a hard limit $0.58 < h < 0.86$ corresponding to the $2\sigma$ range from the HST key
project \cite{Freedman01}, we obtain 
$\Ot=1.030^{+0.029}_{-0.029}$ from WMAP alone, 
$\Ot=1.023^{+0.020}_{-0.033}$ adding SDSS and 
$\Ot=1.010^{+ 0.018}_{-0.017}$ when also adding SN Ia and the $\tau<0.3$ prior.\footnote{
Within the framework of Bayesian inference, such an argument would run as in 
the following example. 
Let us take the current best measurement from above to be $\Ot=1.01\pm 0.02$
and use it to compare an inflation model predicting 
$\Ot=1\pm 10^{-5}$ with a non-inflationary FRW model predicting 
that a typical observer sees $\Ot=1\pm 1$ because of 
anthropic selection effects \cite{Linde95,Vilenkin97,Q}.
% (in the former case, anthropic selection 
% effects are irrelevant, in the latter case galaxy formation is suppressed 
% for $\Ok\ll 1$ and the Universe recollapses by early recollapse of the Universe).
Convolving with the $0.02$ measurement uncertainty, 
our two rival models thus predict that our observed best-fit value 
is drawn from distributions 
$\Ot=1\pm 0.02$ and
$\Ot=1\pm 1$, respectively.
If we approximate these distributions by 
Gaussians $f(\Ot)= e^{-[(\Ot-1)/\sigma]^2/2}/\sqrt{2\pi}\sigma$ with
$\sigma=0.02$ and $\sigma=1$, respectively, we find that
the observed value is about 22 times more likely given inflation.
In other words, if we view both models as equally likely from the outset,
the standard Bayesian calculation
\begin{center}
\begin{tabular}{|l|r|r|r|}
\hline
Explanation	&Prior prob.	&Obs. likelihood	&Posterior prob.\\
\hline
Inflation	&0.5		&17.6			&0.96\\
Anthropic	&0.5		&0.80			&0.04\\
\hline  
\end{tabular}
% set obs   = 0.90
% set obs   = 1.01
% set sdev1 = 0.02
% set sdev2 = 0.50
% set L1    = exp(-((obs-1)/sdev1)**2/2)/(sqrt(2*pi)*sdev1)
% set L2    = exp(-((obs-1)/sdev2)**2/2)/(sqrt(2*pi)*sdev2)
% set P1    = L1/(L1+L2) 
% set P2    = L2/(L1+L2)
% echo $(L1) $(P1)
% echo $(L2) $(P2)
\end{center}   
strongly favors the inflationary model.
Note that it did not have to come out this way: 
observing $\Ot=0.90\pm 0.02$ 
would have given 99.99\% posterior probability for the anthropic model.
}

\section{Testing inflation}

\subsection{The generic predictions}

Two generic predictions from inflation are perfect flatness ($\Ok=0$, \ie, $\Ot\equiv 1-\Ok = 1$)
and approximate scale-invariance of the primordial power spectrum ($\ns\sim 1$).
Tables~{\WMAPtable}-{\AddinfoTable} 
show that despite ever-improving data, inflation still passes both of
these tests with flying colors.\footnote{Further successes, emphasized by
the WMAP team and \cite{Dodel0309057}, are the inflationary predictions
of adiabaticity and phase coherence which account for 
the peak/trough structure observed in the CMB power spectrum.
}

The tables show that although all cases we have explored are consistent 
with $\Ot=\ns=1$, adding priors and non-CMB information shrinks the
error bars by factors around 6 and 4 for $\Ot$ and $\ns$, respectively.

For the flatness test, Table~{\AddinfoTable} shows that $\Ot$ is
within about 20\% of unity with 68\% confidence
%$\Ot\sim 1$ to within about 20\% (68\%)
from WMAP alone without priors (even $\Ot\sim 1.5$ is allowed at the 95\%
confidence contour).  When we include SDSS, the 68\% uncertainty
tightens to 10\%, and the errors shrink impressively to the percent level
with more data and priors: 
$\Ot=1.012^{+ 0.018}_{- 0.022}$ using WMAP, SDSS, SN Ia and $\tau<0.3$.

For the scalar spectral index, Table~{\AddinfoTable} shows that
$\ns\sim 1$ to within about 15\% 
from WMAP alone without priors,
tightening to $\ns=0.977^{+ 0.039}_{- 0.025}$ when adding SDSS and assuming 
the vanilla scenario, so the cosmology community is rapidly approaching 
the milestone where the departures from scale-invariance that most popular inflation
models predict become detectable.

\begin{figure} 
%\vskip\smtopskip
\centerline{\epsfxsize=\figsize\epsffile{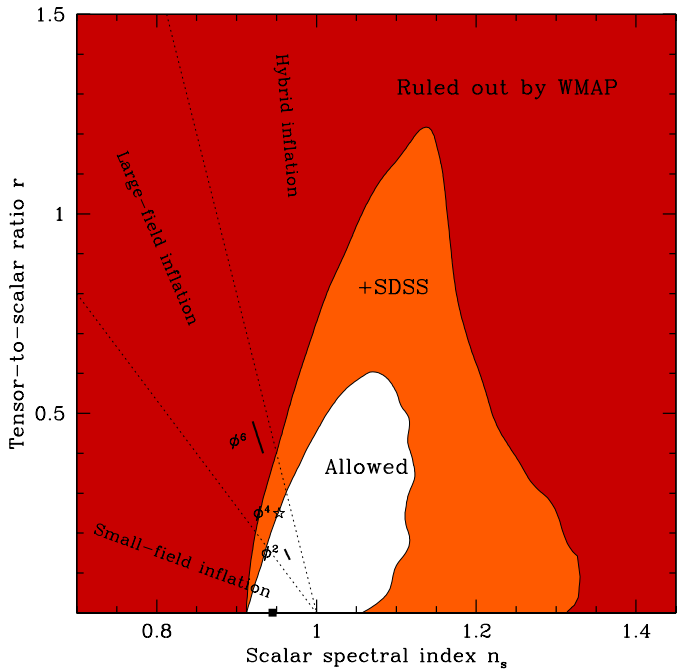}}
%\vskip\smbotskip
\caption[1]{\label{2d_nsr_7parr_fig}\footnotesize%
95\% constraints in the $(\ns,r)$ plane. 
Shaded dark red/grey region is ruled out by WMAP alone
for 7-parameter models (the vanilla models plus $r$).
The shaded light red/grey region is ruled out when adding SDSS information.
The two dotted lines delimit the three classes of inflation models known 
as small-field, large-field and hybrid models.
Some single-field inflation models make highly specific predictions in this plane as indicated.
From top to bottom, the figure shows the predictions for
$V(\phi)\propto\phi^6$ (line segment; ruled out by CMB alone), 
$V(\phi)\propto\phi^4$ (star; a textbook inflation model; on verge of exclusion) and
$V(\phi)\propto\phi^2$ (line segment; the eternal stochastic inflation model; still allowed), and
$V(\phi)\propto 1-(\phi/\phi_*)^2$ (horizontal line segment with $r\sim 0$; still allowed).
These predictions assume 
that the 
number of e-foldings between horizon exit of the observed fluctuations and 
the end of inflation is 64 for the $\phi^4$ model and between 50 and 60 for the others
as per \cite{LiddleLeach03}.
}
\end{figure}

%\begin{figure} 
%%\vskip\smtopskip
%\centerline{\epsfxsize=\figsize\epsffile{2d_nsr_9par_3.ps}}
%%\vskip\smbotskip
%\caption[1]{\label{GammafbFig}\footnotesize%
%95\% constraints in the $(\ns,r)$ plane. 
%Shaded red/grey region is ruled out by WMAP alone
%for 9-parameter models.
%The shaded yellow/light grey region is ruled out when adding SDSS information.
%%and the shaded blue/dark grey region is ruled out when adding SN Ia information
%%as well.
%%Models on the diagonal dotted line are flat, those below are
%%open and those above are closed.
%}
%\end{figure}

\subsection{Tensor fluctuation}

The first really interesting confrontation between theory and observation was predicted to occur in the
$(\ns,r)$ plane (\fig{2d_nsr_7parr_fig}), and the first skirmishes have already begun.
The standard classification of slow-roll inflation 
models \cite{DodelKinneyKolb97,LythRiotto99,LiddleLythBook}
characterized by a single field
inflaton potential $V(\phi)$ conveniently partitions this plane into three parts (\fig{2d_nsr_7parr_fig})
depending on the shape of $V(\phi)$:
\begin{enumerate}
\item Small-field models are of the form expected from spontaneous symmetry breaking,
where the potential has negative curvature $V(\phi)''<0$ and the field $\phi$ rolls down from near 
the maximum, and all predict $r < {8\over 3}(1-\ns)$, $\ns\le 1$.
\item Large-field models are characteristic of so-called chaotic initial conditions,
in which $\phi$ starts out far from the minimum of a potential with positive 
curvature ($V''(\phi)>0$),
and all predict ${8\over 3}(1-\ns) < r < 8(1-\ns)$, $\ns\le 1$.
\item Hybrid models are characterized by a field rolling toward a minimum with 
$V\ne 0$. Although they generally involve more than one inflaton field, they can be 
treated during the inflationary epoch as single-field inflation with $V''>0$ and
predict  $r>{8\over 3}(1-\ns)$, also allowing $\ns>1$.
\end{enumerate}
These model classes are summarized in Table~{\InflationTable} together with a sample of special cases.
For details and derivations of the tabulated constraints, see 
\cite{DodelKinneyKolb97,LythRiotto99,LiddleLythBook,Peiris03,Kinney03,LiddleSmith03,Wands03}. For comparison with other papers, remember
that we use the same normalization convention for $r$ as CMBfast and the WMAP team, where
$r=-8\nt$ for slow-roll models.
The limiting case between small-field and large-field models is the linear potential 
$V(\phi)\propto\phi$, and the limiting case between large-field and hybrid models is
the exponential potential $V(\phi)\propto e^{\phi/\phi_*}$.
The WMAP team \cite{Peiris03} further refine this classification by splitting the hybrid 
class into two: models with $\ns<1$ and models with $\ns>1$.

\begin{table*}
\label{InflationTable}
\noindent 
{\footnotesize
Table~\InflationTable: Sample inflation model predictions.  $N$ is the
number of e-folds between horizon exit of the observed fluctuations
and the end of inflation.
\begin{center}
\begin{tabular}{|llll|}
\hline
Model		&Potential	&$r$						&$\ns$\\
\hline
{\bf Small-field}&$V''<0$		&$r < {8\over 3}(1-\ns)$		&$\ns\le 1$\\
Parabolic	&$V\propto 1-\left({\phi\over\phi_*}\right)^2$	
					&$r=8(1-\ns)e^{-N(1-\ns)}\simlt 0.06$	&$\ns<1$\\ % Cite Peiris
Tombstone	&$V\propto 1-\left({\phi\over\phi_*}\right)^4$	
					&$r\simlt 10^{-3}$			&$\ns=1-{3\over N}\sim 0.95$\\ % Cite Peiris
    		&$V\propto 1-\left({\phi\over\phi_*}\right)^p$, $p>2$	
					&$r\simlt 10^{-3}$			&$\ns=1-{2\over N}{p-1\over p-2}\simgt 0.93$\\ % Cite Peiris
\hline
Linear		&$V\propto\phi$		&$r = {8\over 3}(1-\ns)$		&$\ns\le 1$\\
\hline
{\bf Large-field}&$V''>0$		&${8\over 3}(1-\ns) < r < 8(1-\ns)$	&$\ns\le 1$\\
%Power-law	&$V\propto\phi^p$	&$r = {8p\over p+2}(1-\ns)$		&$\ns = 1 - {{1+p/2}\over N}$\\
%Power-law	&$V\propto\phi^p$	&$r = {4p\over N}$			&$\ns = 1 - {{1+p/2}\over N+(p-1)/2}$\\
Power-law	&$V\propto\phi^p$	&$r = {4p\over N}$			&$\ns = 1 - {{1+p/2}\over N}$\\
Quadratic	&$V\propto\phi^2$	&$r = {8\over N}\sim 0.15$		&$\ns = 1 - {2\over N}\sim 0.96$\\
Quartic		&$V\propto\phi^4$	&$r = {16\over N}\sim 0.29$		&$\ns = 1 - {3\over N}\sim 0.95$\\
Sextic		&$V\propto\phi^6$	&$r = {24\over N}\sim 0.44$		&$\ns = 1 - {4\over N}\sim 0.93$\\
\hline
Exponential	&$V\propto e^{\phi/\phi_*}$	&$r = 8(1-\ns)$			&$\ns\le 1$\\
\hline
{\bf Hybrid}	&$V''>0$		&$r>{8\over 3}(1-\ns)$			&Free\\
\hline  
\end{tabular}
\end{center}     
} 
\end{table*}

Many inflationary theorists had hoped that early data would 
help distinguish between these classes of models, but
\fig{2d_nsr_7parr_fig} shows that all three classes are still
allowed. 
%If the WMAP+SDSS allowed region shown in \fig{2d_nsr_7parr_fig} 
%had been offset to the left or right by 0.2, the current data would 
%already have ruled out small$+$large-field models or hybrid models, respectively.
%Instead, we see that Nature has dashed these optimistic hopes,
%leaving all three classes still allowed.

What about constraints on specific inflation models as opposed to entire classes?
Here the situation is more interesting. 
Some models, such as hybrid ones, allow two-dimensional regions in this plane.
Table~{\InflationTable} shows that many other models predict a one-dimensional
line or curve in this plane. Finally, a handful of models are extremely testable, 
making firm predictions for both $\ns$ and $r$ in terms of $N$, 
the number of e-foldings between horizon exit of the observed fluctuations and 
the end of inflation. Recent work \cite{HuiDodel03,LiddleLeach03}
has shown that
$50\simlt N \simlt 60$ is required for typical inflation models.
The quartic model $V\sim\phi^4$ is an anomaly, requiring $N\approx 64$ with very small uncertainty.
\Fig{2d_nsr_7parr_fig} shows that power law models $V\propto \phi^p$ are ruled out 
by CMB alone for $p=6$ and above. 
% These models are unlikely to be missed, since they are not renormalizable.
\Fig{2d_nsr_7parr_fig} indicates that the textbook $V\propto \phi^4$ model 
(indicated by a star in the figure) 
is marginally allowed. \cite{Peiris03} found it marginally ruled out, but 
this assumed $N=50$ --- the subsequent result $N\approx 64$ \cite{LiddleLeach03}
pushes the model down to the right and make it less disfavored.
$V\propto \phi^2$ has been argued to be the most natural power-law model,
since the Taylor expansion of a generic function near its minimum has
this shape and since there is no need to explain why quantum corrections have not
generated a quadratic term. This potential is used in the stochastic eternal inflation 
model \cite{LindeBook}, and is seen to be firmly in the allowed region, as are the
small-field ``tombstone model'' from Table~{\InflationTable}
and the GUT-scale model of \cite{KyaeShafi0302504} (predicting 
$\ns=1-1/N\approx 0.98$, $r\approx 10^{-8}$).
%For up-to-date details on the $p=2$ and $p=4$ models, see \cite{LiddleSmith}.
%Ah, but their nice fig 2 has curviness just because of the braneworld business that is irrelevant here.

In conclusion, \Fig{2d_nsr_7parr_fig} shows that
observations are now beginning to place interesting constraints
on inflation models in the $(\ns,r)$-plane. As these constraints tighten
in coming years, they will allow us to distinguish between many of the prime contenders.
For instance, the stochastic eternal inflation model predicting 
$(\ns,r)\approx (0.96,0.15)$ will become 
%easily 
distinguishable
from models with negligible tensors, and
in the latter category, small-field models with, say, $\ns\simlt 0.95$, will
become distinguishable from the scale-invariant case $\ns=1$.

\subsection{A running spectral index?}

Typical slow-roll models predict not only negligible spatial curvature, but also
that the running of the spectral index $\alpha$ is unobservably small.
We therefore assumed $\Ok=\al=0$ when testing such models above.

Let us now turn to the issue of searching for departures from a power law primordial power
spectrum. This issue has generated recent interest after the WMAP team claim that
$\alpha<0$ was favored over $\alpha=0$, at least at modest statistical significance, 
with the preferred value being $\alpha\sim -0.07$ \cite{Peiris03,Spergel03}.

Slow-roll models typically predict $|\alpha|$ of order $N^{-2}$; for
these models, $|\alpha|$ is rarely above $10^{-3}$, much smaller than the WMAP-team preferred value.
Those inflation models that do predict such a strong second derivative
of the primordial power spectrum (in log-log space) tend to produce substantial 
third and higher derivatives as well, so that a parabolic curve parametrized
by $\As$, $\ns$ and $\alpha$ is a poor approximation of the model (\eg, \cite{Lidsey03}).
Lacking strong theoretical guidance one way or another, we therefore drop our
priors on $\Ok$ and $r$ when constraining $\alpha$.

Tables~{\WMAPtable} and {\SDSStable} 
show that our best-fit $\al$-values agree with those
of \cite{Peiris03}, but are consistent with $\al=0$, since the 95\%
error bars are of order $0.1$. They show that $\chi^2$ drops by only 5 relative
to vanilla models, which is not statistically significant because a drop of 3 is expected from 
freeing the three parameters $\Ok$, $r$ and $\alpha$.
Moreover, we see that our WMAP-only constraint is similar to our 
WMAP+SDSS constraint, showing that any hint of running comes from the CMB alone, most
likely from the low quadrupole power \cite{Spergel03}; see also \cite{Efstathiou03b,smalluniverse}.
% + Lewis, priv. comm.?
This is at least qualitatively consistent with the WMAP team analysis
\cite{Spergel03}; apart from the low quadrupole,
most of the evidence that $\alpha \ne 0$ comes from
CMB fluctuation data on small scales (\eg, the CBI data \cite{CBI02}) and
measurements of the small-scale fluctuations from the Ly$\alpha$
forest; indeed, including the 2dFGRS data slightly {\it weakens} the
case for running.  
% finds stronger evidence for running when including 2dFGRS and 
%Ly$\alpha$ forest data, so to whatever extent the 2dFGRS data supports running, 
%SDSS does not.
For the Ly$\alpha$ forest case, the key issue is the extent to which the 
measurement uncertainties have been adequately modeled \cite{Seljak03}, and this should
be clarified by the forthcoming Ly$\alpha$ forest measurements from the SDSS.

\begin{figure} 
%\vskip\smtopskip
\centerline{\epsfxsize=\figsize\epsffile{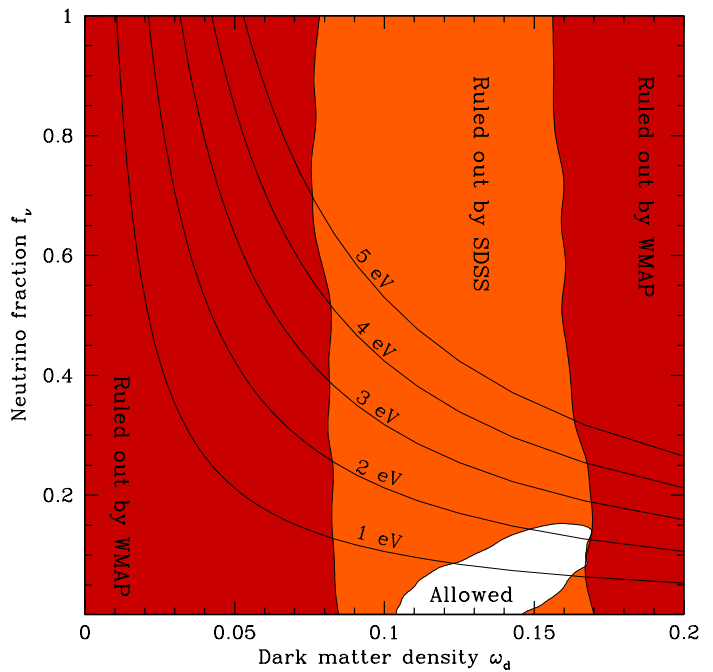}}
%\vskip\smbotskip
\caption[1]{\label{2d_odfn_7parnu_fig}\footnotesize%
95\% constraints in the $(\od,\fn)$ plane. 
Shaded dark red/grey region is ruled out by WMAP alone
when neutrino mass is added to the 6 ``vanilla'' models.
The shaded light red/grey region is ruled out when adding SDSS information.
The five curves correspond to $M_\nu$, the sum of the neutrino masses,
equaling 1, 2, 3, 4 and 5 eV, respectively --- barring sterile neutrinos,
no neutrino can have a mass exceeding $\sim M_\nu/3.$ % \approx 0.6$ eV (95\%).
}
\end{figure}

\section{Neutrino mass}

It has long been known \cite{neutrinos} that galaxy surveys are sensitive probes of 
neutrino mass, since they can detect the suppression of small-scale 
power caused by neutrinos streaming out of dark matter overdensities. 
For detailed discussion of post-WMAP astrophysical neutrino constraints, see 
\cite{Spergel03,Hannestad0303076,ElgaroyLahav0303089,BashinskySeljak03,Hannestad0310133},
and for an up-to-date review of the theoretical and experimental situation, see \cite{King03}.

Our neutrino mass constraints are shown in the $\Mnu$-panel of
\fig{1d_6par_fig}, where we allow our standard 6 ``vanilla''
parameters and $f_\nu$ to be free.  The most favored value is $\Mnu=0$, and obtain
a 95\% 
upper limit $\Mnu<1.7\>$eV.
\Fig{2d_odfn_7parnu_fig} shows that %, contrary to a persistent myth, 
WMAP alone tells us nothing whatsoever about neutrino masses and is consistent 
with neutrinos making up 100\% 
of the dark matter. Rather, the power of WMAP is that it constrains other parameters so strongly 
that it enables large-scale structure data to measure the small-scale $P(k)$-suppression 
that massive neutrinos cause.

The sum of the three neutrino masses (assuming standard freezeout)
is \cite{KolbTurnerBook} $\Mnu\approx(94.4\>{\rm eV})\od\fn.$  The neutrino energy density must be
very close to the standard freezeout density \cite{synch1,synch2,synch3}, given
the large mixing angle solution to the solar neutrino problem
and near maximal mixing from atmospheric results--- see \cite{Kearns02,Bahcall03} for up-to-date reviews.  
Any substantial asymmetries in neutrino density from the standard value would
be ``equilibrated'' and produce a primordial $^4$He abundance inconsistent with
that observed.

Our upper limit is complemented by the lower limit from neutrino oscillation experiments. 
Atmospheric neutrino oscillations
show that there is at least one
neutrino (presumably mostly a linear combination of 
$\nu_\mu$ and $\nu_\tau$) whose mass exceeds a lower limit
around $0.05\>$eV \cite{Kearns02,King03}.
Thus the atmospheric neutrino data corresponds to a lower limit 
% 0.05/94.4:
$\on\simgt 0.0005$, or
%  0.05/(94.4*0.1222)
$\fn\simgt 0.004$.
% Solar: evidence of electron neutrino conversions (presumably into mu)
% Atmospheric: muon neutrino conversions (presumably into tau)
The solar neutrino oscillations occur at a still smaller mass scale, perhaps around $0.008\>$eV
\cite{sno0309004,King03,Bahcall03}.
These mass-splittings 
%indicated by both solar and atmospheric neutrino data 
are much smaller than 1.7 eV, suggesting that all three mass eigenstates would
need to be almost degenerate for neutrinos to weigh in near our upper limit.
Since sterile neutrinos are disfavored from being thermalized in the early
universe \cite{fournu1,fournu2}, it can be assumed that only three neutrino flavors are
present in the neutrino background; this means that none of the three neutrinos
can weigh more than about $1.7/3 = 0.6$ eV.  The mass of the heaviest neutrino
is thus in the range $0.04-0.6$ eV.

%Note that if, as current data suggest, the mixing between
%the flavor eigenstates $\nu_e$, $\nu_\mu$ and $\nu_\tau$ is not small, 
%it is inappropriate to identify the three mass-eigenstates 
%$\nu_1$, $\nu_2$ and $\nu_3$ with these flavor eigenstates.
%For instance, the heaviest eigenstate $\nu_3$ is likely to
%be almost a $50-50$ mixture of $\nu_\tau$ and $\nu_\mu$. 
%The correct way to phrase our upper limit is therefore 
%as a 0.6 eV upper limit on the mass of $\nu_3$.
 
A caveat about non-standard neutrinos is in order. 
To first order, our cosmological constraint probes  
only the {\it mass density} of neutrinos, $\rho_\nu$,
which determines the small-scale power suppression factor,
and the {\it velocity dispersion}, which determines the scale below which 
the suppression occurs. For the low mass range we have discussed,
the neutrino velocities are high and the suppression occurs on 
all scales where SDSS is highly sensitive. We thus measure only the neutrino 
mass density, and our conversion
of this into a limit on the mass sum assumes that the
neutrino number density is known and given by the standard model
freezeout calculation, 112 cm$^{-3}$. In more general scenarios with sterile
or otherwise non-standard neutrinos where the freezeout abundance 
is different, the conclusion to take away is 
an upper limit on the total light neutrino mass density
of $\rho_\nu < 4.8\times 10^{-28}$kg/m$^3$ (95\%).
To test arbitrary nonstandard models, a future challenge will be to
independently measure both the mass density and the velocity dispersion,
and check whether they are both consistent with the same value of $\Mnu$.

The WMAP team obtains the constraint $\Mnu<0.7$ eV \cite{Spergel03} by combining WMAP with the 2dFGRS.
This limit is a factor of three lower than ours because of their
stronger priors,
%The reason that they obtain about three times tighter constraints than we do using a smaller
%galaxy redshift survey is that they make stronger assumptions, 
most importantly that on galaxy bias $b$ 
%that the galaxy bias has been accurately measured
determined using a bispectrum analysis of the 2dF galaxy clustering
data\cite{Verde02}.  
%of non-Gaussian clustering analysis. 
This bias was measured on scales $k\sim 0.2-0.4h$/Mpc
and assumed to be the same on the scales $k<0.2h$/Mpc that were used in the
analysis.
%It has been argued \cite{Scoccimarro03} that the 
%uncertainties in this method may have been substantially underestimated, 
In this paper, we prefer not to include such a prior. 
%we will avoid assuming anything about bias to be conservative.
Since the bias is marginalized over, our SDSS neutrino constraints come 
not from the amplitude of the power spectrum, only from its shape.
This of course allows us to constrain $b$ from WMAP+SDSS directly;
we find values consistent with unity (for $L^*$ galaxies) in almost
all cases (Tables \SDSStable\ and \AddinfoTable). A powerful consistency
test is that our corresponding value $\beta=0.54^{+0.06}_{-0.05}$
from WMAP+SDSS agrees well with the value $\beta\sim 0.5$ measured from
redshift space distortions in \cite{sdsspower}.

\begin{figure} 
%\vskip\smtopskip
\centerline{\epsfxsize=\figsize\epsffile{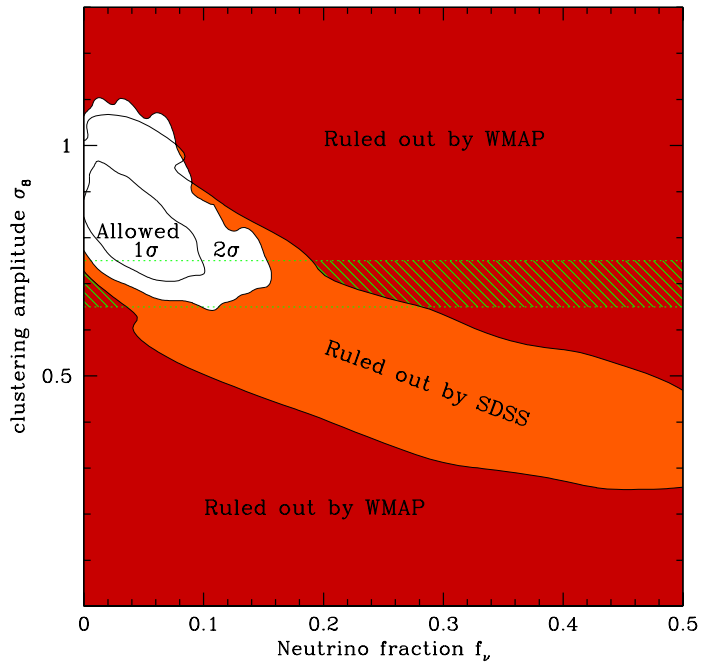}}
%\vskip\smbotskip
\caption[1]{\label{2d_fns8_7parnu_fig}\footnotesize%
Constraints in the $(\fn,\sigma_8)$ plane. 
Shaded dark red/grey region is ruled out by WMAP alone ($95\%$)
when neutrino mass is added to the 6 ``vanilla'' models.
The shaded light red/grey region is ruled out when adding SDSS information.
The recent claim that $\fn>0$ \cite{Allen0306386} hinges on assuming 
that galaxy clusters require low $\sigma_8$-values (shaded horizontal band)
and dissolves when using what we argue are more reasonable uncertainties 
in the cluster constraints.
}
\end{figure}

Seemingly minor assumptions can make a crucial difference for neutrino conclusions, as
discussed in detail in \cite{Spergel03,Hannestad0303076,ElgaroyLahav0303089}. 
A case in point is a recent claim that nonzero neutrino mass has been detected
by combining WMAP, 2dFGRS and galaxy cluster data \cite{Allen0306386}.
Figure 2 in that paper (middle left panel) shows that nonzero neutrino mass is 
strongly disfavored only when including data on X-ray cluster abundance,
which is seen (lower middle panel) to prefer a low normalization of order 
$\sigma_8\approx 0.70\pm 0.05$ (68\%). 
\Fig{2d_fns8_7parnu_fig} provides intuition for the physical origin on the
claimed neutrino mass detection. Since WMAP fixes the normalization at early times before 
neutrinos have had their suppressing effect, 
we see that the WMAP-allowed $\sigma_8$-value drops as the neutrino fraction $\fn$ increases.
A very low $\sigma_8$-value therefore requires a nonzero neutrino fraction.
The particular cluster analysis used by \cite{Allen0306386}
happens to give one of the lowest $\sigma_8$-values in the recent literature.
Table~{\ClusterLensTable} and \fig{2d_Oms8_6par_fig} show a range of
$\sigma_8$ values larger than the individual quoted errors, implying 
the existence of significant systematic effects.  If we 
%Given the range of 
%Since well-known systematic uncertainties are often as important as statistical errors in
%cluster studies, one could opt to 
expand the error bars on the cluster constraints to $\sigma_8=0.8\pm 0.2$, 
to reflect the spread in the recent literature, we find that the
evidence for a cosmological neutrino mass detection goes
away. The sensitivity of neutrino conclusions to 
cluster $\sigma8$ normalization uncertainties was also discussed in \cite{Allen0306386}.

\section{Dark energy equation of state}

\begin{figure} 
%\vskip\smtopskip
\centerline{\epsfxsize=\figsize\epsffile{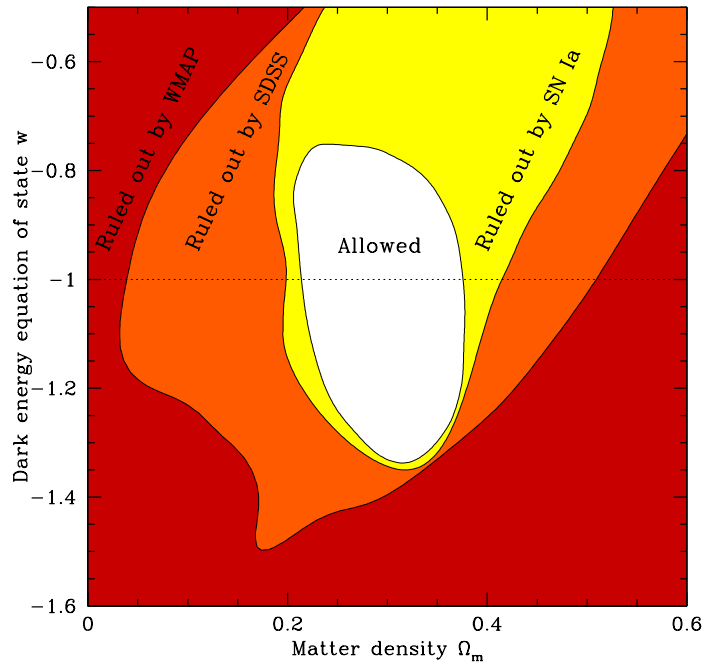}}
%\vskip\smbotskip
\caption[1]{\label{2d_Omw_7parw_fig}\footnotesize%
95\% constraints in the $(\Om,\w)$ plane. 
Shaded dark red/grey region is ruled out by WMAP alone
when equation of state $w$ is added to the 6 ``vanilla'' parameters.
The shaded light red/grey region is ruled out when adding SDSS information,
and the yellow/very light grey region is excluded when including SN Ia
information as well.  
%Note that the contours cross each other, implying
%a slight tension between the data sets regarding the preferred parameter values.
}
\end{figure}

Although we now know its present density fairly accurately, we know precious little 
else about the dark energy, and post-WMAP research is 
focusing on understanding its nature 
\cite{Melchiorri0211522,Barreiro03,dedeo03,CaldwellDoran03,CaldwellKamionWeinberg03,Caldwell03,Balakin03,Weller03,Kunz03}.
Above we have assumed that the dark energy behaves as a cosmological constant with
its density independent of time, \ie, that its equation of state $w=-1$.
\Fig{1d_6par_fig} and \fig{2d_Omw_7parw_fig} show our constraints on $w$, assuming that
the dark energy is homogeneous, \ie, does not cluster\footnote{Dark energy clustering 
can create important modifications of the CMB power spectrum
and can weaken the $w$-constraints by increasing degeneracies \protect\cite{Weller03}.
We have ignored the effect of dark energy clustering since it 
depends on the dark energy sound speed, which is in turn model-dependent and at present
completely unknown. Indeed, all evidence for dark energy so far traces back to the observed
cosmic expansion history $H(z)$ departing from $\Ol=0$ Friedmann equation, and if
this departure is caused by modified gravity rather than some sort of new substance,
then there may be no dark energy fluctuations at all.
}.
Although our analysis adds improved galaxy and SN Ia data to that of the 
WMAP team \cite{Spergel03} and uses different assumptions, 
\fig{2d_Omw_7parw_fig} agrees well with Figure 11 from \cite{Spergel03}
and our conclusions are qualitatively the same: adding $w$ as a free parameter does
not help improve $\chi^2$ for the best fit, and all data are consistent with
the vanilla case $w=-1$, with uncertainties in $w$ at the 20\% 
level. \cite{Melchiorri0211522,Weller03} obtained similar constraints with different data and an $h$-prior.

Tables~{\WMAPtable} and {\SDSStable}
show the effect of
dropping the $w=-1$ assumption on other parameter constraints.
These effects are seen to be similar to those of dropping the flatness assumption, but weaker,
which is easy to understand physically. As long as there are no spatial 
fluctuations in the dark energy (as we have assumed), changing $w$ has only two effects on 
the CMB: it shifts the acoustic peaks sideways by altering the angle-distance relation,
and it modifies the late Integrated Sachs-Wolfe (ISW) effect. Its only
effect effect on the matter power
spectrum is to change its amplitude via the linear growth factor.
The exact same things can be said about the
parameters $\Ol$ and $\Ok$, so the angle-diameter degeneracy becomes a two-dimensional surface
in the three-dimensional space ($\Ok,\Ol,w$), broken only by the late ISW effect.
Since the peak-shifting is weaker for $w$ than for $\Ok$ (for changes generating 
comparable late ISW modification), adding $w$ to vanilla models wreaks less havoc with,
say, $h$ than does adding $\Ok$ to vanilla models (\Sec{CurvatureSec}).

\section{Discussion and conclusions}
\label{DiscussionSec}

We have measured cosmological parameters using the three-dimensional power spectrum $P(k)$
from over 200,000 galaxies in the Sloan Digital Sky Survey (SDSS) in combination with 
WMAP and other data.
% paying particular attention to quantifying the robustness
% of the results to the choice of data sets, parameters used and prior assumptions.
Let us first discuss what we have and have not learned about cosmological 
parameters, then summarize what we have and have not learned about the
underlying physics.

\subsection{The best fit model}

All data we have considered are consistent with a ``vanilla'' flat 
adiabatic $\Lambda$CDM model with no tilt, running tilt, tensor fluctuations, 
spatial curvature or massive neutrinos.  
Readers wishing to choose a concordance model for a calculational
purposes using Ockham's razor can adopt the best fit ``vanilla lite'' model
\beq{5parEq}
% (\tau,\Ol,\od,\ob,\As) = (0.16,0.71,0.12,0.024,0.89)   % 	That's the mean values from the table
%(\tau,\Ol,\od,\ob,\As) = (0.170,0.718,0.121,0.0239,0.894)	 %	That's the best-fit values from bestfit.ps 
(\tau,\Ol,\od,\ob,\As) = (0.17,0.72,0.12,0.024,0.89)	 %	That's the best-fit values from bestfit.ps 
\eeq
% 54608  0.71076580  0.59764099  0.71837928  0.12067137  0.02388562  0.00000000  1.00000000  1.00000000  0.63542517  0.00000000  0.92142547 -1.00000000  0.00000000     -723.94829343  0.00000000  0.28162072  0.14455699  0.71645224  0.17070615 17.29644165  0.89400077  0.00000000  0.53879886 13.39920402  0.94445671  5.07860049  0.45325751  0.46304583  0.63542517  0.00000000
(Table~{\AddinfoTable}, second last column).
Note that this is even simpler than 6-parameter vanilla models, since it has $\ns=1$ and
only 5 free parameters \cite{Easther03}.  
A more theoretically motivated 5-parameter model is that 
of the arguably most testable inflation model, $V\propto\phi^2$ stochastic eternal inflation,
which predicts $(\ns,r)=(0.15,0.96)$ (\Fig{2d_nsr_7parr_fig}) and prefers
\beq{5parLindeEq}
%(\tau,\Ol,\od,\ob,\As) = (0.088,0.685,0.123,0.0226,0.750)	 %	That's the best-fit values from bestfit.ps 
(\tau,\Ol,\od,\ob,\As) = (0.09,0.68,0.123,0.023,0.75)	 %	That's the best-fit values from bestfit.ps 
\eeq
% 58017  0.83867932  0.59502537  0.68475086  0.12300214  0.02257644  0.00000000  0.96000000  0.99325000  0.62874324  0.15000000  1.01031257 -1.00000000  0.00000000     -723.54396789 -0.00000001  0.31524915  0.14557858  0.67955057  0.08796343 11.56997066  0.74968254  0.11245238  0.52296190 13.68865327  0.87823504  4.87162796  0.45601417  0.45060099  0.63375409  0.00000000
(Table~{\AddinfoTable}, second last column).

Note that these numbers are in
substantial agreement with the results of the WMAP team
\cite{Spergel03}, despite a completely independent analysis and 
independent redshift survey data; this is a powerful confirmation of
their results and the emerging standard model of cosmology.
Equally impressive is the fact that we get similar results and error bars
when replacing WMAP by the combined pre-WMAP CMB data (compare the
last columns of Table~{\SDSStable}). In other words, the concordance model 
and the tight constraints on its parameters are no longer dependent on any one
data set --- everything still stands even if we discard either WMAP or pre-WMAP CMB data
and either SDSS or 2dFGRS galaxy data. % Every single data set is dispensable.
No single data set is indispensable.

As emphasized by the WMAP team, it is remarkable that such a broad
range of data are describable by such a small number of parameters.
Indeed, as is apparent from Tables \WMAPtable--\AddinfoTable, $\chi^2$
does not improve significantly upon the addition of further parameters
for any set of data.  However, the ``vanilla lite'' model is not a complete and
self-consistent description of modern cosmology; for example, 
it ignores the well-motivated inflationary arguments for expecting
$\ns\ne 1$.

\subsection{Robustness to physical assumptions}

On the other hand, the same criticism can be leveled against 6-parameter vanilla models,
since they assume $r=0$ even though some of the most popular inflation models predict 
a significant tensor mode contribution. Fortunately, Table~{\SDSStable} shows that 
augmenting vanilla models with tensor modes has little effect on other parameters and their uncertainties, 
mainly just raising the best fit spectral index $\ns$ from 0.98 to 1.01.

Another common assumption is that the neutrino density $\fn$ is negligible, yet
%Another popular yet dubious assumption is that the neutrino density is negligible,
we know experimentally that $\fn>0$ and there is an anthropic argument
for why 
neutrinos should make a small but non-negligible contribution \cite{anthroneutrino}.
The addition of neutrinos changes the slope of the power spectrum on
small scales; in particular, when we allow $\fn$ to be a free
parameter, 
%This assumption is more dangerous, since adding the neutrino density to the vanilla 
%model space has a strong effect on other parameters, mainly lowering 
the value of $\sigma_8$ drops by 10\% and $\Omega_m$ increases by 25\%
(Table~{\WMAPtable}).

We found that the assumption with the most striking implications is that of perfect
spatial flatness, $\Ot=1$ --- dropping it dramatically weakens the limits on the 
Hubble parameter and the age of the Universe, allowing $h=0.5$ and $t_0=18$ Gyr.
Fortunately, this flatness assumption is well-motivated by inflation
theory; while anthropic explanations exist for the near flatness, 
%it is unclear whether they predict
they do not predict
the Universe to be quite as flat as it is now observed to be.
 
Constraints on other parameters are also somewhat weakened by allowing 
a running spectral index $\alpha\ne 0$ and an equation of state $w\ne -1$, 
but we have argued that these results are more difficult to take seriously theoretically.
It is certainly worthwhile testing whether $\ns$ depends on $k$ and whether
$\Ol$ depends on $z$, but parametrizing such departures in terms of 
constants $\alpha$ and $w$ to quantify the degeneracy with other parameters 
is unconvincing, since most inflation models 
predict observably large $|\alpha|$ to depend strongly on $k$ and 
observably large $|w+1|$ can depend strongly on $z$.

It is important to parametrize and constrain possible departures the current cosmological 
framework: any test that could have falsified it, but did not, bolsters its credibility.
Post-WMAP work in this spirit has included constraints on 
the dark energy sound speed \cite{Weller03} and time dependence \cite{WangMukherjee0312192,Alam0311364},
the fine structure constant \cite{Rocha03}, 
the primordial helium abundance \cite{Trotta03,Huey03},
isocurvature modes \cite{Crotty03}
and features in the primordial power spectrum 
\cite{Bridle0302306,Hannestad0311491}.

%Our cosmological parameters agree with those from the WMAP team, but with 
%a slightly smaller Hubble parameter ($h\approx 0.70^{+0.04}_{-0.03}$) and 
%a slightly larger matter density ($\Om=0.30\pm 0.04$) at $1\sigma$. 
%We place particular emphasis on clarifying the physical origin of the constraints,
%\ie, what we do and do not know when using different data sets and prior assumptions.
%For instance, dropping the assumption that space is perfectly flat,
%the WMAP constraints on the Hubble parameter get tightened 
%from $h\approx 0.50^{+0.16}_{-0.13}$ to $h\approx 0.60^{+0.09}_{-0.06}$ $(1\sigma)$ by adding 
%SDSS and SN Ia data.

\subsection{Robustness to data details}
\label{DataRobustnessSec}

How robust are our cosmological parameter measurements to the choice of data
and to our modeling thereof?

For the CMB, most of the statistical power comes from the unpolarized 
WMAP data, which we confirmed by repeating our 6-parameter analysis without polarization information.
The main effect of adding the polarized WMAP data is to give a
positive detection of $\tau$ (\Sec{TauDiscSec} below).  The quantity
$\sigma_8 e^{-\tau}$ determines the amplitudes of acoustic peak
amplitudes, so the positive detection of $\tau$ leads to a value of
$\sigma_8$ 15\% higher than without the polarization data included.
%and as a corollary increases the best-fit normalization
%$\sigma_8$ by about $15\%$ 
%to keep $\sigma_8 e^{-\tau}$ and hence the acoustic peak amplitudes constant. 
% ta < 0.22708 (95%) without polarization

For the galaxy $P(k)$ data, there are options both for what data set to use and how 
to model it. To get a feeling for the quantitative importance of choices, 
we repeat a simple benchmark analysis for a variety of cases.
Let us measure the matter density $\Om$ using galaxy data alone, treating $\As$ as a second 
free parameter and fixing all others at the values 
%$\tau=\Ok=\fn=r=\alpha=0$, $\ob=0.024$, $\ns=1$, $w=-1$, $b=1$ and $h=0.72$.
$\Ok=\fn=\alpha=0$, $\ob=0.024$, $\ns=1$, $w=-1$, $b=1$ and $h=0.72$.
Roughly speaking, we are thus fitting the measured galaxy power spectrum 
to a power spectrum curve that we can shift horizontally (with our ``shape parameter'' $\Om$) and 
vertically (with $\As$). We have chosen this particular example because, as described in 
\Sec{VanillaSec}, it is primarily this shape parameter measurement that breaks the WMAP
banana degeneracy.
The parameters $\tau$ and $r$ of course have no effect on $P(k)$, 
and the remaining two are determined by the matter density via the identities
$\Ol=1-\Om$, $\od=h^2\Om-\ob$.

\begin{table}
%\label{SystematicsTable}
\noindent 
{\footnotesize
Table~\SystematicsTable: Robustness to data and method details.
\begin{center}
\begin{tabular}{|ll|}
\hline
Analysis		&$\Omega_m$\\
\hline
Baseline		&$0.291^{+0.033}_{-0.027}$\\
$\kmax=0.15h/$Mpc	&$0.297^{+0.038}_{-0.032}$\\
$\kmax=0.1h/$Mpc	&$0.331^{+0.079}_{-0.051}$\\
No bias correction	&$0.256^{+0.027}_{-0.024}$\\
Linear $P(k)$		&$0.334^{+0.027}_{-0.024}$\\
2dFGRS			&$0.251^{+0.036}_{-0.027}$\\
\hline  
\end{tabular}
\end{center}     
} 
\end{table}

Our results are summarized in Table~{\SystematicsTable}.
We stress that they should not be interpreted as realistic measurements of $\Om$,
since the other parameters have not been marginalized over. This is why the error bars
are seen to be smaller even than when WMAP was included above (last column of Table~{\AddinfoTable}).

To avoid uncertainties associated with nonlinear redshift space distortions and scale-dependent 
galaxy bias, we have used SDSS measurements of $P(k)$ only for $k\le\kmax$ throughout this paper, chosing 
$\kmax=0.2h/$Mpc as recommended in \cite{sdsspower}.
The WMAP team made this same choice $\kmax=0.2h/$Mpc when analyzing the 2dFGRS data \cite{Verde03}.
An option would be to tighten this cut to be still more cautious. 
Table~{\SystematicsTable}
shows that cutting back to $\kmax=0.15h/$Mpc 
has essentially no effect on the best-fit $\Om$-value and increases error bars by about 20\%.
Cutting back all the way down to $\kmax=0.1h/$Mpc is seen to more than double the baseline
error bars, the baseline measurement lying about $0.6\sigma$ below the new best fit.
%  so there is no indication that reducing $\kmax$ would have affect our results appreciably

As described in \cite{sdsspower}, the SDSS measurements were corrected for luminosity-dependent bias.
Table~{\SystematicsTable} shows that if this were not done, $\Om$ would drop by
about 0.03, or $1\sigma$.
This correction is of course not optional. However, if the correction itself were somehow inaccurate
at say the 10\% 
level, one would expect a bias in $\Om$ around 0.003.

Just like the WMAP team \cite{Spergel03,Verde03}, we have used the nonlinear matter 
power spectrum for all our analysis. 
Table~{\SystematicsTable} shows that if we had used the linear spectrum instead,
then $\Om$ would rise by about 0.04, or $1.3\sigma$.
This happens because the linear power spectrum is redder, with less small-scale power, which can
be roughly offset by raising $\Om$ and hence shifting the curve to the right. 
Like the above-mentioned correction for luminosity-dependent bias,
correction for nonlinearities must be included.   However, given the large uncertainties about how biasing behaves in this
quasilinear regime, it may well be that this correction is only accurate to 25\%, 
say, in which case we would expect an additional uncertainty in $\Om$ at the 0.01 level.

Finally, we have repeated the analysis using an entirely different dataset, the $P(k)$-measurement 
from the 2dFGRS team \cite{Percival01}. Although the WMAP team used $\kmax=0.2h/$Mpc, we used 
the data available online with $\kmax=0.15h/$Mpc here as recommended by the 2dFGRS team \cite{Percival01}.
Table~{\SystematicsTable} shows that 2dFGRS measures a slightly redder power spectrum than 
SDSS, corresponding to $\Om$ down by 0.04, or $1.3\sigma$.
 
In conclusion, we see that a number of issues related to data selection and modeling
can have noticeable effects on the results. Internally to SDSS, such effects could easily change 
$\Om$ by as much as 0.01, and the 2dFGRS difference is about 0.04, or one standard deviation
--- roughly what one would expect with two completely independent data sets.

\begin{figure} 
\vskip\smtopskip
\centerline{\epsfxsize=10.7cm\epsffile{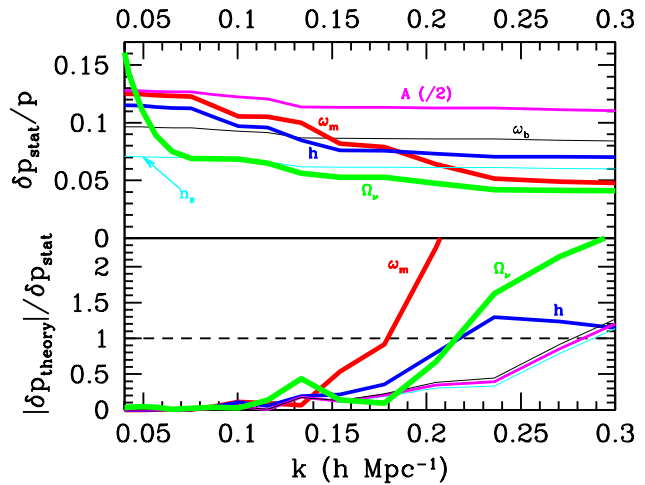}}
\vskip-2cm
\caption[1]{\label{DodelFig}\footnotesize%
Effect of increasing the amount of SDSS data used, given by
the maximum $k$-value used.
Top panel shows how the relative errors on various parameters
shrink as more data is included. For the neutrino density 
$\On\equiv\fn\Od$, the absolute rather than relative error is shown.
Bottom panel shows the ratio of systematic errors to statistical 
errors (from top panel) grows as smaller scales are included.
This is for the extreme case where nonlinear corrections are present but
completely ignored, which we view as a worst-case scenario.
}
\end{figure}

To quantify the effect of systematic uncertainties when both other parameters
and WMAP data are included, we carry out a second testing exercise.
Using the Fisher-matrix technique of \cite{KnoxScocciDodel98},
we compute how our best-fit parameter values shift in response to 
a systematic bias in the theoretically computed power spectrum $P(k)$.
To be conservative, we make the rather extreme assumption that the measurements 
correspond to the nonlinear power spectrum but that the analysis 
ignores nonlinear corrections entirely, simply fitting to the linear power spectrum.
Although we view this as a worst-case scenario, it provides an instructive illustration
of how problems related to nonlinear redshift space distortions and
scale-dependent biasing might scale with $\kmax$, the largest $k$-band included.

Our results are shown in \fig{DodelFig}.
The upper panel shows how the constraints from WMAP alone (on left side of figure)
gradually improve as more SDSS data are included.
The dramatic neutrino improvement seen at small $\kmax$ is due to 
WMAP alone leaving the neutrino fraction unconstrained.
The other parameters where SDSS helps the most are seen to be 
$\om$ and $h$, which can be understood based on our discussion in \Sec{VanillaSec}.
The SDSS power spectrum we have used does not probe to scales much
smaller than $k \sim 0.2$, which is why little further improvement is
seen beyond this value. 
%The reason that little further improvement is seen beyond $\kmax\sim 0.2h/$Mpc is 
%that the SDSS power spectrum measurement we have used \cite{sdsspower} is insensitive 
%to smaller scales.

The lower panel shows the ratio of the above-mentioned systematic error
to the statistical error for each parameter.
We see that the most sensitive parameter is $\om$, 
which justifies our singling it out for special scrutiny 
above in Table~{\SystematicsTable} (where $\om$ is equivalent to $\Om$ since 
we kept $h$ fixed). Although $\ns$ also partially mimics the nonlinear correction
and perhaps scale-dependent bias, it is seen to be somewhat less sensitive.
Our Fisher matrix estimate is seen to be somewhat overly pessimistic for $\Om$,
predicting that neglecting nonlinearities shifts $\om$ by of order
$2\sigma$ for $\kmax=0.2h/$Mpc whereas the 
brute force analysis in
Table~{\SystematicsTable} shows the shift to be only about half as large
even when WMAP is ignored.
The sensitivity to $h$ is linked to the $\Om$-sensitivity 
by the banana in \fig{2d_Omh_6par_fig}. The $\On$-sensitivity comes from 
the small-scale neutrino $P(k)$-suppression being similar to the 
suppression in going from nonlinear to linear $P(k)$-modeling.

In conclusion, as long as errors in 
the modeling of nonlinear redshift distortions and bias are not larger
than the nonlinear correction itself, we expect our uncertainties with
$\kmax=0.2h/$Mpc to be dominated by statistical rather than systematic errors.
The fact that cutting back to $\kmax=0.15h/$Mpc left our results 
virtually unchanged (Table~{\SystematicsTable}) supports this optimistic
conclusion. Indeed, \fig{DodelFig} shows that with $\kmax=0.15h/$Mpc, 
the reader wanting to perform a simple analysis can even use the linear $P(k)$ 
to good approximation.

%Although subdominant systematic effects should be kept in mind, it is important to remember that
However, both statistical errors and the systematic errors we have
discussed in this section are dwarfed by the effects of changing theoretical priors.
For instance, Table~{\WMAPtable} shows that $\Om$ increases by 0.08 when dropping 
either the assumption of negligible neutrinos or the assumption of negligible curvature.
Moreover, to place this in perspective, all Bayesian analysis using Monte Carlo Markov Chains
implicitly assumes a uniform prior on the space of the parameters where the algorithm jumps
around, and different authors make different choices for these parameters, which can make 
a substantial difference.\footnote{For instance, 
we use $\Ap\equiv\As e^{-2\tau}$ where the WMAP team uses $\As$ \cite{Verde03},
and both we and the WMAP team use the CMB peak location parameter $\Th$ where many other
groups use $\Ol$. The difference between these implicit priors is given by the Jacobian of the
transformation, which describes how the volume element changes and generically will have
variations of order unity when a parameter varies by a factor of two. 
For a parameter $p$ that is tightly constrained with a small relative error $|\Delta p/p|\ll 1$,
this Jacobian becomes irrelevant. For weakly constrained parameters like $\tau$, however,
this can easily shift the best-fit value by $1\sigma$.
%For example, changing from a uniform prior on $\Ol$ to a uniform prior on $\Th$ will pile up
%lots of prior likelihood near $\Ol=1$, pulling up the WMAP-only results in 
%\protect\fig{2d_OmOl_7par_fig}.
For example, changing to a uniform prior on 
the reionization redshift $\zion\propto\tau^{2/3}$ as done by \cite{Contaldi03} corresponds to
using a $\tau$-prior $\propto\tau^{-1/3}$, which strongly weights the results
towards low $\tau$.
% Gives prior dF/dtau = |dzion/dtau| dF/dzion ~ |dzion/dtau| ~ tau^{-1/3}
}

A final source of potential uncertainties involves bugs and algorithmic errors in the
analysis software. To guard against this, we performed two completely independent 
analyses for many of the parameter spaces that we have tabulated, one using the
Monte Carlo Markov Chain method described in Appendix A (coded up from scratch) and the second using the
publicly available CosmoMC package \cite{Lewis02} with appropriate modifications.
We found excellent agreement between the two sets of results, with 
all differences much smaller than the statistical errors and prior-related
uncertainties.

\subsection{What have we learned about physics?}
 
The fact that any simple model fits such accurate and diverse measurements
is impressive evidence that the basic theoretical framework of modern cosmology is correct, 
and that we need to take its implications seriously however surprising they may be. 
What are these implications?

\subsubsection{Inflation}

%For {\bf inflation}, 
The two generic predictions of perfect flatness ($|\Ok|\simlt 10^{-5}$) and near scale-invariance
have passed yet another test with flying colors.
We find no evidence for running tilt. We also find no evidence for gravitational waves, 
and are therefore unable to measure the tensor spectral index and test the inflationary
consistency relation $r=-8\nt$. The most interesting 
confrontation between theory and observation is now occurring in the
$(\ns,r)$ plane (\fig{2d_nsr_7parr_fig}). We confirm the conclusion \cite{Peiris03} that 
most popular models are still 
allowed, notably even stochastic eternal inflation with its prediction 
$(\ns,r)\approx (0.96,0.16)$, but modest data improvements over the next few 
years could decimate the list of viable inflationary candidates and rival models 
\cite{Song03}.
% (CITE EKPYROSIS STUFF? LET'S STAY OUT OF THAT BITTER DISPUTE?). 

\subsubsection{Dark energy}

Since its existence is now supported by three independent lines of evidence 
(SN Ia, power spectrum analysis such as ours, the late ISW effect 
\cite{Boughn03,Nolta03,Fosalba0305468,Fosalba03,Scranton03,Afshordi03}) and its current density 
is well known (the last column off Table~{\WMAPtable} gives $\Ol=0.70\pm 0.04$),
the next challenge is clearly to measure whether its density changes with time.
Although our analysis adds improved galaxy and SN Ia data to that of the 
WMAP team \cite{Spergel03},
our conclusions are qualitatively the same: all data are consistent with
the density being time-independent as for a simple cosmological constant ($w=-1$),
with uncertainties in $w$ at the 20\% 
level.

\subsubsection{Cold and hot dark matter}

We measure the density parameter for dark matter to be 
$\od=0.12\pm 0.01$ fairly robustly to theoretical assumptions, 
which corresponds to a physical density of  
$2.3\times 10^{-27}$kg/m$^3$ $\pm 10\%$.
% , with an uncertainty of 10\%.
% = 0.12 * 1.88e-26
Given the WMAP information, SDSS shows that no more than about 
12\%
of this dark matter
can be due to massive neutrinos, giving a 95\% 
upper limit to the sum of the neutrino masses $\Mnu<1.7$ eV.
Barring sterile neutrinos, this means that no neutrino mass can
exceed $\Mnu/3=0.6$ eV.
\cite{Spergel03} quotes a tighter limit by assuming a strong prior on 
%by making a slightly controversial assumption about 
galaxy bias $b$. We show that the recent claim of a neutrino mass detection 
$\Mnu\simgt 0.6$ eV  by Allen {\etal} hinges crucially on 
a particular low galaxy cluster $\sigma_8$ measurement 
and goes away completely when expanding the cluster $\sigma_8$ uncertainty to reflect 
the spread in the literature.

\subsubsection{Reionization and astronomy parameters}
\label{TauDiscSec}

We confirm the WMAP team \cite{Spergel03} measurement of early reionization, 
$\tau=0.12^{+ 0.08}_{- 0.06}$.
This hinges crucially 
on the WMAP polarization data; using only the unpolarized WMAP power spectrum,
our analysis prefers $\tau=0$ and gives an upper limit $\tau<0.23$ (95\%).
% ta < 0.22708 (95%) without polarization

Assuming the vanilla model, our Hubble parameter measurement $h\approx 0.70^{+0.04}_{-0.03}$
agrees well with the HST key project measurement 
$h=0.72\pm 0.07$ \cite{Freedman01}. It is marginally lower than the WMAP
team value $h\approx 0.73\pm 0.03$ because the 
SDSS power spectrum has a slightly bluer slope than that of the 2dFGRS, 
favoring slightly higher $\Om$-values
(we obtain $\Om=0.30\pm 0.04$; the WMAP team quote $\Om=0.26\pm 0.05$ \cite{Spergel03}).

\subsection{What have we not learned?}

The cosmology community has now established the existence of 
dark matter, dark energy and near-scale invariant seed fluctuations.
Yet we do not know why they exist or the physics responsible for generating them.
Indeed, it is striking that standard model physics fails to explain any of the
four ingredients of the cosmic matter budget: it gives too small CP-violation
to explain baryogenesis, does not produce dark matter particles,  
does not produce dark energy at the observed level 
and fails to explain the small yet non-zero neutrino masses.

Fortunately, upcoming measurements will provide much needed guidance for tackling these
issues: constraining 
dark matter properties (temperature, viscosity, interactions, \etc),
dark energy properties (density evolution, clustering), 
neutrino properties (with galaxy and cmb lensing potentially
sensitivity down to the experimental mass limits $\sim 0.05$ eV 
\cite{weaklensnu1,weaklensnu2,weaklensnu3})
and seed fluctuation properties (model-independent measurements of their power spectrum \cite{Bridle0302306}).

The Sloan Digital Sky Survey should be able to make important contributions to many of these
questions. Redshifts have now been measured for about 350{,}000 main-sample galaxies
and 35{,}000 luminous red galaxies, which will allow substantially tighter constraints on even
larger scales where nonlinearities are less important, as will 
analysis of three-dimensional clustering using photometric redshifts \cite{Budavari03}
with orders of magnitude more galaxies.
There is also a wealth of cosmological information to be extracted from analysis of 
higher moments of galaxy clustering, cluster abundance\cite{NBahcall03}, quasar clustering, small-scale galaxy clustering\cite{Zehavi02}, Ly$\alpha$ forest
clustering, dark matter halo properties\cite{Fischer00}, \etc, and using information this to bolster
our understanding the gastrophysics of biasing and nonlinear redshift distortions
will greatly reduce systematic uncertainties associated with galaxy surveys.
In other words, this paper should be viewed not as the final word on SDSS precision cosmology, 
merely as a 
% humble 
promising
beginning.

\bigskip

\bigskip
{\bf Acknowledgements:}
We wish to thank John Beacom, Ed Copeland, Ang{\'e}lica de Oliveira-Costa, Andrew Hamilton, Steen Hannestad, 
% Lam Hui, 
Will Kinney, Andrew Liddle, Ofelia Pisanti, Georg Raffelt, David Spergel, Masahiro Takada, Licia Verde and
Matias Zaldarriaga for helpful discussions
and Dulce de Oliveira-Costa for invaluable help.
We thank the WMAP team for producing such a superb
data set and for promptly making it public via
the Legacy Archive for Microwave Background Data 
Analysis (LAMBDA) at \protect\url{http://lambda.gsfc.nasa.gov}.
We thank John Tonry for kindly providing software evaluating the SN Ia likelihood from
\cite{Tonry03}. We thank the 2dFGRS team for making their power spectrum data
public at \protect\url{http://msowww.anu.edu.au/2dFGRS/Public/Release/PowSpec/}.

Funding for the creation and distribution of the SDSS Archive has been provided
by the Alfred P. Sloan Foundation, the Participating Institutions, the National
Aeronautics and Space Administration, the National Science Foundation, the U.S.
Department of Energy, the Japanese Monbukagakusho, and the Max Planck Society. 
The SDSS Web site is http://www.sdss.org/. 

The SDSS is managed by the Astrophysical Research Consortium (ARC) for the
Participating Institutions.  The Participating Institutions are The University
of Chicago, Fermilab, the Institute for Advanced Study, the Japan Participation
Group, The Johns Hopkins University, Los Alamos National Laboratory, the
Max-Planck-Institute for Astronomy (MPIA), the Max-Planck-Institute for
Astrophysics (MPA), New Mexico State University, University of Pittsburgh, 
Princeton University, the United States Naval Observatory, and the University
of Washington.

MT was supported by NSF grants AST-0071213 \& AST-0134999, NASA grants
NAG5-9194 \& NAG5-11099 and fellowships from the David and Lucile
Packard Foundation and the Cottrell Foundation.  MAS acknowledges
support from NSF grant AST-0307409,
and AJSH from NSF grant AST-0205981 and NASA grant NAG5-10763.

\appendix

\section{Computational issues}

In this Appendix, we briefly summarize the technical details 
of how our analysis was carried out.

\subsection{Monte Carlo Markov Chain summary}

The Monte Carlo Markov chain (MCMC) method is a well-established technique \cite{Metropolis,Hastings,Gilks96} for
constraining parameters from observed data, especially suited for the case when the parameter space
has a high dimensionality. It was recently introduced to the cosmology community by  
\cite{Christensen01} and detailed discussions of its cosmological applications can be found
in \cite{Lewis02,Slosar03,Verde03}.

The basic problem is that we have a vector of cosmological data $\d$ from which we wish to measure
a vector of cosmological parameters $\p$. For instance, $\d$ might be the 1367-dimensional vector 
consisting of the 899 WMAP measurements of the temperature power spectrum $C_\l$ for $\l=2,...,900$, 
the 449 WMAP cross-polarization measurements and the 19 SDSS $P(k)$-measurements we use.
The cosmological parameter vector $\p$ might contain the parameters of \eq{pEq} or some subset thereof. 
Theory $\p$ is connected to data $\d$ by the so-called likelihood function $\L(\p,\d)$,
which gives the probability distribution for observing different $\d$ given a theoretical model $\p$.
In Bayesian analysis, one inserts the actual observed data and reinterprets 
$\L(\p,\d)$ as an unnormalized probability distribution over the cosmological parameters $\p$, optionally after multiplication by a 
probability distribution reflecting prior information.
To place constraints on an single parameter, say $p_7$, one 
needs to marginalize (integrate) over all the others.

Two different solutions have been successfully applied to this problem. 
One is the grid approach (\eg, \cite{Lineweaver98,9par,Lange01}), evaluating $\L(\p,\d)$ on
a grid in the multidimensional parameter space and then marginalizing.
The drawback of this approach is that the number of grid points grows exponentially
with the number of parameters, which has in practice limited this method to 
about 10 parameters \cite{consistent}.
The other is the MCMC approach, where a large set of points $\p_i$,
$i=1,...n$, a {\it chain}, is generated
by a stochastic procedure such that the points have the probability 
distribution $\L(\p,\d)$. Marginalization now becomes trivial: to read off the constraints on
say the seventh parameter, one simply plots a histogram of $p_7$. 

The basic MCMC algorithm is extremely simple, requiring only about ten lines of computer code.
\begin{enumerate}
\item Given $\p_i$, generate a new trial point $\p_*=\p_i+\Delta\p$
      where the jump $\Delta\p$ is drawn from a jump probability distribution $f(\Delta\p)$.
\item Accept the jump (set $\p_{i+1}=\p_*$) or reject the jump (set $\p_{i+1}=\p_i$)
      according to the Metropolis-Hastings rule \cite{Metropolis,Hastings}:
      always accept jumps to higher likelihoods, \ie, if $\L(\p_*,\d)>\L(\p_i,\d)$,
      otherwise accept only with probability $\L(\p_*,\d)/\L(\p_i,\d)$.      
\end{enumerate}
The algorithm is therefore completely specified by 
two entities: the jump function $f(\Delta\p)$ and the likelihood function $\L(\p_*,\d)$.
We describe how we compute $f$ and $\L$ below in sections~\ref{fSec} and~\ref{Lsec},
respectively.

\begin{table*}
\label{ChainTable}
\noindent 
{\footnotesize
Table~\ChainTable: Monte Carlo Markov chains used in the chain. 
The figure of merit for a chain is the effective length 
(the actual length divided by the correlation length). 
Here we have chosen to tabulate correlation lengths for the $\tau$-parameter, since 
it is typically the largest (together with that for $\ns$ and $\ob$, because of the
banana degeneracy of \Sec{VanillaBananaSec}).
The success rate is the percentage of steps accepted. 
``Vanilla'' denotes the six parameters $(\tau,\ob,\od,\Ol,\As,\ns)$. 
In the data column, T denotes the unpolarized power spectrum, X denotes the temperature/E-polarization 
cross power spectrum, and $\tau$ denotes the prior $\tau<0.3$.
\begin{center}
\begin{tabular}{|rrll|rrrr|}
\hline
Chain	&Dim.	&Parameters			&Data			&Length		&Success	&Corr. length	&Eff. length\\
\hline
 1	&9	&Vanilla$+\Ok+r+\alpha$		&WMAP T+X		&189202		&22\% 		&218 		&868\\
 2	&7	&Vanilla$+\fn$			&WMAP T+X		&133361 	&8\% 		&78 		&1710\\
 3	&7	&Vanilla$+w$			&WMAP T+X		&352139 	&3\% 		&135 		&2608\\
 4	&7	&Vanilla$+\Ok$			&WMAP T+X		&101922 	&7\% 		&213 		&479\\
 5	&7	&Vanilla$+r$			&WMAP T+X		&178670 	&13\% 		&29 		&6161\\
 6	&6	&Vanilla			&WMAP T+X		&311391 	&16\% 		&45 		&6920\\
 7	&6	&Vanilla			&WMAP T			&298001 	&15\% 		&25 		&11920\\
 8	&5	&Vanilla$-\ns$			&WMAP T+X		&298001 	&29\% 		&7 		&42572\\
 9	&10	&Vanilla$+\Ok+r+\alpha+b$	&WMAP T+X + SDSS 	&298001 	&4\%		&69		&4319\\
10	&8	&Vanilla$+\fn+b$		&WMAP T+X + SDSS 	&46808  	&18\%		&24		&1950\\
11	&8	&Vanilla$+w+b$			&WMAP T+X + SDSS 	&298002 	&4\% 		&98 		&3041\\
12	&8	&Vanilla$+\Ok+b$		&WMAP T+X + SDSS 	&298001 	&6\%		&83		&3590\\
13	&8	&Vanilla$+r+b$			&WMAP T+X + SDSS 	&298001 	&12\%		&31		&9613\\
14	&7	&Vanilla$+b$			&WMAP T+X + SDSS 	&298001 	&16\%		&18		&16556\\
15	&7	&Vanilla$+b$			&SDSS+WMAP T		&298001 	&16\% 		&17 		&17529\\
16	&6	&Vanilla$-\ns+b$		&WMAP T+X + SDSS 	&298001 	&25\%		&8		&37250\\
17	&6	&Vanilla$-\ns+b$		&WMAP T+X + SDSS +$\phi^2$ &298001 	&25\%		&8		&37250\\
18	&8	&Vanilla$+w+b$			&WMAP T+X + SDSS + SN Ia&298001 	&12\% 		&25 		&11920\\
19	&8	&Vanilla$+r+b$			&WMAP T+X + SDSS + SN Ia&298001 	&5\% 		&89 		&3348\\
20	&8	&Vanilla$+r+b$			&WMAP T+X + SDSS + $\tau$&151045 	&6\% 		&26 		&5809\\
21	&8	&Vanilla$+r+b$			&WMAP T+X + SDSS + SN Ia + $\tau$&68590 &6\% 		&30 		&2286\\
22	&7	&Vanilla$+b$			&Other CMB + SDSS	&315875 	&30\% 		&24 		&13161\\
23	&7	&Vanilla$+b$			&WMAP + other CMB + SDSS&559330 	&20\% 		&10 		&55933\\
24	&2	&$\Om+\As$			&SDSS			&48001 		&41\% 		&6 		&8000\\
25	&2	&$\Om+\As$			&SDSS $\kmax=0.15$	&48001 		&36\% 		&6 		&8000\\
26	&2	&$\Om+\As$			&SDSS $\kmax=0.10$	&48001 		&31\% 		&9 		&5333\\
27	&2	&$\Om+\As$			&SDSS no bias corr.	&48001 		&38\% 		&7 		&6857\\
28	&2	&$\Om+\As$			&SDSS linear $P(k)$.	&48001 		&50\% 		&5 		&19600\\
29	&2	&$\Om+\As$			&2dFGRS			&48001 		&33\% 		&9 		&5333\\
\hline

\end{tabular}
\end{center}     
} 
\end{table*}

% wmappol9par 189202 22% 218 868 17.164
% wmappol7par 101922 7% 213 479 14.634
% wmappol7parr 178670 13% 29 6161 12.547
% wmappol7parnu 133361 8% 78 1710 12.269
% wmappol7parw 352139 3% 135 2608 12.508
% wmappol7parw 156753 21% 95 1650 11.349
% wmappol6par 311391 16% 45 6920 11.496
% wmap6parfit 298001 15% 25 11920 12.622
% wmappol5parTs4 298001 29% 7 42572 10.918
% wmap6parTs3 29991 16% 24 1250 12.120
% sdsswmappol9parTs4 298001 4% 69 4319 16.539
% sdsswmappol7parTs4 298001 6% 83 3590 15.101
% sdsswmappol7parrTs4 298001 12% 31 9613 14.905
% wmappolsdss8parnu 46808 18% 24 1950 14.033
% sdsswmappol7parwTs4 298002 4% 98 3041 19.140
% sdsswmappol6parTs4 299954 16% 18 16664 14.037
% sdsswmap6parTs4 298001 16% 17 17529 14.322
% sdsswmappol5parTs4 298001 25% 8 37250 12.581
% sdsswmappol5parlindeTs4 298001 25% 8 37250 12.276
% snsdsswmappol7parwTs4 298001 12% 25 11920 14.791
% snsdsswmappol7parTs4 298001 5% 89 3348 15.620
% tasdsswmappol7parTs4 151045 6% 26 5809 14.376
% tasnsdsswmappol7parTs4 68590 6% 30 2286 13.724
% cmbsdss6par 315875 30% 24 13161 13.644
% cmbsdsswmappol6parTs4 559330 20% 10 55933 13.682
% sdssnonlin2par 48001 41% 6 8000 6.210
% sdssnodebias2par 48001 38% 7 6857 6.185
% sdsslin2par 98001 50% 5 19600 5.858
% sdsskmax0.1_2par 48001 31% 9 5333 7.046
% sdsskmax0.15_2par 48001 36% 6 8000 6.184
% sdsskmax0.3_2par 48001 42% 6 8000 6.232
% 2dfthx2par 48001 30% 9 5333 6.133
% 2dfpercival2par 48001 33% 9 5333 6.287

%Include Licia chains here? Include Kev chains here?

Table~{\ChainTable} lists the chains we used and their basic properties:
dimensionality of the parameter space, parameters used, data $\d$ 
used in likelihood function, number of steps $n$ (i.e., the length of
the chain), the success rate 
(fraction of attempted jumps that were accepted according to the above-mentioned Metropolis-Hastings rule),
the correlation length (explained below) 
and the effective length.
We typically ran a test chain with about 10000 points to optimize our choice
of jump function $f$ as described in \Sec{fSec}, then 
used this jump function to run about 40 independent chains with different 
randomly generated starting
points $\p_1$. In total, this used about 30 CPU-years of Linux workstation time.
Each chain has a period of ``burn-in'' in the beginning, before it
converges to the allowed region of parameter space: 
we computed the median likelihood of all chains combined, then defined the end of the burn-in
for a given chain as the first step where its likelihood exceeded this value.
Most chains burned in within 100 steps, but a small fraction of them failed to burn in at all 
and were discarded, having started in a remote and unphysical part of parameter space and become
stuck in a local likelihood maximum.
After discarding the burn-in, we merged these independent chains to produce those listed in Table~{\ChainTable}.
This standard procedure of concatenating independent chains preserves their Markov character,
since they are completely uncorrelated with one another.

\subsection{The jump function $f$}
\label{fSec}

\begin{figure} 
\vskip\smtopskip
\centerline{\epsfxsize=\figsize\epsffile{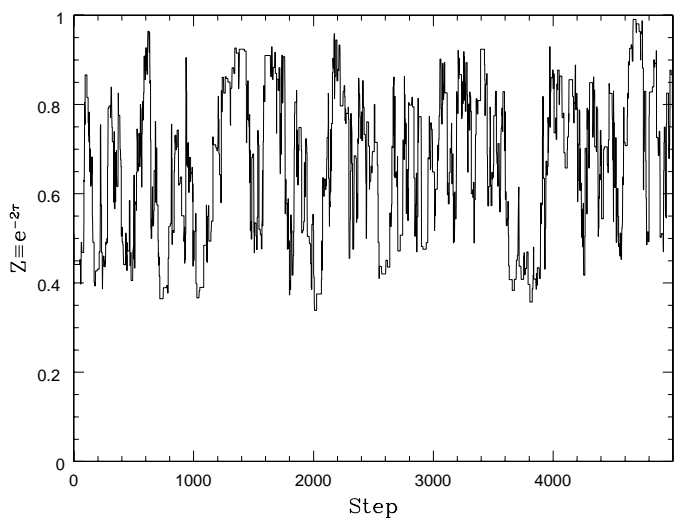}}
\vskip\smbotskip
\caption[1]{\label{timestreamFig}\footnotesize%
The reionization parameter $Z$ as a function of the MCMC step.
This example is for chain 6 from Table~{\ChainTable}.
% This example is for the chain_wmappol6par.dat
}
\end{figure}

\begin{figure} 
\vskip\smtopskip
\centerline{\epsfxsize=\figsize\epsffile{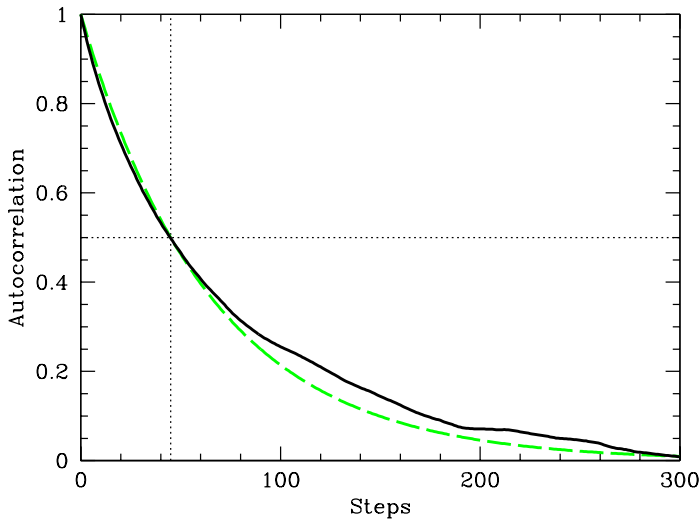}}
\vskip\smbotskip
\caption[1]{\label{correlationFig}\footnotesize%
The autocorrelation function (solid curve) 
for the example in \fig{timestreamFig} is seen to be approximately fit by
an exponential (dashed curve), dropping to 50\% 
at a correlation length of 45 steps as indicated by the dotted lines.
}
\end{figure}

As illustrated in \fig{timestreamFig}, consecutive points $\p_i$, $i=1,...$ of a MCMC are correlated.
We quantify this by the dimensionless autocorrelation function $c$, 
shown for the reionization parameter $\tau$ in \fig{correlationFig} and defined by
\beq{CorrFuncEq}
c_j \equiv {\expec{\tau_i\tau_{i+j}} - \expec{\tau_i}^2 \over \expec{\tau_i^2} - \expec{\tau_i}^2},
\eeq
where averages are over the whole chain.
The correlation is by definition unity at zero lag, 
and we define the {\it correlation length} as the number of steps over which the correlation 
drops to 0.5.
The figure of merit for a chain is its effective length $N$, defined as 
the number of steps divided by the correlation length.
% its length in units of the correlation length.
Since $N$ is roughly speaking the number of independent points, the MCMC technique
measures statistical quantities such as the standard deviation $\sigma_p$ and
the mean $\expec{p}$ for cosmological parameters 
to an accuracy of order $\sigma_p/N^{1/2}$. Unless $N\gg 1$,
the results are useless and misleading, a problem referred to as insufficient mixing in
the MCMC literature \cite{Gilks96}.

We attempt to minimize the correlation length by 
tailoring the jump function to the structure of the likelihood function.
Consider first a toy model with a one-dimensional parameter space and 
a Gaussian likelihood $\L(p)\propto e^{-p^2/2}$ and 
a Gaussian jump function $f(\Delta p)\propto e^{-\Delta p^2/2\sigma^2}$.
What is the best choice of the characteristic jump size $\sigma$?
In the limit $\sigma\to\infty$, all jumps will fail; $p_1=p_2=...$,
$c_j=1$ for all $j$ and the correlation length becomes infinite.
In the opposite limit $\sigma\to 0$, almost all steps succeed and we obtain Brownian motion
with the rms value $|p_i|\sim\sigma i^{1/2}$, so it takes of order 
$\sigma^{-2}\to\infty$ steps to wander from one half of the distribution to the other, again 
giving infinite correlation length. This implies that there must be an optimal jump size
between these extremes, and numerical experimentation shows that $\sigma\sim 1$ minimizes the correlation
length.

In the multiparameter case, strong degeneracies can cause a huge correlation length if the jump size is chosen
independently for each parameter, with the chain taking a very long time to wander from one end of
the banana to the other. A clever choice of parameters that reduces degeneracies
therefore reduces the correlation length. For this reason, 
the WMAP team used the parameters suggested by \cite{Kosowsky02},
and we do the same with the minor improvement of replacing $\As$ by $\Ap$ as the 
scalar normalization parameter as in \cite{Knox03}\footnote{
When imposing a flatness prior $\Ok=0$, we retained $\Th$ as a free parameter and dropped $\Ol$.
When additionally imposing a prior on $h$ (for the last 6 chains in Table~{\ChainTable}),
we dropped both $\Th$ and $\Ol$ as free parameters, setting $\Ol=1-(\od+\ob)/h^2$.
}.
To minimize the remaining degeneracies, we compute the
parameter covariance matrix
$\C\equiv\expec{\p\p^t}-\expec{\p}\expec{\p}^t$ from the chain itself, diagonalize
it as $\C=\R\LL\R^t$, $\R\R^t=\R^t\R=\I$, $\LL_{ij}=\delta_{ij}\lambda_i^2$, 
and work with the transformed parameter vector $\p'\equiv \LL^{-1/2}\R^t[\p-\expec{\p}]$ which 
has the benign properties $\expec{\p'}=\zero$, $\expec{\p'\p'^t}=\I$. 
Inspired by the above-mentioned one-dimensional example, we then use the simple
jump function $f(\Delta\p')\propto e^{-|\Delta\p'|^2/2\sigma^2}$.
We use $\sigma=1$ for all chains except number 1 in Table~{\ChainTable}, 
% That's wmappol9par
where we obtain a shorter correlation length using $\sigma=0.7$.
When running our test chains to optimize $f$, we start by guessing a diagonal $\C$
and after the burn-in, we update our estimates of both $\C$ and the 
eigenbasis every 100 steps.
%Essentially the same 
A very similar 
approach is used in other recent MCMC codes, \eg, 
\cite{Lewis02,ZahnZalda03}.

The WMAP team perform extensive testing to confirm that their chains are properly 
mixed \cite{Verde03}, and we have followed the WMAP team in using the 
Gelman and Rubin $R$-statistic \cite{GelmanRubin92} to verify that our chains are sufficiently 
converged and mixed. 
Indeed, we find that the above-mentioned eigenbasis technique helps further improve
the mixing by cutting our correlation length by 
about an order of magnitude relative to that obtained with the WMAP jump function,
hence greatly increasing the effective length of our chains.

\subsection{The likelihood function $\L$}
\label{Lsec}

For a detailed discussion of how to compute cosmological likelihood functions,
see \cite{BJK,Verde03}. Our calculation of $\L(\p,\d)$ does little more 
than combine public software described in other papers, so the details in
this brief section are merely of interest for the reader interested in 
exactly reproducing our results.

The total likelihood $\L$ is simply a product of likelihoods corresponding 
to the data sets used, \eg, WMAP, SDSS and SN Ia.
For the CMB, we compute theoretical power spectra using 
version 4.3 of CMBfast \cite{cmbfast}, with both the
``RECfast'' and ``PRECISION'' options turned on.
We compute the WMAP likelihood corresponding to these spectra
using the public software provided by the WMAP team \cite{Verde03}.
Since this software is designed for physically reasonable models,
not for crazy models that may occur during our burn-in,
we augment it to produce large negative likelihoods for unphysical models where it
would otherwise give negative $\chi^2$-values.
For some of the WMAP+SDSS chains, we evaluate the WMAP likelihood $\L_{\rm WMAP}$
by fitting a quartic polynomial to $\ln \L_{\rm WMAP}$ 
from the corresponding WMAP-only chains. 
For this fit, we replace $\od$ by $\om$, $\ob$ by $H_2$, $\ns$ by $H_3$
and $\Ap$ by $\Apivot$ inspired by the normal parameter method of \cite{Knox03}.
This approach, described in detail in \cite{cmbfit}, is merely a numerical 
tool for accelerating the computations, and we 
verify that it has negligible impact on our results.

\begin{figure} 
\vskip\smtopskip
\bigskip\bigskip\smallskip
\centerline{\epsfxsize=\figsize\epsffile{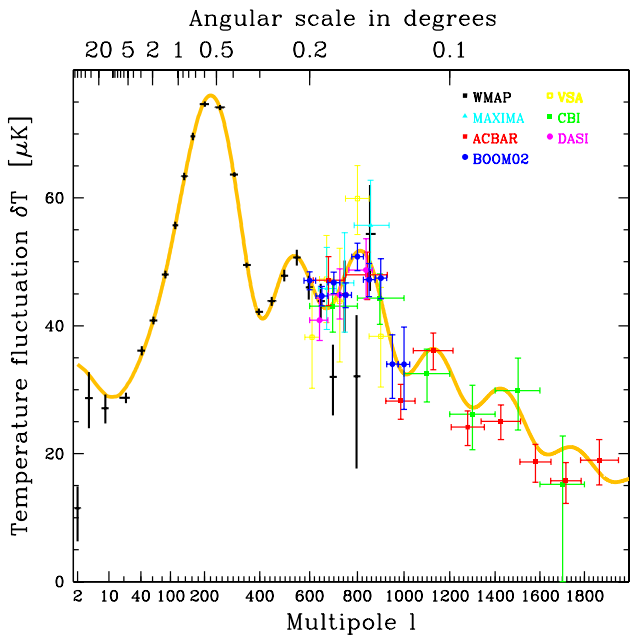}}
\vskip\smbotskip
\caption[1]{\label{cmbdataFig}\footnotesize%
CMB data used. Error bars show do not include the calibration and beam uncertainties that we include
as described in the text. Solid curve corresponds to the ``vanilla lite'' model of \eq{5parEq}.
}
\end{figure}

When combining non-WMAP CMB data with WMAP, we include the latest band-power detections from
Boomerang \cite{Ruhl02} (madcap), DASI \cite{Halverson01}, MAXIMA \cite{Lee01}, VSA \cite{VSA02},
CBI (mosaic, even binning) \cite{Sievers02} and ACBAR \cite{Kuo02} with probing effective multipoles
$\l\ge 600$ (where they are collectively more sensitive than WMAP) 
and $\l\le 2000$ (to avoid complications related to reported small-scale
excess, which may be due to secondary anisotropies or non-CMB effects), 
which corresponds to the $9+3+3+4+6+9=34$ data points plotted in \fig{cmbdataFig}.
The pre-WMAP data has been shown to be consistent both internally and
with WMAP \cite{GriffithsLineweaver03}.
We marginalize over the quoted calibration uncertainties of 
10\% for Boomerang, 4\% for MAXIMA and DASI, 5\% for CBI, 3.5\% for VSA and 10\% for ACBAR
as well as over quoted beam uncertainties of 15\% for Boomerang, 5\% for ACBAR and 14\% for MAXIMA
(this last number provides a good fit to the combined beam and pointing uncertainties 
for the three measurements used from Table 1 of \cite{Lee01}).
We make the approximation that all experiments are uncorrelated with each other and with
WMAP, which should be quite accurate both since sample variance correlations are 
negligible (given their small sky coverage relative to WMAP) and 
since the wmap errors are dominated by detector noise for $\l\ge 600$.
When using non-CMB data without WMAP, we use all 151 pre-WMAP band power
measurements compiled in \cite{cmblsslens}. For the non-WMAP data, we computed the CMB power spectra
with the DASh package \cite{DASh}.
% Mention windows?

For SDSS, we compute the likelihood by fitting $b^2$ times the nonlinear power spectrum $P(k)$
to the first 19 band power measurements 
(for $k<\kmax\sim 0.2h/$Mpc) using the window functions and likelihood
software provided by that paper \cite{sdsspower}. For more details about the SDSS data, 
see \cite{BlantonTiling,EisensteinLRGselection,FukugitaFilters,GunnCamera,HoggPt,PierAstrometry,SmithPhotometry,StraussTargetSelection}.
We compute the nonlinear $P(k)$ using the method and software provided by \cite{Smith03}.
This software takes the linear power spectrum $P(k)$ as input, and we compute it
using the fitting software provided by \cite{Novosyadlyj99} for the transfer function,
the approximation of \cite{Carroll92} for the linear growth factor
and the approximation that $P(k)$ is as a product of these
two quantities as per equation (C3) from \cite{concordance}.
For $k\simgt 0.2$, this typically agrees with CMBfast 4.3 to better than a few percent.
% Don't bother showing Pcomparison_generic.ps,
% Pcomparison_lmdm.ps, etc.
In the absence of massive neutrinos ($\fn=0$), the separability approximation becomes exact
and the transfer function fits \cite{Novosyadlyj99} become 
identical to those of Eisenstein \& Hu \cite{Eisenstein99}.
%The reason that we cannot use CMBfast is that the nonlinear $P(k)$ package \cite{Smith03}.
%requires linear $P(k)$ out to 

For SN Ia, we use the 172 redshifts and corrected peak magnitudes compiled and uniformly 
analyzed by Tonry {\etal} \cite{Tonry03} and compute the likelihood with software kindly
provided by John Tonry. This likelihood depends only on 
$(\Ol,\Ok,w)$, and is marginalized over the corrected SN Ia ``standard candle'' 
absolute magnitude.

\subsection{Confidence limits and likelihood plots}

\begin{figure} 
%\vskip\smtopskip
\centerline{\epsfxsize=\figsize\epsffile{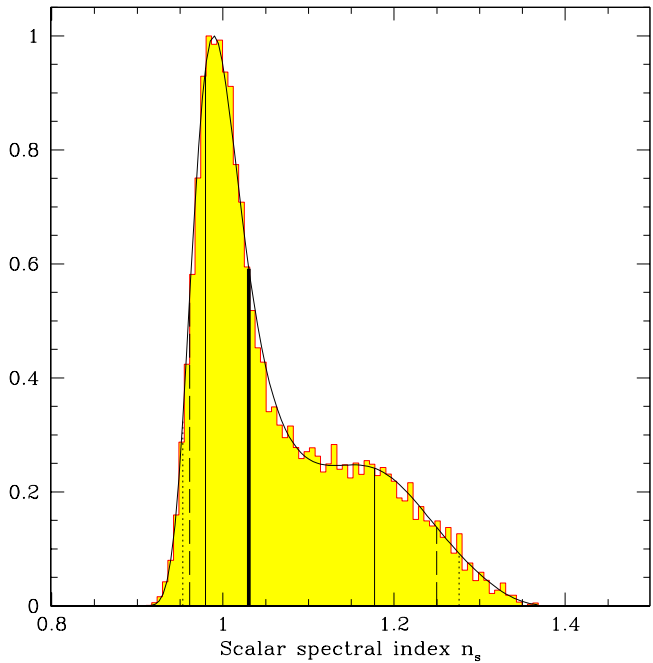}}
%\vskip\smbotskip
\caption[1]{\label{1dfit_example_fig}\footnotesize%
Example of the likelihood fitting technique we use for plotting.
The shaded histogram shows the distribution of the 311391 $\ns$-values 
from chain 6 in Table~{\ChainTable}.. 
The black curve shows our fit $e^{P(\ns)}$ for the 6th order $P(\ns)$
maximizing the Poisson likelihood as described in the text.
The vertical lines show the quantiles of the distribution that we use to 
quote confidence intervals: the median (heavy line), the central 
$68\%$ (between thin solid lines), the central 90\% (between dashed lines)
and the central 95\% (between dotted lines) of the distribution.
}
\end{figure}

All confidence limits quoted in the tables and text of this paper are quantiles,
as illustrated in \fig{1dfit_example_fig}.
For instance, our statement in Table~{\SDSStable} that
$\Mnu <1.74$eV at 95\% confidence simply means that 
95\% of the $\Mnu$-values in that chain are smaller than $1.74$eV.
Similarly, the entry
$\ns=0.972^{+0.041}_{-0.027}$
in the same column means that 
the distribution of $\ns$-values has median $0.972$, that
erfc$(2^{-1/2})/2\approx 15.87\%$ 
of the values lie below $0.972-0.027$
and that
15.87\% 
of the values lie above $0.972+0.041$, so that 
68.27\%
lie in the range $\ns=0.972^{+0.041}_{-0.027}$.
There is thus no assumption about the distributions being Gaussian.
In a handful of cases involving $r$, $\fn$ and $\Mnu$, the distribution
(see \fig{1d_6par_fig} for examples) peaks at zero rather than near the median;
in such cases, we simply quote an upper limit. 

When plotting 1-dimensional distributions $f(p)$ in \fig{1d_6par_fig}, we fit 
each histogram to a smooth function of the form $f(p)=e^{P(p)}$
where $P(p)$ is the 6th order polynomial that maximizes the likelihood
%\beq{Poission}
$\prod_{i=1}^n f(p_i)$
%\eeq
that the points $p_1,...,p_n$ in the chain are drawn from the distribution $f(p)$,
subject to the constraint that $\int f(p)dp=1$.
We found that these smooth curves visually match the raw histograms extremely well
(see \fig{1dfit_example_fig} for a rather non-Gaussian example)
and have the advantage of avoiding both the Poisson jaggedness and the excessive 
smoothing inherent in a histogram.

Our 2-dimensional contours are plotted where the point density has dropped by
$e^{-\Delta\chi^2/2}$ from its maximum, where
$\Delta\chi^2=6.18$ as recommended in \S 15.6 of \cite{NumRecipes}.
These contours would enclose 95\% 
of the points if the distribution were Gaussian.
When computing the point density, there is tradeoff between insufficient smoothing
(giving contours dominated by Poisson noise) and
excessive smoothing (which artificially broadens the contours, particularly in the narrow
direction of a degeneracy banana).
We found that this was alleviated by computing the contours in the linearly transformed 2-dimensional
space defined in \Sec{fSec} where the covariance matrix is the identity matrix.

%%%%%%%%%%%%%%%%%%%%%%%%%%%%%%%%%%%%%%%%%%%%%%%%%%%%%%%%%%%%
%%%%%%%%%%%%%%%%%%%%%% REFERENCES: %%%%%%%%%%%%%%%%%%%%%%%%%

%\clearpage
%\end{multicols}

%\vskip-1.0cm

\end{document}